\documentclass[conference]{IEEEtran}
\ifCLASSINFOpdf
\else
\fi
\usepackage{amsmath,amsthm,amsfonts,amssymb,bm}
\usepackage[table,xcdraw]{xcolor}
\usepackage{diagbox}
\usepackage{graphicx,marvosym}
\usepackage{xspace}
\usepackage{pifont}
\usepackage{multirow}
\usepackage{subcaption}
\usepackage{comment}
\usepackage{amsmath}

\usepackage{booktabs}
\newcommand{\cmark}{\ding{51}}%
\newcommand{\xmark}{\ding{55}}%

\let\labelindent\relax
\usepackage{enumitem}
\setlist{leftmargin=4mm}

\usepackage{tcolorbox}

\makeatletter
  \newcommand\figcaption{\def\@captype{figure}\caption}
  \newcommand\tabcaption{\def\@captype{table}\caption}
\makeatother

\newcommand{\attackname}{\textsc{SLMIA-SR}\xspace}
\newcommand{\simi}{{\tt sim}}
\newcommand{\stat}{{\tt stat}}

\usepackage{verbatim}
\usepackage{hyperref}
\usepackage{balance}

\usepackage{colortbl}

\usepackage{flexisym}
\usepackage{wrapfig,lipsum}

\usepackage{fancyhdr}
\newcommand{\distance}{2mm}
\usepackage{cleveref}
\crefname{section}{\S}{\S}
\Crefname{section}{\S}{\S}

\usepackage{caption}
\usepackage{subcaption}

\usepackage[ruled,linesnumbered,vlined,noend]{algorithm2e}
\makeatletter
\newcommand{\removelatexerror}{\let\@latex@error\@gobble}
\makeatother
\usepackage[noend]{algpseudocode}

\usepackage{makecell}

\usepackage[flushleft]{threeparttable}

\begin{document}
%
\title{\attackname: Speaker-Level Membership Inference Attacks against Speaker Recognition Systems\vspace*{-10mm}}

\author{\IEEEauthorblockN{Guangke Chen}
\IEEEauthorblockA{ShanghaiTech University\\chengk@shanghaitech.edu.cn
}
\and
\IEEEauthorblockN{Yedi Zhang}
\IEEEauthorblockA{National University of Singapore \\ yd.zhang@nus.edu.sg}
\and
\IEEEauthorblockN{Fu Song\textsuperscript{\Letter}}
\IEEEauthorblockA{Institute of Software, Chinese Academy of Sciences \\ University of Chinese Academy of Sciences \\ songfu@ios.ac.cn}}


%


\IEEEoverridecommandlockouts
\makeatletter\def\@IEEEpubidpullup{6.5\baselineskip}\makeatother
\IEEEpubid{\parbox{\columnwidth}{
    Network and Distributed System Security (NDSS) Symposium 2024\\
    26 February - 1 March 2024, San Diego, CA, USA\\
    ISBN 1-891562-93-2\\
    https://dx.doi.org/10.14722/ndss.2024.241323\\
    www.ndss-symposium.org
}
\hspace{\columnsep}\makebox[\columnwidth]{}}

\maketitle

 


\newcommand{\nathaniel}[1]{{\leavevmode\color{blue}[#1]}}

\begin{abstract}

    Membership inference attacks allow adversaries to determine whether
    a particular example was contained in the model's training dataset.
    While previous works have confirmed the feasibility of such attacks in various applications,
    none has focused on speaker recognition (SR), a promising voice-based biometric recognition technique.
    In this work, we propose \attackname,
    the first membership inference attack tailored to SR.
    In contrast to conventional \emph{example}-level attack,
    our attack features \emph{speaker}-level membership inference, i.e., determining if \emph{any} voices of a given speaker,
    either the same as or different from the given inference voices, have been involved in the training of a  model.
    It is particularly useful and practical since the training and inference voices are usually distinct,
    and it is also meaningful considering the open-set nature of SR, namely, the recognition speakers were often not present in the training data.
    We utilize
    intra-similarity and inter-dissimilarity,
    two training objectives of SR,
    to characterize the differences between training and non-training speakers
    and quantify them with two groups of features driven by carefully-established feature engineering to mount the attack.
    To improve the generalizability of our attack,
    we propose a novel mixing ratio training strategy to train attack models.
    To enhance the attack performance,
    we introduce voice chunk splitting to cope with the limited number of inference voices
    and propose to train attack models dependent on the number of inference voices.
    Our attack is versatile and can work in both white-box and black-box scenarios.
    Additionally, we propose two novel techniques to reduce the number of black-box queries while maintaining the attack performance.
    Extensive experiments demonstrate the effectiveness of \attackname.
\end{abstract}

\section{Introduction}\label{sec:intro}

Speaker recognition (SR) is a promising biometric recognition technique that recognizes the identity of a person based on her/his voices~\cite{MASBZ23}.
SR is already deployed in a wide variety of realistic applications, such as identity verification of bank customers
during telephone-communication~\cite{Citi-Bank},
password-free and voice-based payment~\cite{TMall-Genie-payment},
device access control in smart home~\cite{RenSYS16},
and service personalization in voice assistants~\cite{amazon-alexa-voiceid}.

Modern and state-of-the-art speaker recognition systems (SRSs) are often built upon deep neural networks (DNNs)~\cite{sr-overview},
which are trained on increasingly sensitive data.
However,
it is known that DNNs tend to memorize their training data because of overfitting~\cite{model-inversion,inversion-speaker,data-extraction-LLM}.
As a result,
an adversary given access to trained DNNs is able to
recover representative views of a subset of training data~\cite{model-inversion,inversion-speaker}
or even reconstruct verbatim training data~\cite{data-extraction-LLM}, leading to privacy leakage.
Thus, it is crucial to evaluate the privacy risks of DNNs
prior to deployment so that actions can be taken to enhance the privacy level based on the evaluation results.
Nowadays, membership inference attacks (MIA) that are able to determine if a sample is involved in the training of a DNN 
are the de facto standard for assessing DNNs' privacy risks~\cite{membership-first,first-principle,loss-trajectory,MIA-Multi-Exit-Networks}.

While numerous studies have confirmed the feasibility of MIA in various applications of DNNs,
including image classification~\cite{membership-first, membership-survey, enhanced-membership, first-principle,
evaluation-membership, loss-trajectory,MIA-Semi}, speech recognition~\cite{speech-membership-2, speech-membership-1}, language models~\cite{MIA-LM},
and generative models~\cite{membership-VAE, membership-GAN},
MIA against SRSs has not been considered yet.
Considering the wide spread of voices across social media platforms,
online meetings, and voice-enabled smart devices,
users' voice data may be collected
and used for training SRSs without their consent.
For instance,  Amazon was sued as Alexa recorded children's voiceprint
without their parents' permission~\cite{amazon-sued},
probably used to improve its Voice ID~\cite{amazon-alexa-voiceid}
that recognizes users' identity to provide personalized services.
It violates data protection regulations, e.g., GDPR~\cite{GDPR}.
Hence, ordinary users are increasingly eager to know if their voices
were used for training SRSs without their permission.
There are also regulations in place regarding artificial intelligence (AI) systems,
e.g., the Blueprint for an AI Bill of Rights introduced by the White House~\cite{AI-Bill-of-Rights}.
This regulation requires companies to continuously inspect their systems to mitigate any unsafe outcomes that exceed their intended use.
Given that many companies offer speaker recognition as a machine learning as a service (MLaaS),
e.g., Microsoft~\cite{microsoft-azure-vpr} and Nuance~\cite{Nuance-VPR},
they need to 
evaluate the privacy level of their SRSs before making them publicly available.
This evaluation is necessary to prevent queries to their systems from potentially disclosing sensitive information about the training data.
These urge us to design MIA against SRSs.
Also, understanding such attacks benefits further studies
towards building more secure and privacy-preserving SRSs.

We first study the applicability of prior MIA to SRSs, which
targeted conventional classification\footnote{
    By conventional classification, we refer to the supervised learning paradigm
    that defines a set of target classes
    and directly trains a model to recognize them using labeled examples,
    e.g., MNIST. Fine-tuning after pretraining
    and few-shot learning-based facial recognition are out of the scope.},
generative, regression, or embedding models (cf.~\cite{membership-survey} for a  survey).
We find that: (1) The first three own distinct training paradigms and architectures from SRSs,
so their MIA cannot be easily ported to SRSs.
Take conventional classification models as examples
on which most prior MIA focused~\cite{membership-first, membership-survey, enhanced-membership, first-principle}.
They are typically trained by minimizing the cross entropy loss on training samples,
which requires appending a final fully connected layer to models.
Thus, the outputs of the final layer are utilized to mount MIA~\cite{membership-first, membership-survey, enhanced-membership, first-principle,
evaluation-membership, loss-trajectory}.
In contrast, the training of SR models
aimed at deriving a voice embedding extractor
utilizes either verification- or classification-based loss functions~\cite{sr-overview}.
Verification-based loss functions (e.g.,
angular prototypical loss~\cite{prototypical}
and generalized end-to-end loss~\cite{GE2E})
are computed on voice embeddings,
and when minimized, the embeddings of two voices are close to each other (called  \emph{intra-similarity})
if they are uttered by the same speaker, otherwise far from each other (called  \emph{inter-dissimilarity}). 
SR models trained in this paradigm do not have a final fully connected layer.
Though SRSs may be trained by classification-based loss (e.g., the cross entropy loss)
in a similar paradigm as conventional classification,
the final fully connected layer will be dropped after training due to the open-set nature,
i.e., recognized speakers are enrolled speakers instead of training speakers.
In contrast, conventional classification is of closed-set nature, as
classes to be predicted are predefined and should be involved in training.
(2) When prior MIA targeting embedding models~\cite{person-re-identification-MIA,EncoderMI,self-supervised-speech-membership,FACE-AUDITOR} are applied to speaker recognition,
they lead to unsatisfactory performance,
e.g., no more than 2\% True Positive Rate (TPR) at 0.1\% False Positive Rate (FPR),
because these MIA are not tailored to speaker recognition.
Details refer to Appendix~\ref{sec:oursvspriorMIA}.

Motivated by the above study results, in this work, we propose and design
the \emph{first} membership inference attack tailored to speaker recognition.
Our attack is \emph{speaker-level} 
while most prior MIA are \emph{example}-level,
namely, determining if a given sample was contained in the training dataset or not.
In the context of speaker recognition, example-level MIA is indeed
a \emph{voice}-level MIA, which becomes less practical, less useful, and thus less interesting.
We argue that in the real world, it is of low probability that the voices of a training speaker
provided for membership inference were used in training,
hence it is more important to determine if \emph{any} voices of a given speaker,
either the same as or different from the voices provided for membership inference were used in training.

To perform speaker-level MIA against SRSs, we build an attack model
that takes as input a few voices of a target speaker and outputs
a binary decision indicating training (i.e., member) or non-training (i.e., non-member) speakers,
by leverage the widely-adopted shadow training method~\cite{membership-first, membership-survey, enhanced-membership, first-principle},
i.e., training a shadow SRS to approximate the behavior of the target SRS.
To train the attack model in a supervised manner,
we need to address the following technical challenge:

\smallskip \noindent CH-1: {\it How to comprehensively characterize the differences between training and non-training speakers?}

Observing that the training objectives of SRSs
are \emph{intra-similarity} and \emph{inter-dissimilarity},
we hypothesize that training speakers enjoy better intra-similarity
and inter-dissimilarity than non-training speakers.
We then design features to quantify intra-similarity and inter-dissimilarity
by using two types of similarities, four types of distances,
four different statistics, and different arrangements of similarities and distances.
Through this carefully-established feature engineering process,
we totally design a set of 103 features,
which are expected to comprehensively characterize the differences
between training and non-training speakers in a complementary way. 
Compared with previous MIAs against embedding models, except for the concurrent work~\cite{FACE-AUDITOR}, they only design a small set of features to characterize the intra-similarity.

Compared to example-level MIA, speaker-level MIA faces the following technical challenges:

\smallskip \noindent CH-2: {\it How to ensure reliably reaching the ``member'' decision regardless of the number of voices provided for MIA and the ratio of provided voices that are used in training?}

Regarding the number of inference voices, 
to improve the attack performance, we propose to train voice-number-dependent attack models
and propose a voice chunk splitting approach to artificially increase the number of inference voices. To realize the generalizability over different ratios, we split the voices of the shadow SRS's training speakers and propose the mixing ratio training strategy to train the attack model.

\smallskip \noindent CH-3: {\it How to reduce the number of black-box queries probed to the target SRS, given that the query budget may be limited?}

Compared with previous MIAs against embedding models, we are the first considering this aspect by proposing two novel techniques, namely, group enrollment and enrollment voice concatenation, with negligible impact on attack effectiveness.

We implement our approach in a tool, called \attackname,
and thoroughly evaluate its performance on three voice datasets and six SRSs including a commercial SRS under two settings regarding the inference voices. 
\attackname can achieve an average TPR of 10.2\% at an extremely low FPR of 0.1\%
when there are only ten inference voices and none of them were used in training for training speakers.
\attackname also outperforms previous MIA targeting embedding models,
e.g., increasing the TPR at 0.2\% FPR from 4.8\% to 46.7\%.
We also conduct experiments to confirm the effectiveness of the approches
to improve the attack performance and generalizability, and reduce the number of queries.
For instance, our voice chunk splitting can boost the TPR at 0.1\% FPR by 13\%,
and group enrollment and enrollment voice concatenation can reduce the queries
from 400 to 30 without sacrificing accuracy.
Finally, we perform ablation studies to study the effect
of dataset distribution and architecture shift on \attackname.
Unsurprisingly,  performance may degrade,
but \attackname
still remains effective with more than 2\% TPR at 0.1\% FPR in the worst case.

To summarize, we make the following major contributions:
\begin{itemize}
 \item We propose the first speaker-level membership inference attack
 \attackname for auditing privacy risks of speaker recognition systems.

\item Through carefully-established feature engineering,
we design 103 diverse features
to quantify intra-similarity and inter-dissimilarity, and characterize
the differences between training and non-training speakers in a comprehensive manner.

\item We propose a mixing ratio training strategy to improve the generalizability, 
enabling \attackname to determine if any voices of a speaker were used in training
regardless of the ratio of provided inference voices that were used in training.

\item To enhance the attack performance, we propose to build  voice-number-dependent attack models
and propose a voice chunk splitting approach  to cope with the limited number of inference voices.

    \item We propose two techniques, group enrollment and enrollment voice concatenation,
    to significantly reduce the number of queries probed to the target SRS in the black-box scenario,
    with no or little impact on the attack performance.
\end{itemize}

For convenient reference, we summarize the main notations used in this work in \tablename~\ref{tab:notations}. 
Our code is available at \cite{Code-of-SLMIA-SR}.

\section{Background \& Related Works}

\subsection{Speaker Recognition Systems}\label{sec:srs}
\begin{figure}[t]
    \centering
    \includegraphics[width=0.4\textwidth]{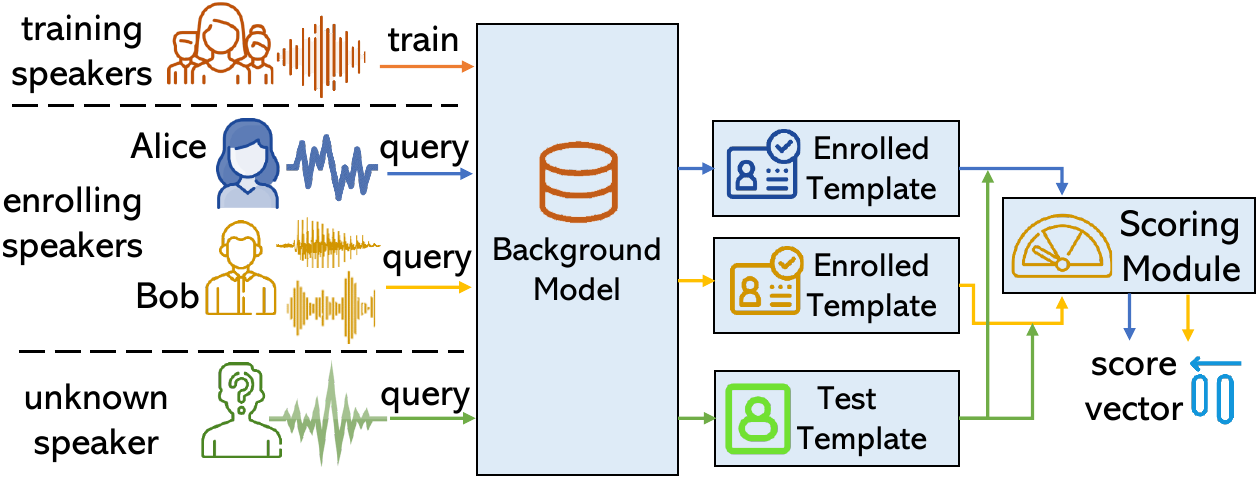}
    \caption{Overview of SRSs.}
    \label{fig:srs}\vspace*{-2mm}
\end{figure}

The overview of generic speaker recognition systems (SRSs) is shown in \figurename~\ref{fig:srs},
comprising three phases: \emph{training}, \emph{enrollment}, and \emph{recognition}.
The training phase trains a background model using lots of voices from numerous training speakers~\cite{sr-overview}
by minimizing a either classification- or verification-based loss function~\cite{sr-overview}.
The background model learns a mapping $E(\cdot)$ from voices $v$ to embeddings $E(v)$
such that the voice embeddings of the same speaker are pulled together,
while the voice embeddings of distinct speakers are pushed away.
Classification-based losses, e.g., cross entropy (CE) loss~\cite{sr-overview},
aim to maximize the classification accuracy of the training speakers
and require appending a fully connected layer such that the number of neurons is the same as the number of training speakers.
But, different from conventional classification,
the fully connected layer will be dropped after training and not used in later phases,
since speakers in later phases are not necessarily involved in the training phase.
Verification-based losses, e.g., angular prototypical (AP) loss~\cite{prototypical}
and generalized end-to-end loss (GE2E)~\cite{GE2E}, aim to minimize
the Equal Error Rate (EER) of the training trials\footnote{A trail is a pair of voices
uttered by the same or distinct speakers.
When the ground truth is the same (resp. distinct), but SRS concludes the opposite,
SRS commits false rejection (resp. false acceptance).
EER is the case where the false rejection rate is equal to the false acceptance rate on all trials.}.
As such losses are directly computed on the voice embeddings, the fully connected layer is not needed.
Typically, verification-based losses explicitly penalizes
the distance among voice embeddings of the same speaker
and the similarity between voice embeddings of distinct speakers.
In contrast, cross entropy loss only explicitly separates different speakers,
and there are some variants of cross entropy loss
by explicitly regularizing the distance among voice embeddings of the same speaker,
e.g., additive angular margin softmax loss (AAM)~\cite{AAM-loss-2}.

\begin{table}[t]
    \centering 
    \caption{Main Notations.} \setlength\tabcolsep{3pt}  \vspace*{-1mm}
  \scalebox{0.88}{ 
    \begin{tabular}[htp]{c|c|c|c|c|c}
         \hline
         {\bf Category}  & {\bf Notation} & {\bf Meaning}  & {\bf Category} & {\bf Notation} & {\bf Meaning} \\
         \hline
         \multirow{6}[8]{*}{\bf Target} & $SR^t$ & \makecell[c]{target SRS} & \multirow{6}[8]{*}{\bf Shadow} & $SR^s$ & \makecell[c]{shadow SRS} \\
       \cline{2-3}\cline{5-6}
        & $\mathcal{S}^{t}_{tr}$ & \makecell[c]{training speakers} & & $\mathcal{S}^{s}_{tr}$ & \makecell[c]{training speakers} \\
        \cline{2-3}\cline{5-6}
        & $\mathcal{V}^{t}_{tr}$ & \makecell[c]{training voices} & & $\mathcal{V}^{s}_{tr}$ & \makecell[c]{training voices} \\
        \cline{2-3}\cline{5-6}
        & $\mathcal{V}^{t}_{ntr,tr}$ & \makecell[c]{non-training  voices \\ of training speakers} & & $\mathcal{V}^{s}_{ntr,tr}$ & \makecell[c]{non-training voices\\of training speakers} \\
        \cline{2-3}\cline{5-6}
        & $\mathcal{S}^{t}_{ntr}$ & \makecell[c]{non-training\\speakers} & & $\mathcal{S}^{s}_{ntr}$ & \makecell[c]{non-training\\speakers} \\
        \cline{2-3}\cline{5-6}
        & $\mathcal{V}^{t}_{ntr,ntr}$ & \makecell[c]{non-training voices\\ of  non-training\\ speakers} & & $\mathcal{V}^{s}_{ntr,ntr}$ & \makecell[c]{non-training voices  \\ of non-training \\ speakers} \\
         \hline
        \multirow{2}{*}{\bf Imposter} & $\mathcal{S}^{im}$ & \makecell[c]{imposters} & \multirow{2}{*}{\bf Auxiliary} & $\mathcal{S}^a=\mathcal{S}^{s}_{tr} \cup$ & \multirow{2}{*}{\makecell[c]{auxiliary set \\ of speakers}} \\
        \cline{2-3}
        & $\mathcal{V}^{im}$ &  imposters voices & & $\mathcal{S}^{s}_{ntr} \cup \mathcal{S}^{im}$ & \\
        \hline
        \multirow{3}[6]{*}{\bf Approach} & VCS & \makecell[c]{voice chunk\\ splitting} & \multirow{3}[6]{*}{\bf Approach} & Intra-Ens & \makecell[c]{ensemble of \\ intra-features} \\
        \cline{2-3}\cline{5-6}
        & VND & \makecell[c]{voice-number\\-dependent \\ attack models} & & Inter-Ens & \makecell[c]{ensemble of \\ inter-features} \\
        \cline{2-3}\cline{5-6}
        & VNID & \makecell[c]{voice-number\\-independent \\ attack models} & & \attackname & \makecell[c]{ensemble of \\ intra- and inter-\\ features} \\
        \hline
    \end{tabular}
    }
    \label{tab:notations}
\end{table}

In the enrollment phase, an \emph{enrolled template} $E(\mathbf{v}^e)$
is registered using the enrollment voice(s) $\mathbf{v}^e$ of an enrolling speaker.
Note that the same speaker may be enrolled repeatedly
to better characterize the speaker,
so $\mathbf{v}^e$ can be multiple voices.
When $|\mathbf{v}^e|>1$, $E(\mathbf{v}^e)$ is the centroid of the embeddings $E(v)$ for $v\in{\bf v}$.
An SRS may allow only one speaker (speaker verification)
or multiple speakers (speaker identification) to enroll,
leading to one or multiple enrolled templates, respectively.
In the recognition phase, the \emph{test template} $E(v)$ of a given voice $v$ is first retrieved,
and then the scoring module measures the similarity
between each enrolled template $E(\mathbf{v}^e)$ and the test template $E(v)$,
producing a (recognition) score $S(v|\mathbf{v}^e)$.

\subsection{Security of Speaker Recognition Systems}\label{sec:adver-attack}
Various security implications of SRSs have been unveiled.
Audio adversarial example attacks~\cite{QFA2SR, chen2019real, AS2T, SpeakerGuard, li2020adversarial,
du2020sirenattack, Kenansville, Occam, LiW00020} craft an adversarial voice
from a voice uttered by a \emph{source} speaker
such that the SRS misclassifies it as a \emph{target} speaker,
but ordinary users do not.
Hidden voice attacks~\cite{AbdullahGPTBW19} perturb a voice uttered by a target speaker
such that the resulting voice is perceived as mere noise by humans,
but is still correctly classified as the target speaker by the SRS.
Audio deepfake attacks~\cite{wu2015spoofing},
including speech synthesis~\cite{wenger2021hello} and voice conversion~\cite{wenger2021hello},
create a voice such that it is recognized as the target speaker by both SRSs and humans.
Dictionary attacks~\cite{DA-attack-1} create a master voice that matches the identity of a large population
instead of one specific target speaker.
All these attacks are aimed at bypassing the authentication of target speakers using crafted voices.
In contrast,
this work reveals the privacy implication of SRSs.
We show that the adversary can infer whether any voices of a given speaker were contained in the training of an SRS
by querying the SRS and leveraging the feedback,
thus obtaining extra information about the training speakers.

\subsection{Membership Inference Attack}\label{sec:review-MIA}
Membership inference attack (MIA) aims to determine whether a given example is contained in the training of a model~\cite{membership-first, membership-survey},
posing privacy risks
since it provides the adversary with extra information about the training data.
The feasibility of MIA lies in the overfitting to the training data.
The key to an effective MIA is the characterization of differences between training and non-training data via designated features,
which should be tailored to the architecture and training paradigm of the specific task.

\noindent {\bf MIA on embedding models.}
This type of MIA targeted contrastive learning~\cite{EncoderMI},
speech self-supervised learning~\cite{self-supervised-speech-membership},
metric learning-based person re-identification~\cite{person-re-identification-MIA},
and few-shot learning-based facial recognition~\cite{FACE-AUDITOR}.
The first attack is example-level while the others are user- or speaker-level.
\cite{self-supervised-speech-membership} utilized the average pairwise cosine similarity
among embeddings of examples from the target speaker as the feature,
with the assumption that this similarity is higher for the training speaker than non-training speakers.
It also improved the attack by replacing the predefined cosine similarity
with a similarity metric learned by neural networks.
The same feature is also used in the example-level MIA EncoderMI~\cite{EncoderMI}
which computed the feature on the embeddings of the given example and its augmented versions,
denoted by EncoderMI-T(hreshold).
EncoderMI also utilized the set of similarities contributing to the feature
and the sorted set (i.e., vector) as features, denoted by EncoderMI-S(et) and EncoderMI-V(ector), respectively.
In addition to the pairwise similarity, \cite{person-re-identification-MIA}
also utilize the average similarity between the centroid embedding
and embeddings contributing to the centroid embedding.

The closest work to ours is FaceAuditor~\cite{FACE-AUDITOR} which was available online
when we were preparing this manuscript.
Different from above attacks that assumed the accessibility of embeddings,
FaceAuditor only relies on the recognition score derived from embeddings.
It designed different features for different facial recognition networks.
For SiameseNet, it utilized the same feature as EncoderMI-V,
i.e., the vector of pairwise similarities among the facial images of the target user.
For ProtoNet and RelationNet, it utilized the set of scores in which
each score is the similarity between one facial image of the target user
and the ``prototype'' of the target user or the ``prototype'' of other supplemented users.

While these MIA could be applied to SRSs,
besides the recognition task, our attack \attackname differs from them in the following aspects:
(1) Our threat model is more systematic.
They assumed that the adversary has access to either the embeddings or recognition scores but not both,
while our attack applies to both (cf.~\ref{sec:threat-model}).
(2) Our designed features used for membership inference are more comprehensive.
We design the features from both intra-similarity and inter-dissimilarity,
quantified by using two types of similarities, four types of distances, four different statistics,
and different arrangements of similarities and distances (cf.~\ref{sec:feature-extractor}).
Through this carefully-established feature engineering, we design in total 103 features
which strictly cover the features used in these previous MIA.
Our designed features are found to complement each other,
leading to a more expressive characterization of the differences
between training and non-training speakers.
(3) We propose a training strategy to improve  attack generalizability.
For speaker-/user-level MIA, the ratio of examples of the training speaker/user
provided for membership inference that are used in training
is unknown and may vary with the target speaker.
\cite{person-re-identification-MIA,FACE-AUDITOR}
used only the non-training examples of training speakers/users (i.e., the ratio is $0$)
to train an attack model, leading to low generalizability to different ratios.
Hence, we propose a mixing ratio training strategy to enhance the generalizability (cf.~\ref{sec:generalization-attack-model}).
(4) We propose four approaches that are tailored to speaker recognition,
which either effectively enhance the attack performance
or significantly reduce the number of queries (cf.~\ref{sec:VND-attack-model}, \ref{sec:VCS}, and \ref{sec:reduce-query}).
(5) Our attack outperforms the previous MIA for speaker-level MIA on speaker recognition
under all the three datasets and five models (cf.~\ref{sec:overall-performance}).

\noindent {\bf MIA on other models.}
Other MIA targeted conventional classification~\cite{membership-first, enhanced-membership, first-principle,
evaluation-membership, loss-trajectory,MIA-Semi,MIA-Multi-Exit-Networks},
speech recognition~\cite{speech-membership-2, speech-membership-1}, generative~\cite{membership-VAE, membership-GAN},
regression models~\cite{membership-survey}, etc.
These MIA cannot be ported to speaker recognition
due to the distinct training paradigm and architecture of speaker recognition.
For instance, the training of speech recognition models minimizes
Word Error Rate (WER) of the training utterances,
so WER is used as a characterization feature.
However, the training of SRSs focuses on speaker characteristics
rather than the speech text or command.
The training of conventional classification models minimizes the cross entropy loss of the training examples,
so probability vector, class confidence, and entropy, derived from the final fully connected layer, are used as characterization features.
However, SRSs may be trained in the paradigm of using verification-based loss functions without a final fully connected layer.
Even if it is trained in a similar manner,
the fully connected layer will be dropped after training.
Thus, all the above features
cannot be used to characterize the differences between training and non-training speakers.

\section{Overview of \attackname}\label{sec:overview}

\subsection{Problem Formulation}\label{sec:problem-formulation}
Assume the target SRS $SR^t$ is trained on a set of training speakers $\mathcal{S}^{t}_{tr}$
(drawn from some underlying distribution $\mathbb{S}$)
and their voices $\mathcal{V}^{t}_{tr}$.
Fix a set of voices $\mathbf{v} = \{v_1,\cdots,v_N\}$ ($N\geq 1$) of a target speaker ${s}$.
The \emph{speaker-level membership inference attack} is defined as:
   $ \mathcal{A}: SR^t, s, \mathbf{v} \rightarrow \{0, 1\}$,
where ``1'' (resp. ``0'') means
$s \in \mathcal{S}^{t}_{tr}$ (resp. $s\not\in \mathcal{S}^{t}_{tr}$).
Intuitively,
a speaker-level MIA takes as input a set of voices from one speaker $s$
and determines whether \emph{any} voices uttered by the speaker $s$
is contained in the training of the model.

We emphasize that for the ``member'', i.e., $s\in \mathcal{S}^{t}_{tr}$,
we do not require that the inference voices $\mathbf{v}$ are involved in training, i.e.,
$\mathbf{v}\cap \mathcal{V}^{t}_{tr}$ may be $\emptyset$.
This makes the attack more practical since
the inference voices of training speakers are less likely to be used in training.
Thus, a speaker-level MIA should be effective even if none of the inference voices of a training speaker has been used for training.

\subsection{Threat Model}\label{sec:threat-model}
\noindent {\bf Adversarial purpose.}
The adversary of speaker-level MIA may be users, regulators (e.g., government), SRS developers, and adversarial attackers. 
Users may want to identify whether their voices
have been used for training SRSs without their permission
given the wide spread of voices across social media platforms and online meetings.
Regulators can check if SRSs are compliant with their published data protection rules,
e.g., GDPR~\cite{GDPR}.
SRS developers can evaluate the privacy risk of their SRSs before publishing
in case of being punished or sued due to privacy violations.
Adversarial attackers can train a better shadow SRS using common training speakers to
improve transferability of adversarial examples~\cite{AS2T}.

\noindent {\bf Adversarial capacity.}
We consider both white- and black-box scenarios with different capacities.
The white-box adversary has access to the background model of the target SRS, and thus can obtain the embedding of any given voice.
Note that the parameters of the target SRS are still unknown.
The black-box adversary only has access to the enrollment and recognition APIs exposed by the target SRS
and has to invoke them sequentially to obtain similarity scores.
In practice, white-box adversaries (e.g., regulators or SRS developers)
are constrained by the number of queries to background models,
while box-box adversaries (e.g., users and adversarial attackers)
attempt to perform speaker-level MIA using as few queries as possible
due to the charge or cost of queries. 
Note that in both scenarios, the adversary may either possess or lack knowledge of the target SRS's architecture and training dataset distribution.

\noindent {\bf Adversarial knowledge.}
We assume the adversary has an auxiliary speaker dataset $\mathcal{S}^{a}$
that is sampled from the same distribution $\mathbb{S}$ as the training speaker dataset $\mathcal{S}^{t}_{tr}$ of the target SRS.
In addition, we assume the adversary knows the architecture of the target SRS.
These two assumptions follow the standard setting of most previous MIAs~\cite{membership-first,
membership-survey, enhanced-membership, first-principle, evaluation-membership, loss-trajectory}.
In ablation studies (cf. \cref{sec:ablation-study}),
we show that our attack remains effective,
although the performance may degrade,
when these two assumptions are relaxed.
However, different from \cite{self-supervised-speech-membership}
which assumed that the adversary is aware of a dataset
in which each data is the non-member of the target SRS, we do not assume such knowledge.
The reason is that although this assumption can free the adversary from training shadow SRSs,
it also makes the attack less practical.

\subsection{Pipeline of \attackname}
\begin{figure}[t]
    \centering
    \includegraphics[width=0.47\textwidth]{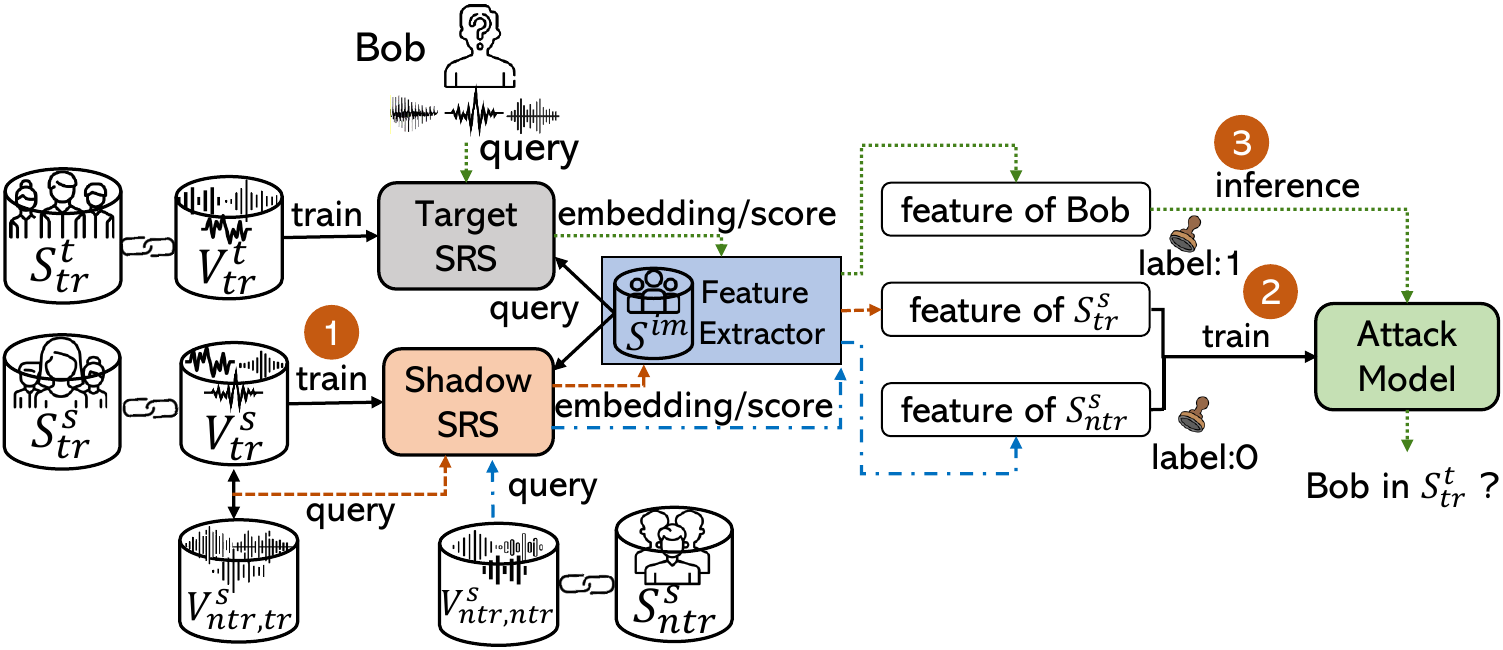}
    \caption{The working pipeline of \attackname.}
    \label{fig:overview-attack}\vspace*{-1mm}
  \end{figure}

Membership inference is indeed a binary classification task (\emph{member} or \emph{non-member}),
so we can build an attack model to do the classification.
Following the common pipeline of prior MIAs,
the working pipeline of \attackname is illustrated in \figurename~\ref{fig:overview-attack},
consisting of three stages: \emph{shadow SRS training}, \emph{attack model building}, and \emph{membership inference}.
We first briefly describe \emph{feature extractor}, a key module used in two stages.

\noindent {\bf Feature extractor.}
During the attack model building (resp. membership inference),
feature extractor queries the shadow (resp. target) SRS with a set of voices
to obtain their outputs (embeddings for white-box scenario and scores for black-box scenario),
based on which it produces features for the characterizing differences between training and non-training speakers.

\noindent {\bf Shadow SRS training.}
Due to the lack of knowledge about the training and non-training speakers of the target SRS,
it is impossible for the adversary to directly build the attack model in a supervised manner.
To tackle this problem, we train a shadow SRS as a proxy of the target SRS
upon which we build the attack model.
We first sample an auxiliary speaker dataset $\mathcal{S}^{a}$ and their corresponding voices
from some underlying distribution $\mathbb{S}$
and partition $\mathcal{S}^a$ into two sets $\mathcal{S}^s$ and $\mathcal{S}^{im}$.
The speakers in the set $\mathcal{S}^{im}$, called \emph{imposters}, will be used in attack model building and membership inference.
The speaker set $\mathcal{S}^{s}$ is further partitioned into two sets with the same number of speakers:
the set of training speakers ($\mathcal{S}^{s}_{tr}$)
and the set of non-training speakers ($\mathcal{S}^{s}_{ntr}$).
Since the inference voices of a training speaker are not necessarily used in training,
we also partition the voices of the training speakers $\mathcal{S}^{s}_{tr}$ into two sets, $\mathcal{V}^{s}_{tr}$ and $\mathcal{V}^{s}_{ntr,tr}$,
each of which has the same number of voices per speaker and belongs to ``member".
Finally, we train a shadow SRS using the training speakers $\mathcal{S}^{s}_{tr}$ and their voices $\mathcal{V}^{s}_{tr}$
with some chosen architecture and training algorithm.

\noindent {\bf Attack model building.}
To build an attack model,
we query the shadow SRS using the voices $\mathcal{V}^{s}_{tr}\cup \mathcal{V}^{s}_{ntr,tr}$ of the shadow SRS's training speakers $\mathcal{S}^{s}_{tr}$
 and obtain their outputs (embeddings for white-box scenario and scores for black-box scenario).
Note that the voices $\mathcal{V}^{s}_{tr}$ are used in training the shadow SRS  while the voices $\mathcal{V}^{s}_{ntr,tr}$ are not.
The outputs are then fed into the feature extractor
to extract the features of the training speakers $\mathcal{S}^{s}_{tr}$, which are labeled as ``member''.
The same is done for the voices $\mathcal{V}^{s}_{ntr,ntr}$ of the shadow SRS's the non-training speakers $\mathcal{S}^{s}_{ntr}$,
except that the extracted features are labeled as ``non-member''.
Finally, the features of $\mathcal{S}^{s}_{tr}$ and $\mathcal{S}^{s}_{ntr}$ are used
to build an attack model (cf. \cref{sec:attack-model-type}).

\noindent {\bf Membership inference.}
To determine the speaker-level membership of a given speaker,
we first query the target SRS
with the available inference voices of the given speaker,
then forward the output of the target SRS to the feature extractor,
and finally feed the extracted features  to the attack model,
which makes a ``member'' or ``non-member'' decision.

\section{Methodology of \attackname}\label{sec:methodology}
In this section, we 
first elaborate in detail the feature extractor,
then present the attack model,
and finally propose an approach to boost the performance of \attackname
when the target speaker provides a limited number of voices
as well as two techniques to reduce the number of queries to the target SRS in the black-box scenario.

\subsection{Feature Extractor}\label{sec:feature-extractor}
\begin{figure}[t]
  \centering
  \begin{subfigure}[b]{0.235\textwidth}
      \centering
      \includegraphics[width=.95\textwidth]{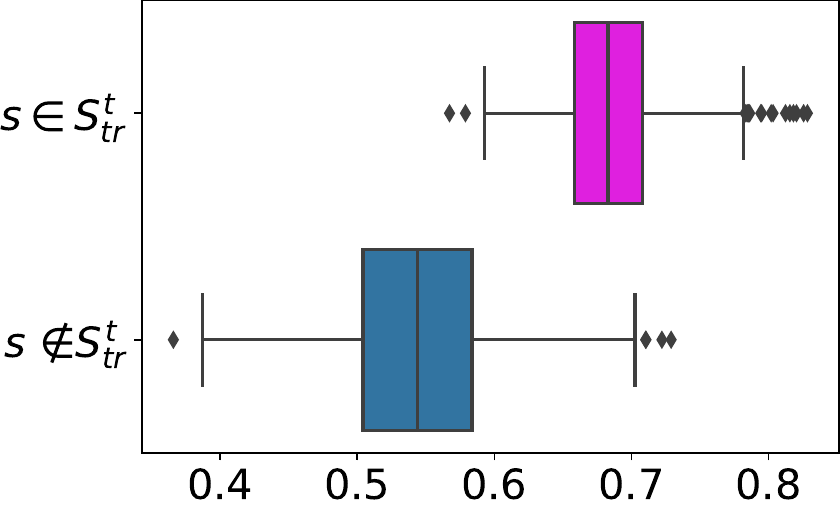}
   \caption{intra-similarity}
      \label{fig:mov-intra}
  \end{subfigure}
  \hfill
  \begin{subfigure}[b]{0.235\textwidth}
      \centering
      \includegraphics[width=.95\textwidth]{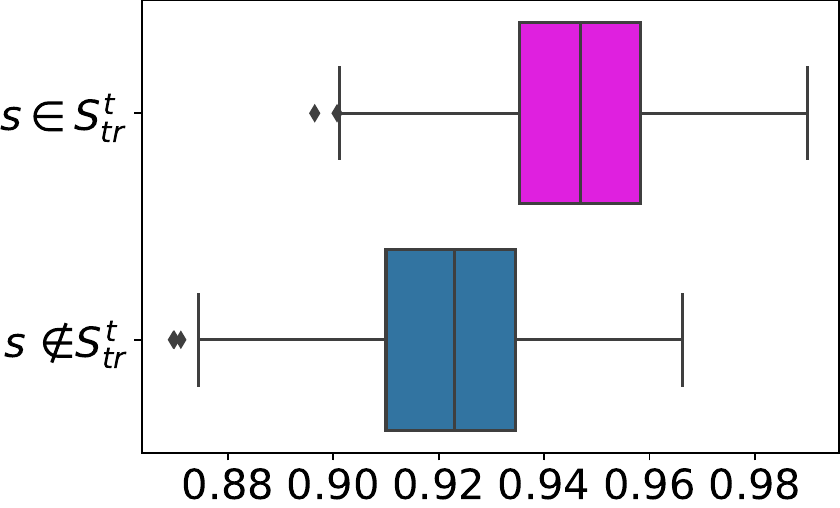}
    \caption{inter-dissimilarity}
      \label{fig:mov-inter}
  \end{subfigure}\vspace*{-3mm}
      \caption{The comparison of intra-similarity and inter-dissimilarity between training and non-training speakers.}
     \label{fig:mov} \vspace*{-1mm}
\end{figure}

Feature extractor takes  the embeddings or scores of voices as input
and outputs features that can effectively characterize the differences
between training and non-training speakers.
As mentioned in \cref{sec:review-MIA}, existing MIA features are not applicable to SR
due to its unique architecture and training paradigm.
Hence, we turn to analyze and understand the training objectives of SR.
Since SR is used to recognize the identity of individual speakers from their voices,
a model is trained with the following two objectives regardless of the training paradigms (classification- and verification-based losses):
(1)   intra-similarity: the embeddings of two voices from the same speaker are close to each other,
 and (2) inter-dissimilarity: embeddings of two voices from two distinct speakers are far enough from each other.
Thus, we hypothesize that training speakers enjoy better intra-similarity
and inter-dissimilarity than non-training speakers,
i.e., the embeddings of a training speaker's voices are closer to each other
and farther from the embeddings of the other speakers' voices, than that of a non-training speaker.
To give a first impression,
we quantify the intra-similarity and inter-dissimilarity
by the average pairwise cosine {\it similarity} among the embeddings of the same speaker's voices
and by the average of maximal cosine {\it distance} from distinct speakers' voices, respectively.
The box charts of intra-similarity and inter-dissimilarity are depicted
in \figurename~\ref{fig:mov-intra} and \figurename~\ref{fig:mov-inter}, respectively.
We can observe a significant statistic difference between the training and non-training speakers,
supporting our hypothesis.

Based on the above hypothesis, we design two groups of features:
intra-features and inter-features.
During the membership inference for a target speaker,
the intra-features quantify the similarity of voice embeddings of a target speaker (i.e., intra-similarity),
while the inter-features quantify the distance of voice embeddings between the target speaker
and other speakers (i.e., inter-dissimilarity).

To introduce the designed features, we first define the following function
which takes as input two sets of voices ($\{v_1,\cdots,v_m \}$ and $\{v^e_1,\cdots, v^e_n\}$) and produces the similarity
between the centroid embeddings of the two sets:
\begin{align}{\small
    \omega \langle v_1,\cdots,v_m | v^e_1,\cdots, v^e_n\rangle
    = \simi \langle \frac{1}{m}\sum_{i=1}^{m}E(v_i),\frac{1}{n}\sum_{i=1}^{n}E(v^e_i) \rangle \nonumber}
\end{align}
where $\simi \langle a,b\rangle$ is the similarity
between two embeddings $a$ and $b$ and $m,n\geq 1$.
However, in the black-box scenario,
the adversary has no access to embeddings $E(\cdot)$.
To solve this issue, we propose an alternative approach to compute the similarity $\omega$.
We first register an enrolled template with voices $v^e_1,\cdots,v^e_n$ in the enrollment phase
and then obtain the recognition score $S(v_1|v^e_1,\cdots,v^e_n)$ of the test voice $v_1$
w.r.t. the enrolled template in the recognition phase (cf.~\cref{sec:srs}).
Then $\omega \langle v_1|v^e_1,\cdots,v^e_n\rangle$ is computed as $S(v_1|v^e_1,\cdots,v^e_n)$.
Note that $m$ should be $1$ in the black-box scenario,
since an SRS accepts only one voice per query in the recognition phase.

\subsubsection{Intra-Features}
We design intra-features
by leveraging centroid-based similarity and pairwise similarity.
Fix a set of voices $\{v_1,\cdots,v_N\}$ of the target speaker
and let $[n]$ denote the set $\{1,\cdots, n\}$ for an integer $n$.

\noindent {\bf Centroid-based similarity} measures the closeness
between the embeddings and their centroid,
namely, the set of centroid-based similarities is
${\tt F}_{\tt c} = \{\omega \langle v_i|v_1,\cdots,v_N \rangle \mid i\in[N]\}$.

\noindent {\bf Pairwise similarity} measures the closeness between each pair of  embeddings.
The set of pairwise similarities
is defined as ${\tt F}_{\tt p} = \{\omega \langle v_j|v_i \rangle \mid i,j\in [N],i<j\}$.
The set ${\tt F}_{\tt p}$ can be refined as follows:
for each voice $v_i$, we first compute the similarities between $v_i$ and other voices as
${\tt F}_{\tt \tilde{p}}^{i} = \{\omega \langle v_j|v_i \rangle\mid j\in[N], j\neq i\}$,
then compute the statistic of ${\tt F}_{\tt \tilde{p}}^{i}$ as $\stat({\tt F}_{\tt \tilde{p}}^{i})$,
and finally define the refined set of similarities as
${\tt F}_{\tt \tilde{p}}=\{\stat({\tt F}_{\tt \tilde{p}}^{i})\mid i\in[N]\}$.
We will instantiate $\stat$ by average, negative standard derivative, maximum, and minimum statistics,
leading to the sets ${\tt F}_{\tt \tilde{p},avg}$, ${\tt F}_{\tt \tilde{p},std}$,
${\tt F}_{\tt \tilde{p},max}$, and ${\tt F}_{\tt \tilde{p},min}$, respectively.

The intra-features are defined as the statistics
(i.e., average, negative standard derivative, maximum, and minimum)
of the above six sets of similarities (${\tt F}_{\tt c}$, ${\tt F}_{\tt p}$,
${\tt F}_{\tt \tilde{p},avg}$, ${\tt F}_{\tt \tilde{p},std}$,
${\tt F}_{\tt \tilde{p},max}$, and ${\tt F}_{\tt \tilde{p},min}$).
Let $\Theta_{\tt x}^{\tt y}$ denote the intra-feature
which is defined as the statistic $\tt y$ of the set ${\tt F_x}$.
After de-duplicating three pairs of equivalent intra-features
($\Theta_{\tt p}^{\tt avg}$ and $\Theta_{\tt \tilde{p},avg}^{\tt avg}$,
$\Theta_{\tt p}^{\tt max}$ and $\Theta_{\tt \tilde{p},max}^{\tt max}$,
$\Theta_{\tt p}^{\tt min}$ and $\Theta_{\tt \tilde{p},min}^{\tt min}$),
there are ($6\times 4-3=21$) unique intra-features.
Note that we use negative standard derivative instead of standard derivative
since we expect the features of training speakers to be larger than that of non-training speakers.

\subsubsection{Inter-Features}
Inter-features make use of an additional set of $M$ imposters
$\mathcal{S}^{im}=\{s^{im}_{1},\cdots, s^{im}_M\}$ and their voices
$\mathcal{V}^{im} = \bigcup_{j=1}^{M}\mathcal{V}^{im}_{j}$,
where 
$\mathcal{V}^{im}_{j}= \{v^{im, j}_{1},\cdots, v^{im, j}_{K_j}\}$ is the set of voices of the imposter $s_j^{im}$.
Also let $\mathcal{V}^{im} = \{v^{im}_1\cdots, v^{im}_Q\}$
with $Q=\sum_{j=1}^{M}K_j$.
We design four types of distances.

\noindent {\bf Centroid-centroid distance} measures the distance
between the \underline{centroid} of the voice embeddings of the target speaker
and the \underline{centroid} of the voice embeddings of all the imposters.
Thus, we define the set of distances
${\tt F}_{\tt cc}=\{-\omega \langle v_1^{im,j},\cdots,v_{K_j}^{im,j}
|v_1,\cdots,v_N \rangle \mid j\in [M]\}$.
Recall that
$K_j$ should be $1$ for this distance in the black-box scenario.
In~\cref{sec:reduce-query}, we propose an enrollment voice concatenation technique,
allowing
the adversary to obtain {\it one} concatenated and longer voice for both the target speaker and each imposter
with multiple voices (i.e., $K_j\geq 1$).
It does not lead to obvious performance gap
between the two scenarios (cf. Appendix~\ref{sec:c-c-w-b-comp}).

 \noindent {\bf Centroid-voice distance} measures the distance
between the \underline{centroid} of the voice embeddings of the target speaker
and the \underline{voice} embeddings of all the imposters.
The set of distances is defined as
${\tt F}_{\tt cv}=\{-\omega \langle v^{im}|v_1,\cdots,v_N\rangle \mid v^{im}\in \mathcal{V}^{im}\}$.
The set ${\tt F}_{\tt cv}$ can be refined as follows:
for each imposter $s^{im}_{j}$,
we first compute the set of distances between the centroid of the voice embeddings of the target speaker
and the voice embeddings of this imposter, i.e.,
${\tt F}_{\tt c\tilde{v}}^{j}=\{-\omega \langle v^{im}|v_1,\cdots,v_N\rangle\mid v^{im}\in \mathcal{V}^{im}_{j}\}$,
then compute the statistic of ${\tt F}_{\tt c\tilde{v}}^{j}$ as $\stat({\tt F}_{\tt c\tilde{v}}^{j})$,
and finally define the refined set as
${\tt F}_{\tt c\tilde{v}}=\{\stat({\tt F}_{\tt c\tilde{v}}^{j})\mid j\in[M]\}$.

 \noindent {\bf Voice-centroid distance} measures the distance between the \underline{voice} embeddings
of the target speaker and the \underline{centroid} of the voice embeddings of all the imposters.
The set of distances is
${\tt F}_{\tt vc}=\{-\omega \langle v_i|v_1^{im,j},\cdots,v_{K_j}^{im,j}\rangle \mid i\in[N],j\in[M]\}$.
The set ${\tt F}_{\tt vc}$ can also be refined in two ways:
(i) for each voice $v_i$ of the target speaker,
we first compute the distance between the embedding of $v_i$ and the centroid of the voice embeddings of all the imposters
as ${\tt F}_{\tt \tilde{v}c}^{i} = \{-\omega \langle v_i|v_1^{im,j},\cdots,v_{K_j}^{im,j}\rangle \mid j\in[M]\}$,
then compute the statistic of ${\tt F}_{\tt \tilde{v}c}^{i}$ as $\stat({\tt F}_{\tt \tilde{v}c}^{i})$,
and finally define the refined set  
${\tt F}_{\tt \tilde{v}c}=\{\stat({\tt F}_{\tt \tilde{v}c}^{i})\mid i\in [N]\}$;
(ii) for each imposter $s^{im}_{j}$, we first compute the distance between the centroid of voice embeddings
of this imposter and the voice embeddings of the target speaker, i.e.,
${\tt F}_{\tt v\tilde{c}}^{j} = \{-\omega \langle v_i|v_1^{im,j},\cdots,v_{K_j}^{im,j}\rangle \mid i\in[N]\}$,
then compute the statistic of ${\tt F}_{\tt v\tilde{c}}^{j}$ as $\stat({\tt F}_{\tt v\tilde{c}}^{j})$,
and finally define the refined set  ${\tt F}_{\tt v\tilde{c}}=\{\stat({\tt F}_{\tt v\tilde{c}}^{j})\mid j\in [M]\}$.

 \noindent {\bf Voice-voice distance} measures the distance between the \underline{voice} embeddings
of the target speaker and the \underline{voice} embeddings of all the imposters.
Formally, the set of distances is defined as
${\tt F}_{\tt vv} = \{ -\omega \langle v_i|v^{im}\rangle \mid i\in[N], v^{im}\in \mathcal{V}^{im}\}$.
Similarly, two refined sets of distances can be defined accordingly, namely,
${\tt F}_{\tt \tilde{v}v} = \{ \stat({\tt F}_{\tt \tilde{v}v}^{i}) \mid i\in [N]\}$
and ${\tt F}_{\tt v\tilde{v}} = \{ \stat({\tt F}_{\tt v\tilde{v}}^{k}) \mid k\in[Q]\}$
where ${\tt F}_{\tt \tilde{v}v}^{i} =\{ -\omega \langle v_i|v^{im}\rangle \mid v^{im}\in \mathcal{V}^{im} \}$
and ${\tt F}_{\tt v\tilde{v}}^{k} =\{ -\omega \langle v_i|v^{im}_k\rangle \mid i\in[N]\}$.

The same to intra-features, we instantiate $\stat$  by average, negative standard derivative, maximum, and minimum statistics,
leading to 24 sets of distances, namely,
$\tt F_{cc}$, $\tt F_{cv}$,
$\tt F_{c\tilde{v},avg}$, $\tt F_{c\tilde{v},std}$, $\tt F_{c\tilde{v},max}$, $\tt F_{c\tilde{v},min}$,
$\tt F_{vc}$,
$\tt F_{\tilde{v}{c},avg}$, $\tt F_{\tilde{v}{c},std}$, $\tt F_{\tilde{v}{c},max}$, $\tt F_{\tilde{v}{c},min}$,
$\tt F_{v\tilde{c},avg}$, $\tt F_{v\tilde{c},std}$, $\tt F_{v\tilde{c},max}$, $\tt F_{v\tilde{c},min}$,
$\tt F_{vv}$,
$\tt F_{\tilde{v}{v},avg}$, $\tt F_{\tilde{v}{v},std}$, $\tt F_{\tilde{v}{v},max}$, $\tt F_{\tilde{v}{v},min}$,
$\tt F_{v\tilde{v},avg}$, $\tt F_{v\tilde{v},std}$, $\tt F_{v\tilde{v},max}$, and $\tt F_{v\tilde{v},min}$.

The inter-features are defined as the statistics of the above 24 sets of distances.
Let $\Phi_{\tt x}^{\tt y}$ denotes the inter-feature
which is defined as the statistic $\tt y$ of the set $\tt F_x$.
After de-duplicating two pairs and six triples of equivalent features
($\Phi_{\tt cv}^{\tt max}$ and $\Phi_{\tt c\tilde{v},max}^{\tt max}$,
$\Phi_{\tt cv}^{\tt min}$ and $\Phi_{\tt c\tilde{v},min}^{\tt min}$,
$\Phi_{\tt vc}^{\tt avg}$ and $\Phi_{\tt \tilde{v}c,avg}^{\tt avg}$ and $\Phi_{\tt {v}\tilde{c},avg}^{\tt avg}$,
$\Phi_{\tt vc}^{\tt max}$ and $\Phi_{\tt \tilde{v}c,max}^{\tt max}$ and $\Phi_{\tt v\tilde{c},max}^{\tt max}$,
$\Phi_{\tt vc}^{\tt min}$ and $\Phi_{\tt \tilde{v}c,min}^{\tt min}$ and $\Phi_{\tt v\tilde{c},min}^{\tt min}$,
$\Phi_{\tt vv}^{\tt avg}$ and $\Phi_{\tt \tilde{v}v,avg}^{\tt avg}$ and $\Phi_{\tt v\tilde{v},avg}^{\tt avg}$,
$\Phi_{\tt vv}^{\tt max}$ and $\Phi_{\tt \tilde{v}v,max}^{\tt max}$ and $\Phi_{\tt v\tilde{v},max}^{\tt max}$,
$\Phi_{\tt vv}^{\tt min}$ and $\Phi_{\tt \tilde{v}v,min}^{\tt min}$ and $\Phi_{\tt v\tilde{v},min}^{\tt min}$),
there are ($24\times 4 - 2\times 1 - 6\times 2=82$) unique inter-features.

We remark that inter-features can exploit a set of imposters which is fully controllable by the adversary,
thus are applicable when the target speaker only has one voice (i.e., $N=1$).
In contrast, intra-features requires $N\geq 2$.

\subsubsection{Property of Feature Extractor}\label{sec:property-feature-extractor}
Here we discuss some notable properties of the proposed feature extractor.

\noindent {\bf Unity under different scenarios.}
The features defined above work in both white- and black-box scenarios,
only differing in the way of obtaining the similarity $\omega$.
Consequently,
in the white-box scenario, the embedding $E(v)$ of each voice $v$ can be computed by querying
the background model once, based on which all the intra- and inter-features can be computed.
Thus, the total number of queries is $N+Q$.
However, in the black-box scenario, as shown in \tablename~\ref{tab:number-query} (Baseline),
the enrollment and recognition APIs have to be queried multiple times
to obtain the required scores for computing a feature, although
these scores can be reused for computing the features in the same group.
To reduce the number of queries to the target SRS, we will propose two techniques,
group enrollment and enrollment voice concatenation (cf. \cref{sec:reduce-query}).
The reduced numbers of queries are also shown in \tablename~\ref{tab:number-query}.

\noindent {\bf Comprehensiveness.}
In total,
we design 103 features from two different aspects (intra-similarity and inter-dissimilarity)
using two types of similarities, four types of distances, four statistics,
and different arrangements of similarities or distances
(original v.s. refined sets of similarities or distances),
aiming to effectively quantify the differences between training and training speakers
from different perspectives in a complementary way.
 
\begin{table}[t]
  \centering \renewcommand\arraystretch{1.06} \setlength\tabcolsep{4pt} 
  \caption{The number of queries in the black-box scenario.} \vspace{-1mm}
\scalebox{0.88}{\begin{threeparttable}
    \begin{tabular}{c|c|c|c|c|c}
    \hline
    \multicolumn{2}{c|}{\multirow{2}{*}{\textbf{Feature}}}& \textbf{Feature} & \multicolumn{3}{c}{\textbf{\#Query}} \\
\cline{4-6}    \multicolumn{2}{c|}{}& \textbf{Computation} & \textbf{Enrollment} & \textbf{Recognition} & \textbf{Total} \\
    \hline
    \multirow{3}{*}{\textbf{Intra}} & \multirow{2}{*}{\textbf{Centroid-based}} & \textbf{Baseline} & N     & N     & 2N \\
\cline{3-6}          &       & \textbf{Concat} & 1     & N     & 1+N \\
\cline{2-6}          & \textbf{Pairwise} & \textbf{Baseline} & N-1   & N(N-1)/2 & (N-1)(N+2)/2 \\
    \hline
    \multirow{10}{*}{\textbf{Inter}} & \multirow{2}{*}{\textbf{Centroid-centroid}} & \textbf{Baseline} & N     & M     & N+M \\
\cline{3-6}          &       & \textbf{Concat} & 1     & M     & 1+M \\
\cline{2-6}          & \multirow{2}{*}{\textbf{Centroid-voice}} & \textbf{Baseline} & N     & Q     & N+Q \\
\cline{3-6}          &       & \textbf{Concat} & 1     & Q     & 1+Q \\
\cline{2-6}          & \multirow{4}{*}{\textbf{Voice-centroid}} & \textbf{Baseline} & Q     & NM    & Q+NM \\
\cline{3-6}          &       & \textbf{Group} & Q     & N     & Q+N \\
\cline{3-6}          &       & \textbf{Concat} & M     & NM    & M+NM \\
\cline{3-6}          &       & \textbf{Group+Concat} & M     & N     & M+N \\
\cline{2-6}          & \multirow{2}{*}{\textbf{Voice-voice}} & \boldmath{}\textbf{Baseline}\unboldmath{} & N     & QN    & (1+Q)N \\
\cline{3-6}          &       & \boldmath{}\textbf{Group}\unboldmath{} & N     & Q    & N+Q \\
    \hline
    \multicolumn{2}{c|}{\multirow{1}{*}{\bf All}} & \makecell[c]{{\bf Concat+Group$^\sharp$} \\ {\bf +Share}} & 1+N & N+M+Q & 2N+M+Q+1 \\ 
    \hline
    \end{tabular}
  \begin{tablenotes}
    \small
  \item Note: (i) ``Concat'' and ``Group'' are short for ``Enrollment Voice Concatenation''
  and ``Group Enrollment'', respectively. (ii) Baseline does not utilize the two techniques. (iii) ``Share" denotes enrolled templates sharing across different groups of features, avoiding duplicated registered templates (cf.~\cref{sec:reduce-query}). (iv) $\sharp$: Using ``Group Enrollment" for both intra-group and inter-group features (cf.~\cref{sec:reduce-query}). 
  \end{tablenotes}
  \end{threeparttable}}
  \label{tab:number-query}
\end{table}

\subsection{Attack Model}\label{sec:attack-model}
\subsubsection{Attack Model Configuration}\label{sec:attack-model-type}
To utilize multiple features,
\attackname adopts a classifier-based attack model.
Specifically, we train a binary classifier $f$ as the attack model in a
supervised fashion based on the shadow SRS
using the features of the voices of the training speakers $\mathcal{S}^{s}_{tr}$
and non-training speakers $\mathcal{S}^{s}_{ntr}$.
Then, $\mathcal{A}(SR^t, \mathbf{v}, s)$ is implemented as $\mathbb{I}[f(\Psi(\mathbf{v}))>0.5]$
where $\Psi(\mathbf{v})$ denotes the features of the inference voices $\mathbf{v}$ of a target speaker
produced by the feature extractor via querying the target SRS $SR^t$ and $f$
gives the probability that the speaker $s$ is one of the training speakers of the target SRS $SR^t$.

To demonstrate the complementarity of features,
we also include a threshold-based attack model in our experiments,
which makes decisions by thresholding a single feature.
The attack $\mathcal{A}(SR^t,\mathbf{v},s)$ is implemented as $\mathbb{I}[\psi(\mathbf{v})> \tau]$
where $\mathbb{I}$ is the indicator function, $\tau$ is a threshold tuned on the shadow SRS by maximizing the classification accuracy
between the training speakers $\mathcal{S}^{s}_{tr}$ and non-training speakers $\mathcal{S}^{s}_{ntr}$,
and $\psi(\mathbf{v})$ is the feature of the inference voices $\mathbf{v}$ produced by the feature extractor via querying the target SRS $SR^t$.
Namely, a speaker $s$ is regarded as a training speaker of $SR^t$
if the feature of the provided voices $\mathbf{v}$ is larger than the threshold $\tau$.

\subsubsection{Attack Model Generalization}\label{sec:generalization-attack-model}
\begin{figure}[t]
  \centering
  \begin{subfigure}[b]{0.235\textwidth}
      \centering
      \includegraphics[width=\textwidth]{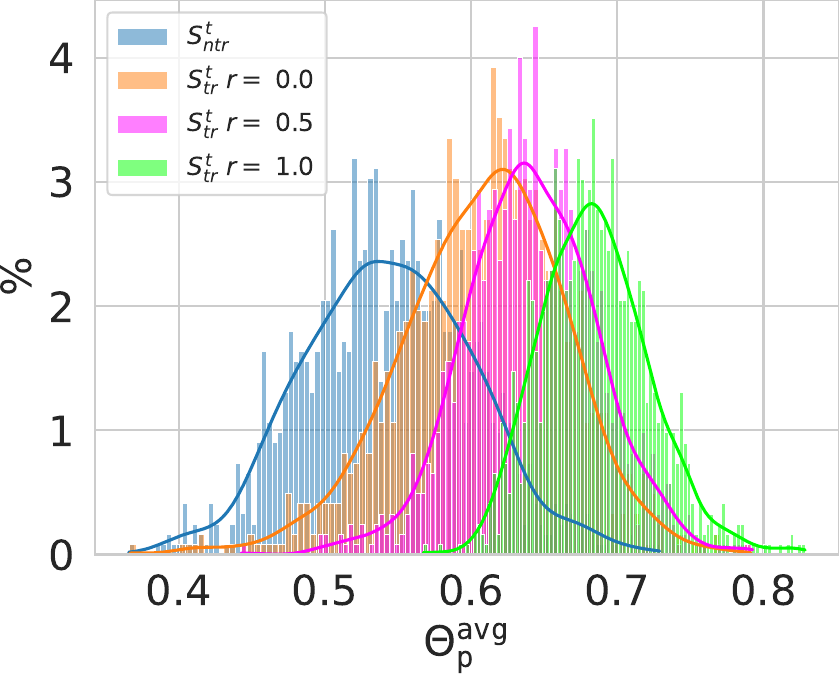}
        \vspace*{-6mm}
        \caption{intra-similarity}
      \label{fig:alpha-intra}
  \end{subfigure}
  \hfill
  \begin{subfigure}[b]{0.235\textwidth}
      \centering
      \includegraphics[width=\textwidth]{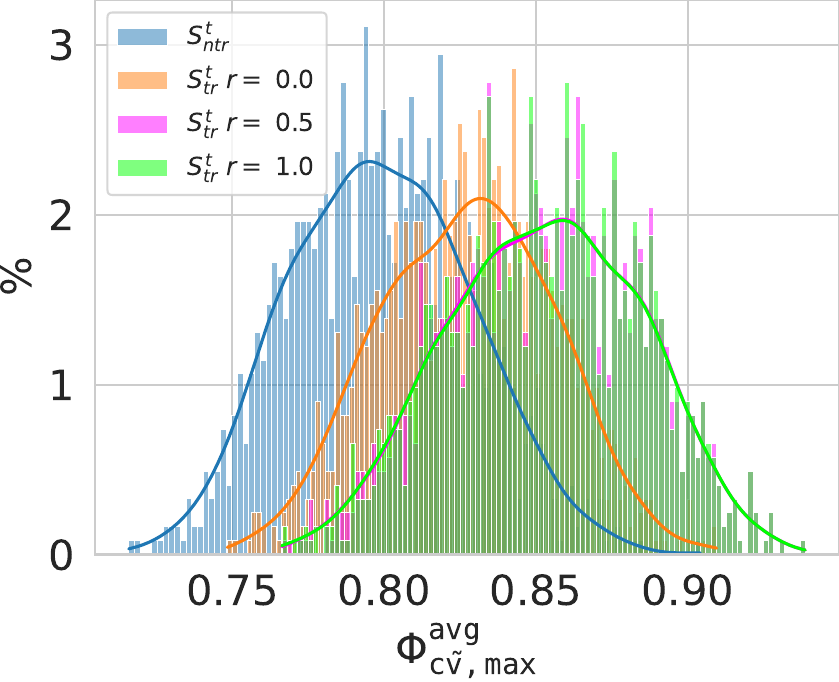}
    \vspace*{-6mm}
    \caption{inter-dissimilarity}
      \label{fig:alpha-inter}
  \end{subfigure}\vspace*{-3mm}
      \caption{Comparison of features with different $r$.}
     \label{fig:alpha}\vspace*{-1mm}
\end{figure}

For a target speaker $s$,
let $r$ denote the ratio of the target speaker's inference voices ${\bf v}$
that are included in the training of the target SRS.
Since the ratio $r$ is unknown in practice, it is expected that
a speaker-level MIA should be effective for unknown $r$ even if $r=0$, i.e., none of the inference voices is used in the training
of the target SRS. However, as depicted in \figurename~\ref{fig:alpha},
the distribution of features of training speakers varies with the ratio $r$.
Thus, the attack model trained with one fixed $r$
is more likely to generalize poorly on other $r$.
To address this issue, we propose a \emph{mixing ratio training strategy} which utilizes features of
the training voices $\mathcal{V}^{s}_{tr}$ ($r=1$) and non-training voices $\mathcal{V}^{s}_{ntr,tr}$ ($r=0$) of the training speakers as  ``member'',
and the non-training speakers' voices $\mathcal{V}^{s}_{ntr,ntr}$ as  ``non-member'',
to train the attack model.
Though the attack model is only trained with $r=1$ and $r=0$,
it generalizes for membership inference with different $r$.
We could include more diverse $r$ when training the attack model.
However, it will introduce more overhead to compute the features for different $r$
and train the attack model on a larger dataset, thus not considered in this work.

\begin{figure}[t]
  \centering
  \begin{subfigure}[b]{0.23\textwidth}
    \centering
    \includegraphics[width=\textwidth]{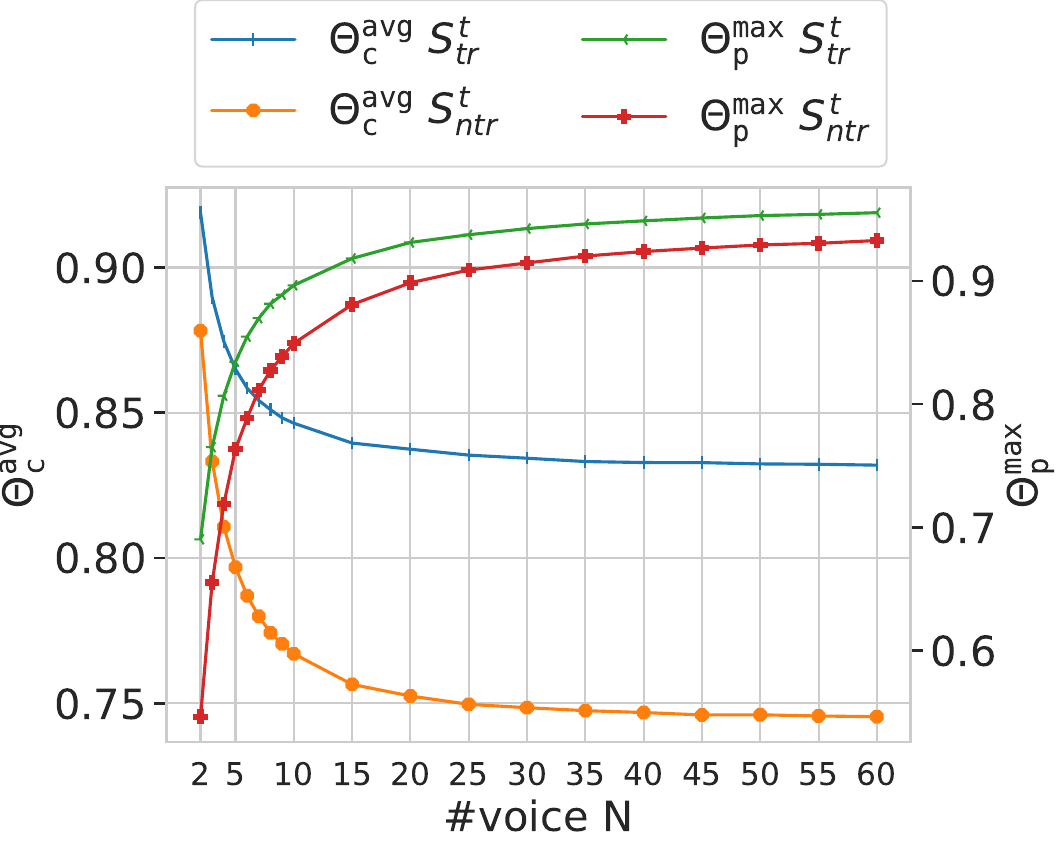}
      \vspace*{-5mm}
      \caption{intra-similarity}
  \end{subfigure}$~$
  \begin{subfigure}[b]{0.24\textwidth}
    \centering
    \includegraphics[width=\textwidth]{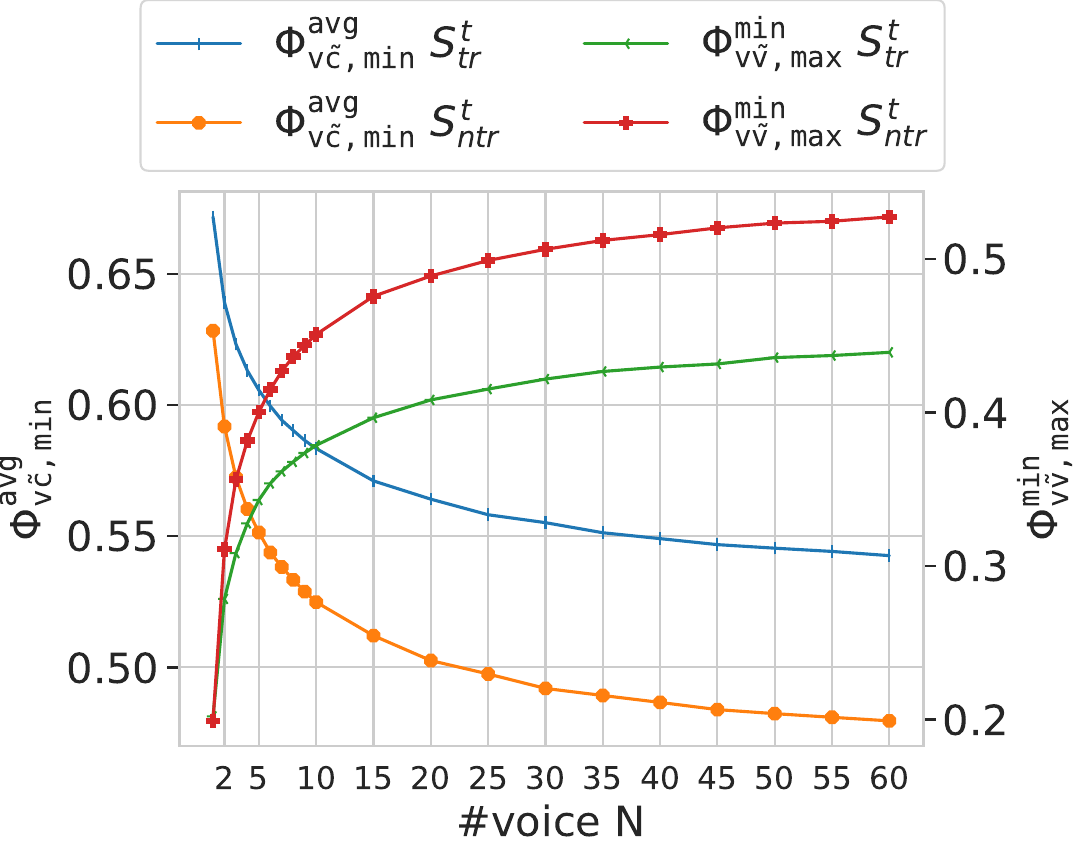}
       \vspace*{-5mm}
       \caption{inter-dissimilarity}
  \end{subfigure} \vspace*{-6mm}
  \caption{Comparison of features with different $N$.}
    \label{fig:num-voice-feature-value} \vspace*{-1mm}
\end{figure}

\begin{table*}[htbp]
  \centering\setlength{\tabcolsep}{6pt}
  \caption{Details of dataset partition} \vspace*{-1mm}
  \resizebox{.99\textwidth}{!}{\begin{threeparttable}
    \begin{tabular}{cc|c|c|c|c|c}
    \hline
    \multicolumn{2}{c|}{\multirow{3}{*}{\bf Dataset}}    &     \multirow{3}{*}{\textbf{Total}}  & \textbf{Imposters} & \multicolumn{3}{c}{\textbf{Shadow / Target SRS}}\\
    \cline{4-7}
     &  & & \makecell[c]{{\bf Speakers} \boldmath{}\textbf{$\mathcal{S}^{im}$}\unboldmath{}}
    & \multicolumn{2}{c|}{\makecell[c]{\textbf{Training} \boldmath{}\textbf{speakers $\mathcal{S}^{s}_{tr}$ / $\mathcal{S}^{t}_{tr}$}\unboldmath{}}}
    & \makecell[c]{\textbf{Non-training} \boldmath{}\textbf{speakers $\mathcal{S}^{s}_{ntr}$ / $\mathcal{S}^{t}_{ntr}$}\unboldmath{}} \\
\cline{4-7}
       &   &       & \makecell[c]{{\bf Voices}  \boldmath{}\textbf{$\mathcal{V}^{im}$}\unboldmath{}}
& \makecell[c]{\textbf{Training Voices} \boldmath{}\textbf{$\mathcal{V}^{s}_{tr}$ / $\mathcal{V}^{t}_{tr}$}\unboldmath{}}
& \makecell[c]{\textbf{Non-training voices} \boldmath{}\textbf{$\mathcal{V}^{s}_{ntr,tr}$ / $\mathcal{V}^{t}_{ntr,tr}$}\unboldmath{}}
& \makecell[c]{\textbf{Non-training voices} \boldmath{}\textbf{ $\mathcal{V}^{s}_{ntr,ntr}$ / $\mathcal{V}^{t}_{ntr,ntr}$}\unboldmath{}} \\
    \hline
    \multicolumn{1}{c}{\multirow{2}[1]{*}{\textbf{VoxCeleb-2}}} & \textbf{\#speakers} & {6112} & {1222} & {1222} / 1222 & {1222} / 1222 & {1222} / 1222 \\
 \multicolumn{1}{c}{}      & \textbf{\#voices} & {$>$1 million} &    101,572   &    113,617 / 115,780   &   113,055 / 115,212  &   111,083 / 108,879 \\
    \hline
    \multicolumn{1}{c}{\multirow{2}[1]{*}{\textbf{LibriSpeech}}} & \textbf{\#speakers} &    2484   &   400    &   521 / 521    &   521 / 521    &    521 / 521 \\
\multicolumn{1}{c}{}       & \textbf{\#voices} &    $\approx$ 0.3 million   &   46,140    &   30,928 / 30,734   &  30,648 / 30,746    &   30,892 / 30,992 \\
    \hline
     \multicolumn{1}{c}{\multirow{2}[1]{*}{\textbf{KeSpeech}}} & \textbf{\#speakers} & {6110} & {1222} & {1222} / 1222 & {1222} / 1222 & {1222} / 1222 \\
 \multicolumn{1}{c}{}      & \textbf{\#voices} & {$\approx$ 0.2 million} &    18,665   &    19,553 / 19,091   &   18,941 / 18,481  &   19,145 / 18,694 \\
    \hline
    \end{tabular}%
    \begin{tablenotes}
      \footnotesize
      \item Note: (i) For each dataset, 
      we first partition the speakers into five approximately equal and \emph{disjoint} parts,
      denoted by $\mathcal{S}^{im}$,
      $\mathcal{S}^{s}_{tr}$, $\mathcal{S}^{s}_{ntr}$,
      and $\mathcal{S}^{t}_{tr}$, and $\mathcal{S}^{t}_{ntr}$, respectively.
      (ii) $\mathcal{S}^{im}$ contains the imposters used to compute inter-features.
      (iii) $\mathcal{S}^{s}_{tr}$ and $\mathcal{S}^{s}_{ntr}$ are the training
      and non-training speakers for the shadow SRS.
      $\mathcal{S}^{t}_{tr}$ and $\mathcal{S}^{t}_{ntr}$ are the training
      and non-training speakers for the target SRS.
      (iv) For each speaker in $\mathcal{S}^{s}_{tr}$,
      we partition his/her voices into two nearly equal and \emph{disjoint} parts: $\mathcal{V}^{s}_{tr}$ and $\mathcal{V}^{s}_{ntr,tr}$.
      The same is applied to $\mathcal{S}^{t}_{tr}$, leading to $\mathcal{V}^{t}_{tr}$ and $\mathcal{V}^{t}_{ntr,tr}$.
      The shadow SRS and target SRS are trained on
      $\mathcal{V}^{s}_{tr}$ and $\mathcal{V}^{t}_{tr}$, respectively.
      (v) The training dataset of the attack model is derived from $\mathcal{V}^s_{tr}$ (label ``member''),
      $\mathcal{V}^s_{ntr,tr}$ (label ``member''), and $\mathcal{S}^s_{ntr}$ (label ``non-member'').
      (vi) The testing dataset used to evaluate the membership inference attack includes $\mathcal{V}^t_{tr}$ (label ``member''),
      $\mathcal{V}^t_{ntr,tr}$ (label ``member''), and $\mathcal{S}^t_{ntr}$ (label ``non-member'').
    \end{tablenotes}
  \end{threeparttable}
}
  \label{tab:dataset-split} 
  \end{table*}

\subsubsection{Voice-Number-Dependent Attack Model}\label{sec:VND-attack-model}
As shown in \figurename~\ref{fig:num-voice-feature-value}, features vary significantly with  the number $N$ of the inference voices,
building one voice-number-independent (VNID) attack model
will achieve sub-optimal performance.
Thus, we propose to build a voice-number-dependent (VND) attack model
for each number $N$.
However, there is no upper bound of $N$,
and it is impossible to build infinite attack models.
Based on the observation that features converge after some $N'$ in \figurename~\ref{fig:num-voice-feature-value},
we can set an upper bound $N'$ and build $N'$ VND attack models.
During membership inference, the adversary simply discards some voices when $N>N'$.
Note that this bound can be decided by the adversary on the shadow SRS with the auxiliary dataset,
and we propose a T-test~\cite{two-sample-T-test} based algorithm in Alg.~\ref{al:decide-attack-model-num} in Appendix~\ref{sec:bound-num-attack-models}.

\subsection{Voice Chunk Splitting}\label{sec:VCS}
The performance of an attack model decreases with the decrease in the number of inference voices,
because features become less precise.
It is possible to create a sufficient number of inference voices for some adversaries (e.g., users),
but may not be feasible for other adversaries (e.g., regulators).
A straightforward solution to address this issue is to create augmented voices
by applying voice augmentation (e.g., Gaussian white noise).
However, we find that this solution cannot effectively improve and even sometimes worsen the performance,
since the target SRS does not necessarily utilize augmentation during training.
Instead, we propose an approach, named voice chunk splitting, based on the observation that due to diverse voice duration,
training voices are always divided into short segments with fixed duration to form a training batch.
We apply a sliding window with certain window size $w$ and window step $s$
to split a voice into multiple overlapped chunks,
thus increasing the number of voices and improving the precision of features.
We note that voice chunk splitting is applied in both the attack model building
and the membership inference.

\subsection{Reducing Queries in the Black-Box Scenario}\label{sec:reduce-query}
The black-box adversary often attempts to perform membership inference using as few queries as possible.
We present two techniques to reduce the number of queries,
thus reducing testing time and resources.
The comparison of the number of queries with/without the two techniques
is shown in \tablename~\ref{tab:number-query}.

\begin{wrapfigure}{r}{0.21\textwidth}[before=\vspace{-\baselineskip}]
      \includegraphics[width=.21\textwidth]{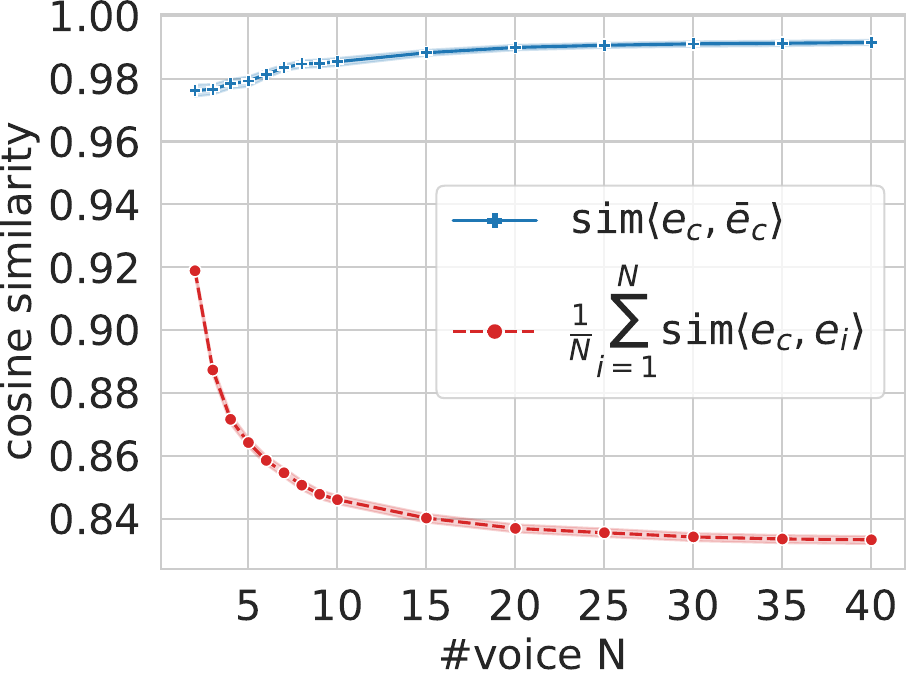}
 \caption{The similarity between $e_c$
      and $\bar{e}_c$ varying $N$.}\vspace*{-3mm}
      \label{fig:bb-concat}
  \end{wrapfigure}

\noindent {\bf Enrollment voice concatenation.}
To compute the features that rely on the centroid embedding of the voices ${\bf a}=\{a_1,\cdots,a_p\}$ of a speaker,
the enrolled template of the speaker has to be registered via
$p$ enrollment queries.
Observing that the embedding is extracted from the {\it frame-wise} acoustic feature\footnote{
Due to the time-varying non-stationary property, voices are not resilient enough to noises
and other variations, and waveforms fail to effectively represent speaker characteristics.
Hence, to achieve better performance, a raw voice is often
transformed into a two-dimensional time-frequency representation via frequency analysis,
called acoustic feature, where one coordinate at the time-axis represents a frame,
a short segment of the voice.},
we conjecture that the centroid embedding $e_c$ computed as the average of embeddings of the voices ${\bf a}$
is similar to the embedding $\bar{e}_c$ of the concatenation ${\tt concat}({\bf a})$ of the voices ${\bf a}$, for which  one enrollment query is sufficient.
\figurename~\ref{fig:bb-concat} shows the cosine similarity between $e_c$ and $\bar{e}_c$,
which is very close to 1.0 and
much higher than the similarity between the centroid embedding
and embeddings of the voices ${\bf a}$,
regardless of the number $N$ of enrollment voices, confirming our conjecture.
Hence, we apply enrollment voice concatenation to reduce the number of enrollment queries
when computing centroid-related intra-features and inter-features.
Specifically, we register an enrolled template using the voice ${\tt concat}({\bf a})$,
and approximate the recognition score of a voice $v$ w.r.t. the centroid embedding
$S(v|a_1,\cdots,a_p)$ by the one w.r.t. $S(v|{\tt concat}({\bf a}))$.
This reduces the number of enrollment queries from $p$ to $1$.

\noindent {\bf Group enrollment.}
Speaker identification allows multiple speakers to be enrolled each of which has an enrolled template,
and a vector of the scores w.r.t. these enrolled templates can be obtained via {\it one} recognition query.
In contrast, speaker verification allows only one speaker to be enrolled,
so to obtain the scores of multiple speakers,
these speakers have to be enrolled individually
and multiple recognition queries are required. 
We utilize speaker identification to reduce the number of recognition queries in the following ways:
\begin{itemize}
    \item Group enrollment for intra-group features. To obtain the voice-centroid distance based inter-features, we enroll all imposters and compute the scores of a voice of the target speaker w.r.t. all imposters' enrolled templates via one query. It reduces the number of recognition queries from $N\times M$ to $N$. Similarly, to obtain the voice-voice distance based inter-features, we enroll each of the target speaker's voices as an enrolled template and compute the scores of a voice of imposters w.r.t. all the target speaker's enrolled templates via one query. It reduces the number of recognition queries from $Q\times N$ to $Q$.
\item Group enrollment for inter-group features.
The recognition voices for the centroid-based similarity and the pairwise similarity based intra-features are the target speaker's voices, thus, we can obtain the scores of each recognition voice for both types of features in one query, reducing the the number of recognition queries from $N+\frac{N(N-1)}{2}$ to $N$. Similarly, we can reduce the number of recognition queries for the centroid-voice and voice-voice (resp. centroid-centroid and voice-centroid) distance based inter-features from $2\times Q$ to $Q$ (resp. from $2\times M$ to $M$ when enrollment voice concatenation is applied).
\end{itemize}

Besides the above two techniques, we can also save the enrollment queries with enrollment template sharing by avoiding duplicated enrolled templates. Specifically, the enrolled templates for centroid-based similarity based intra-features, centroid-centroid and centroid-voice distance based inter-features are the same, thus only need to be registered once. 
The same can be applied for the pairwise similarity based intra-features, the voice-centroid and voice-voice distance based inter-features.

\section{Evaluation of \attackname}\label{sec:evaluation}

\subsection{Experimental Setting}
\noindent {\bf Datasets.}
We use three widely-used voice datasets: two English dataset VoxCeleb-2~\cite{resnet34} and LibriSpeech~\cite{panayotov2015librispeech}, and one Chinese dataset KeSpeech~\cite{KeSpeech}.  
VoxCeleb-2 contains more than 1 million voices from 6,112 speakers,
while LibriSpeech contains approximately 0.3 million voices from 2,484 speakers. For KeSpeech, we randomly select approximately 0.2 million voices from 6,110 speakers saying mandarin Chinese. 
The partition of these datasets is summarized in \tablename~\ref{tab:dataset-split}.

\noindent {\bf SRSs.}
We use five SRSs: LSTM-GE2E, TDNN-CE, Raw-AAM, Res-AP, and VGG-GE2E.
These SRSs are the combinations of five model architectures and four training losses, as shown in \tablename~\ref{tab:srs-info}.
We trained them in their default settings with the number of training epochs ranging from 50 to 1500 depending on the number of trainable parameters. 
Regarding the recognition tasks of SRSs, \attackname exhibits the same effectiveness for both the speaker identification and verification tasks except that more black-box queries are required for verification (cf.~\ref{sec:reduce-query} and \tablename~\ref{tab:reduce-query-exper} for details). 
We also consider the commercial SRS of Microsoft Azure~\cite{microsoft-azure-vpr} in \cref{sec:add_eva}.

\begin{table}[h]  
        \centering\setlength{\tabcolsep}{2pt}
    \caption{The information and performance of SRSs}\vspace*{-1mm}
    \resizebox{.49\textwidth}{!}{\begin{threeparttable}
      \begin{tabular}{c|c|c|c|c|c|c|c|c|c}
      \hline
      \multirow{3}{*}{\textbf{Name}} & \multirow{3}{*}{\textbf{Archi}} & \multicolumn{2}{c|}{\multirow{2}{*}{\textbf{Training}}} & \multicolumn{6}{c}{\textbf{Performance in terms of EER}} \\
  \cline{5-10}          &       & \multicolumn{2}{c|}{} & \multicolumn{2}{c|}{\textbf{VoxCeleb-2}} & \multicolumn{2}{c|}{\textbf{LibriSpeech}} &  \multicolumn{2}{c}{\textbf{KeSpeech}} \\
  \cline{3-10}          &       & \textbf{paradigm} & \textbf{loss} & \textbf{Training} & \textbf{Testing} & \textbf{Training} & \textbf{Testing} & \textbf{Training} & \textbf{Testing} \\
      \hline
      \textbf{LSTM-GE2E} & LSTM~\cite{GE2E,SV2TTS-github}  & verification & GE2E~\cite{GE2E}  & 9.66\% & 18.19\% & 0.31\% & 8.08\% & 0.76 \% & 14.53\% \\
      \hline
      \textbf{TDNN-CE} & TDNN~\cite{ECAPA-TDNN,ECAPA-TDNN-sb}  & classification & CE~\cite{sr-overview}    & 1.43\% & 6.70\% & 0.36\% & 2.39\% & 0.40\% & 7.61\% \\
      \hline
      \textbf{Raw-AAM} & RawNet3~\cite{jung2022pushing,chung2020in} & classification & AAM~\cite{AAM-loss-2}   & 2.5\% & 6.23\% & 0.20\% & 2.56\% & 0.10\% &  5.40\% \\
      \hline
      \textbf{Res-AP} & ResNetSE34V2~\cite{he2016deep,chung2020in} & verification & AP~\cite{prototypical}    & 2.11\% & 6.70\% & 0.15\% & 6.84\% & 0.20\% & 8.46\% \\
      \hline
      \textbf{VGG-GE2E} & VGGVox40~\cite{nagrani2017voxceleb,chung2020in} & verification & GE2E~\cite{GE2E}  & 4.18\% & 9.43\% & 0.24\% & 5.66\% & 2.72\% & 9.23\% \\
      \hline
      \end{tabular}%
      \begin{tablenotes}
        \small
      \item Note: Training EER and testing EER are calculated on 4,000 randomly chosen trials from $\mathcal{V}_{tr}^t$ and $\mathcal{V}_{ntr,ntr}^t$, respectively.
      \end{tablenotes}
    \end{threeparttable}
    }
    \label{tab:srs-info}%
  \end{table}

\begin{table*}
  \centering\setlength{\tabcolsep}{8pt} 
\caption{The effectiveness of \attackname in the Setting-1 when the ratio $r=0$.}\vspace*{-1mm}
  \resizebox{0.9\textwidth}{!}{
    \begin{tabular}{c|c|ccc|ccc|ccc|ccc}
      \hline
      \multicolumn{2}{c|}{\multirow{2}[6]{*}{}} & \multicolumn{3}{c|}{\textbf{Accuracy}} & \multicolumn{3}{c|}{\textbf{AUROC}} & \multicolumn{3}{c|}{\textbf{TPR @ x\% FPR}} & \multicolumn{3}{c}{\textbf{TPR @ 1\% FPR}}  \\
  \multicolumn{2}{c|}{} & {\textbf{VC-2}} & {\textbf{LS}} & {\textbf{KS}} & {\textbf{VC-2}} & {\textbf{LS}} & {\textbf{KS}} & \makecell[c]{{\textbf{VC-2}} \\ {\textbf{(x=0.1)}}} & \makecell[c]{{\textbf{LS}} \\ {\textbf{(x=0.2)}}} & \makecell[c]{{\textbf{KS}} \\ {\textbf{(x=0.1)}}} & {\textbf{VC-2}} & {\textbf{LS}} & {\textbf{KS}} \\ 
      \hline
      \multirow{4}{*}{\textbf{LSTM-GE2E}} & \textbf{LRL-MIA} & 0.698 & 0.895 & 0.669 & 0.789 & 0.952 & 0.767 & 1.5\% & 31.7\% & 1.7\% & 8.1\% & 41.7\% & 8.4\% \\
  & \textbf{EncoderMI-T} & 0.7   & 0.877 & 0.592 & 0.792 & 0.952 & 0.768 & 1.6\% & 31.7\% & 1.8\% & 7.8\% & 41.7\% & 8.5\% \\
  & \textbf{TLK-MIA} & 0.7   & 0.877 & 0.592 & 0.792 & 0.952 & 0.768 & 1.6\% & 31.7\% & 1.8\% & 7.8\% & 41.7\% & 8.5\% \\
  & \textbf{\attackname} & {\bf 0.894} & {\bf 0.974} & {\bf 0.785} & {\bf 0.958} & {\bf 0.994} & {\bf 0.880} & {\bf 33.5\%} & {\bf 66.5\%} & {\bf 10.5\%} & {\bf 58.7\%} & {\bf 83.5\%} & {\bf 28.1\%} \\
      \hline
      \multirow{4}{*}{\textbf{TDNN-CE}} & \textbf{LRL-MIA} & 0.82  & 0.723 & 0.595 & 0.906 & 0.791 & 0.676 & 20.1\% & 0.6\% & 1.8\% & 46.6\% & 6.7\% & 4.6\% \\
  & \textbf{EncoderMI-T} & 0.779 & 0.713 & 0.583 & 0.904 & 0.791 & 0.668 & 20.8\% & 0.6\% & 0.8\% & 45.9\% & 6.7\% & 4.0\% \\
  & \textbf{TLK-MIA} & 0.779 & 0.713 & 0.583 & 0.904 & 0.791 & 0.668 & 20.8\% & 0.6\% & 0.8\% & 45.9\% & 6.7\% & 4.0\% \\
  & \textbf{\attackname} & {\bf 0.891} & {\bf 0.83} & {\bf 0.679} & {\bf 0.965} & {\bf 0.897} & {\bf 0.761} & {\bf 33.8\%} & {\bf 11.1\%} & {\bf 5.2\%} & {\bf 64.4\%} & {\bf 21.9\%} & {\bf 11.1\%} \\
      \hline
      \multirow{4}{*}{\textbf{Raw-AAM}} & \textbf{LRL-MIA} & 0.705 & 0.679 & 0.622 & 0.786 & 0.732 & 0.676 & 1.6\% & 0.8\% & 0.4\% & 9.6\% & 2.7\% & 6.1\% \\
  & \textbf{EncoderMI-T} & 0.703 & 0.661 & 0.601 & 0.785 & 0.732 & 0.656 & 1.9\% & 0.8\% & 0.2\% & 9.8\% & 2.7\% & 1.9\% \\
  & \textbf{TLK-MIA} & 0.703 & 0.661 & 0.601 & 0.785 & 0.732 & 0.656 & 1.9\% & 0.8\% & 0.2\% & 9.8\% & 2.7\% & 1.9\% \\
  & \textbf{\attackname} & {\bf 0.749} & {\bf 0.783} & {\bf 0.689} & {\bf 0.856} & {\bf 0.856} & {\bf 0.754} & {\bf 5.6\%} & {\bf 6.8\%} & {\bf 3.0\%} & {\bf 18.3\%} & {\bf 12.7\%} & {\bf 9.0\%} \\
      \hline
      \multirow{4}{*}{\textbf{Res-AP}} & \textbf{LRL-MIA} & 0.756 & 0.924 & 0.627 & 0.842 & 0.974 & 0.740 & 8.8\% & 6.6\% & 1.0\% & 24.4\% & 64.5\% & 7.4\% \\
  & \textbf{EncoderMI-T} & 0.747 & 0.887 & 0.606 & 0.841 & 0.974 & 0.740 & 8.8\% & 6.6\% & 1.0\% & 24.3\% & 64.7\% & 7.4\% \\
  & \textbf{TLK-MIA} & 0.747 & 0.887 & 0.606 & 0.841 & 0.974 & 0.740 & 8.8\% & 6.6\% & 1.0\% & 24.3\% & 64.7\% & 7.4\% \\
  & \textbf{\attackname} & {\bf 0.799} & {\bf 0.956} & {\bf 0.699} & {\bf 0.892} & {\bf 0.986} & {\bf 0.796} & {\bf 12.5\%} & {\bf 14.4\%} & {\bf 5.1\%} & {\bf 40.2\%} & {\bf 72.3\%} & {\bf 11.0\%}  \\
      \hline
      \multirow{4}{*}{\textbf{VGG-GE2E}} & \textbf{LRL-MIA} & 0.714 & 0.847 & 0.592 & 0.783 & 0.916 & 0.634 & 5.6\% & 9.8\% & 0.2\% & 17.2\% & 15.4\% & 2.8\% \\
  & \textbf{EncoderMI-T} & 0.711 & 0.827 & 0.574 & 0.785 & 0.916 & 0.624 & 5.5\% & 9.8\% & 0.1\% & 17.4\% & 15.4\% & 2.2\% \\
  & \textbf{TLK-MIA} & 0.711 & 0.827 & 0.574 & 0.785 & 0.916 & 0.624 & 5.5\% & 9.8\% & 0.1\% & 17.4\% & 15.4\% & 2.2\% \\
  & \textbf{\attackname} & {\bf 0.743} & {\bf 0.914} & {\bf 0.648} & {\bf 0.835} & {\bf 0.968} & {\bf 0.700} & {\bf 16.6\%} & {\bf 22.1\%} & {\bf 1.6\%} & {\bf 26.4\%} & {\bf 45.9\%} & {\bf 5.0\%} \\
      \hline
      \end{tabular}%
  }
  \label{tab:overall-comp-s1}%
\end{table*}%

\noindent {\bf Classifier-based attack model.}
The classifier-based attack model is a multilayer perceptron with one hidden layer
comprising 64 neurons and ReLU as the activation function.
It is trained by the Adam optimizer with 1e-3 learning rate for 1,000 epochs.
We utilize the cosine similarity to measure the similarity between two embeddings.
To reduce randomness, the training is repeated independently ten times
and the average results are reported.

\noindent {\bf Evaluation metrics.}
Following most prior work on MIA, we adopt two aggregate metrics: accuracy and AUROC.
Since it is highlighted in \cite{first-principle} that
a practical MIA should yield a high True Positive Rate (TPR)
at sufficiently low False Positive Rate (FPR), we also consider
TPR at 0.1\% (for VoxCeleb-2), 0.2\% (for LibriSpeech, each subset of which contains less than 1,000 speakers), and 1\% FPR.
When calculating these metrics, the number of training speakers
is set to be the same as that of non-training speakers, as accuracy is highly sensitive to the ratio between the numbers of positive and negative examples.

\noindent {\bf Baselines.}
We consider recent promising baselines designed for embedding models:
Li et al.~\cite{person-re-identification-MIA} (named LRL-MIA),
Tseng et al.~\cite{self-supervised-speech-membership}  (named TKL-MIA),
EncoderMI~\cite{EncoderMI},
and FaceAuditor~\cite{FACE-AUDITOR},
where EncoderMI has two variants EncoderMI-T and EncoderMI-V (the lest effective variant EncoderMI-S is not considered),
and FaceAuditor has two variants FaceAuditor-S and FaceAuditor-P/R.
Thus, there are six baselines.
Details of these attacks refer to \cref{sec:review-MIA} and Appendix~\ref{sec:oursvspriorMIA}.

\noindent {\bf Experimental designs.} Following the setting in recent works~\cite{first-principle, FACE-AUDITOR, loss-trajectory,MIA-Multi-Exit-Networks},
we will first assume in \cref{sec:evaluation} that the adversary can adopt the same architecture as the target SRS for the shadow SRS
and owns an auxiliary dataset following the same distribution as the target SRS's training dataset,
and then conduct ablation studies to investigate the effect of the dataset distribution shift and the architecture shift in \cref{sec:ablation-study}.

\subsection{Experimental Results}

\subsubsection{Overall Performance}\label{sec:overall-performance}
We evaluate the effectiveness of \attackname by comparing with the baselines in two settings:

\noindent {\bf Setting-1.}
We use all the voices of a target speaker in either $\mathcal{S}^{t}_{tr}$ or $\mathcal{S}^{t}_{ntr}$ for computing features,
simulating the scenario where the target speaker provides sufficient voices for MIA.
Also, all the imposters in $\mathcal{S}^{im}$ and all their voices in $\mathcal{V}^{im}$ are used to compute the inter-features,
assuming the number of queries to the target SRS is unconstrained (e.g., in the white-box scenario).
We do not build voice-number-dependent  attack models in this setting since each target speaker has a different number of voices,
and do not utilize the voice chunk splitting strategy since the number of voices is sufficient.

\noindent {\bf Setting-2.}
We use a voice-number-dependent (VND) attackmodel, randomly choose 10 voices per target speaker,
20 imposters, and 10 voices per imposter to compute the inter-features, and apply the voice chunk splitting strategy.

\attackname utilizes the mixing ratio training strategy in both settings.
We compare with LRL-MIA, EncoderMI-T,
and TKL-MIA in Setting-1,
because the other three baselines 
use classifier-based attack models requiring fixed number of voices per target speaker and fixed number of imposters,
thus only compared in Setting-2.
Notice that the following results apply for both white-box and black-box scenarios,
due to the unity of the feature extractor (cf.~\ref{sec:property-feature-extractor}).

\noindent {\bf Results in Setting-1.}
The results are reported in \tablename~\ref{tab:overall-comp-s1} 
where we only report results of $r=0$. \attackname performs better when $r=1$ (cf. \figurename~\ref{fig:alpha-exper}).
Overall, \attackname achieves the best performance
across the five target SRSs and the three datasets in terms of all the four metrics,
e.g., 83.5\%-99.4\% AUROC and 5.6\%-33.8\% TPR at 0.1\% FPR when $r=0$, i.e.,
all the voices used for membership inference of training speakers are different from their training voices.

We observe that the performance of \attackname varies with the target SRSs and datasets.
TDNN-CE and Raw-AAM are generally the least vulnerable to \attackname, in particular, on the dataset LibriSpeech.
It is because they were trained with the classification-based loss functions CE and AAM, respectively,
while the CE loss do not explicitly constrain the intra-similarity
and the constraint of AAM may be weaker than that of verification-based losses.
Regarding datasets, \attackname generally performs better on LibriSpeech than on VoxCeleb-2 when $r=0$.
It is probably because LibriSpeech contains fewer speakers than VoxCeleb-2,
hence the target SRSs trained with LibriSpeech are more likely to memorize the training speakers. 
Also, \attackname performs better on the English dataset VoxCeleb-2 than on the Chinese dataset KeSpeech. We conjure that it is because although they have the same number of training speakers, KeSpeech has much less training voices per training speaker than VoxCeleb-2, and a training speaker with less training voices is less likely to be memorized.

Comparing with the baselines, \attackname outperforms them on all target SRSs and all datasets in terms of all the metrics.
For example, on the target SRS LSTM-GE2E and the dataset VoxCeleb-2, when $r=0$,
\attackname improves the Accuracy, AUROC, TPR at 0.1\% FPR,
TPR at 1\% FPR by 19.4\%, 16.6\%, 31.9\%, and 50.6\%, respectively, compared to the most effective baseline.
The improvement mainly comes from two aspects. First, \attackname utilizes features that characterize both intra-similarity and inter-dissimilarity,
while these baselines only consider intra-similarity.
Second, \attackname utilizes much more features
and these features driven by carefully-established feature engineering
are comprehensive and complementary,
providing a better characterization of the differences between training and non-training speakers.

\begin{table}[t]
  \centering\setlength{\tabcolsep}{3pt}\vspace*{-1mm}
  \caption{The effectiveness of \attackname in Setting-2.
  }\vspace*{-2mm}
  \resizebox{0.48\textwidth}{!}{\begin{threeparttable}
    \begin{tabular}{c|c|ccc|ccc|cccccc}
    \hline
    \multicolumn{2}{c|}{\multirow{3}[4]{*}{}} & \multicolumn{3}{c|}{\multirow{2}{*}{\textbf{Accuracy}}} & \multicolumn{3}{c|}{\multirow{2}{*}{\textbf{AUROC}}} & \multicolumn{6}{c}{\textbf{TPR @ x\% FPR}} \\
    \multicolumn{2}{c|}{} & \multicolumn{3}{c|}{}  & \multicolumn{3}{c|}{} & {\bf x=0.1} & {\bf x=0.2} & {\bf x=0.1} & {\bf x=1} & {\bf x=1} & {\bf x=1} \\
\multicolumn{2}{c|}{} & \textbf{VC-2} & \textbf{LS} & \textbf{KS} & \textbf{VC-2} & \textbf{LS} & \textbf{KS} & \textbf{VC-2} & \textbf{LS} & \textbf{KS} & \textbf{VC-2} & \textbf{LS} & \textbf{KS}  \\
    \hline
    \multirow{4}{*}{\textbf{LSTM-GE2E}}
& \textbf{EncoderMI-V} & 0.649 & 0.866 & 0.632 & 0.72  & 0.932 & 0.770 & 2.0\% & 19.2\% & 4.4\% & 6.6\% & 35.2\% & 10.5\% \\
& \textbf{FaceAuditor-S} & 0.655 & 0.842 & 0.698 & 0.714 & 0.932 & 0.768 & 1.2\% & 16.8\% & 1.6\% & 5.3\% & 33.0\% & 7.3\% \\
& \textbf{FaceAuditor-P/R} & 0.614 & 0.768 & 0.615 & 0.691 & 0.863 & 0.773 & 1.6\% & 3.8\% & 2.8\% & 6.2\% & 14.5\% & 9.9\% \\
& \textbf{\attackname} & {\bf 0.785} & {\bf 0.976} & {\bf 0.794} & {\bf 0.861} & {\bf 0.994} & {\bf 0.885} & {\bf 7.2\%} & {\bf 62.1\%} & {\bf 13.6\%} & {\bf 24.4\%} & {\bf 82.7\%} & {\bf 24.4\%} \\
    \hline
     \multirow{4}{*}{\textbf{TDNN-CE}}
 & \textbf{EncoderMI-V} & 0.724 & 0.703 & 0.603 & 0.81  & 0.772 & 0.681 & 19.6\% & 1.1\% & 1.0\% & 28.2\% & 5.6\% & 4.8\% \\
 & \textbf{FaceAuditor-S} & 0.784 & 0.666 & 0.604 & 0.866 & 0.742 & 0.639 & {20.1\%} & 2.1\% & 0.1\% & 34.2\% & 4.9\% & 1.6\% \\
& \textbf{FaceAuditor-P/R} & 0.772 & 0.578 & 0.512 & 0.866 & 0.628 & 0.562 & 11.9\% & 0.3\% & 0.2\% & 30.9\% & 1.4\% & 1.9\% \\
& \textbf{\attackname} & {\bf 0.839} & {\bf 0.773} & {\bf 0.661} & {\bf 0.92} & {\bf 0.856} & {\bf 0.733} & {\bf 23.6\%} & {\bf 2.8\%} & {\bf 1.7\%}  & {\bf 42.9\%} & {\bf 8.1\%} & {\bf 6.6\%} \\
    \hline
     \multirow{4}{*}{\textbf{Raw-AAM}}
& \textbf{EncoderMI-V} & 0.657 & 0.657 & 0.606 & 0.709 & 0.708 & 0.658 & 2.7\% & 0.9\% & 0.2\% & 8.7\% & 2.0\% & 2.3\%  \\
 & \textbf{FaceAuditor-S} & 0.636 & 0.64 & 0.598 & 0.686 & 0.702 & 0.640 & 0.2\% & 0.2\% & 0.3\% & 2.3\% & 2.0\% & 1.8\% \\
 & \textbf{FaceAuditor-P/R} & 0.663 & 0.592 & 0.514 & 0.732 & 0.659 & 0.584 & 3.5\% & 0.4\% & 0.2\% & 6.7\% & 2.8\% & 1.3\% \\
      & \textbf{\attackname} & {\bf 0.697} & {\bf 0.764} & {\bf 0.650} & {\bf 0.774} & {\bf 0.827} & {\bf 0.701} & {\bf 3.8\%} & {\bf 3.1\%} & {\bf 1.0\%} & {\bf 8.8\%} & {\bf 5.1\%} & {\bf 2.8\%} \\
    \hline
     \multirow{4}{*}{\textbf{Res-AP}}
        & \textbf{EncoderMI-V} & 0.712 & 0.87 & 0.599 & 0.789 & 0.948 & 0.730 & 4.9\% & 33.1\% & 3.0\% & 13.7\% & 41.3\% & 6.4\%  \\
       & \textbf{FaceAuditor-S} & 0.722 & 0.885 & 0.678 & 0.794 & 0.961 & 0.750 & 4.9\% & 28.3\% & 3.1\% & 14.9\% & 52.5\% & 8.8\% \\
       & \textbf{FaceAuditor-P/R} & 0.672 & 0.697 & 0.549 & 0.744 & 0.771 & 0.630 & 4.2\% & 4.2\% & 0.6\% & 11.6\% & 8.7\% & 3.1\% \\
        & \textbf{\attackname} & {\bf 0.763} & {\bf 0.932} & {\bf 0.692} & {\bf 0.841} & {\bf 0.982} & {\bf 0.782} & {\bf 13.6\%} & {\bf 43.0\%} & {\bf 4.6\%} & {\bf 29.0\%} & {\bf 60.7\%} & {\bf 10.9\%} \\
    \hline
     \multirow{4}{*}{\textbf{VGG-GE2E}}
      & \textbf{EncoderMI-V} & 0.692 & 0.797 & 0.584 & 0.756 & 0.878 & 0.634 & 1.0\% & 4.7\% & 0.5\% & 11.8\% & 9.9\% & 2.0\% \\
        & \textbf{FaceAuditor-S} & 0.671 & 0.797 & 0.560 & 0.728 & 0.89 & 0.579 & 0.4\% & 4.8\% & 0.1\% & 5.9\% & 14.4\% & 0.6\% \\
        & \textbf{FaceAuditor-P/R} & 0.605 & 0.648 & 0.527 & 0.685 & 0.71 & 0.566 & 2.1\% & 1.1\% & 0.5\% & 7.3\% & 2.8\% & 1.9\% \\
      & \textbf{\attackname} & {\bf 0.708} & {\bf 0.934} & {\bf 0.637} & {\bf 0.776} & {\bf 0.98} & {\bf 0.686} & {\bf 6.3\%} & {\bf 46.7\%} & {\bf 0.8\%} & {\bf 15.5\%} & {\bf 65.8\%} & {\bf 3.5\%} \\
    \hline
    \end{tabular}%
    \begin{tablenotes}
      \item Note: VC-2, LS, and KS denote the dataset VoxCeleb-2, LibriSpeech, and KeSpeech, respectively. The ratio $r=0$.
    \end{tablenotes}
  \end{threeparttable}
  }
  \label{tab:overall-comp-s2}%
\end{table}%
\noindent {\bf Results in Setting-2.}
The results are reported in \tablename~\ref{tab:overall-comp-s2}.
where the results of $r=1$ is omitted since it is less challenging than $r=0$. 
Overall,
\attackname outperforms the baselines on all target SRSs and all datasets in Setting-2 as well, especially in terms of TPR.
For example, it improves the TPR at 0.2\% FPR from 4.8\% to 46.7\% on the target SRS VGG-GE2E and dataset LibriSpeech.
We conjecture that the improvement is mainly brought by our diverse features and the voice chunk splitting strategy.

We find that in most cases, the  performance of \attackname degrades compared to Setting-1,
which is possibly due to the reduced number of voices used for membership inference of target speakers in Setting-2.
More voices of a target speaker contribute to a more precise approximation
of the centroid embedding for that speaker, more precise statistics, and hence more precise features.
However, on the target SRSs Res-AP (for both VoxCeleb-2 and LibriSpeech)
and VGG-GE2E (for LibriSpeech only),
we find that \attackname achieves much higher TPR at 0.1\% or 0.2\% FPR in Setting-2 than in Setting-1.
This is probably attributed to the two strategies used in Setting-2,
voice-number-dependent attack models and voice chunk splitting strategy
which artificially increases the number of voices used for membership inference of target speakers.

\subsubsection{Additional Evaluation}\label{sec:add_eva}
Here we first study the difference of difficulty in inferring the membership between male and female speakers and then evaluate the effectiveness of \attackname on commercial SRSs. 

\begin{figure*}\centering
    \begin{minipage}{0.24\textwidth}
      \includegraphics[width=.95\textwidth]{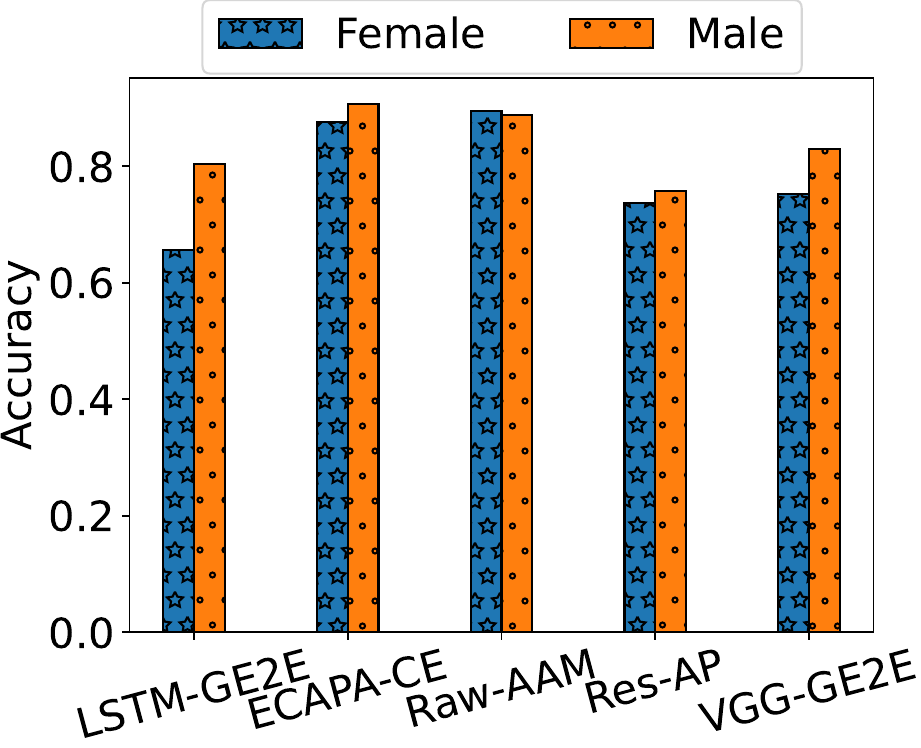}
    \end{minipage}
    \begin{minipage}{0.24\textwidth}
      \includegraphics[width=.95\textwidth]{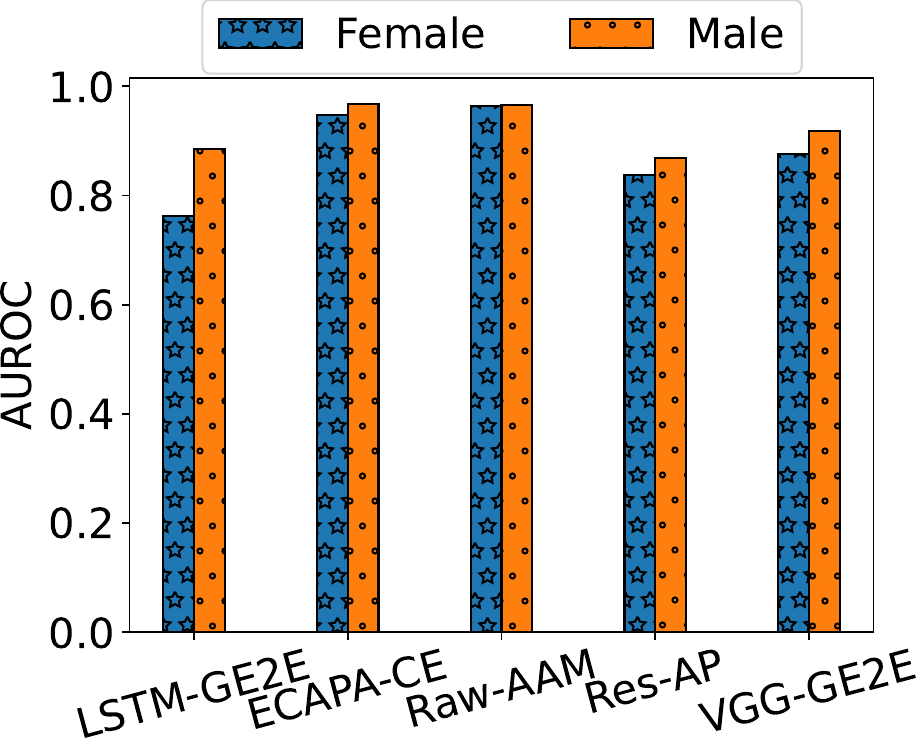}
    \end{minipage}
    \begin{minipage}{0.24\textwidth}
      \includegraphics[width=.95\textwidth]{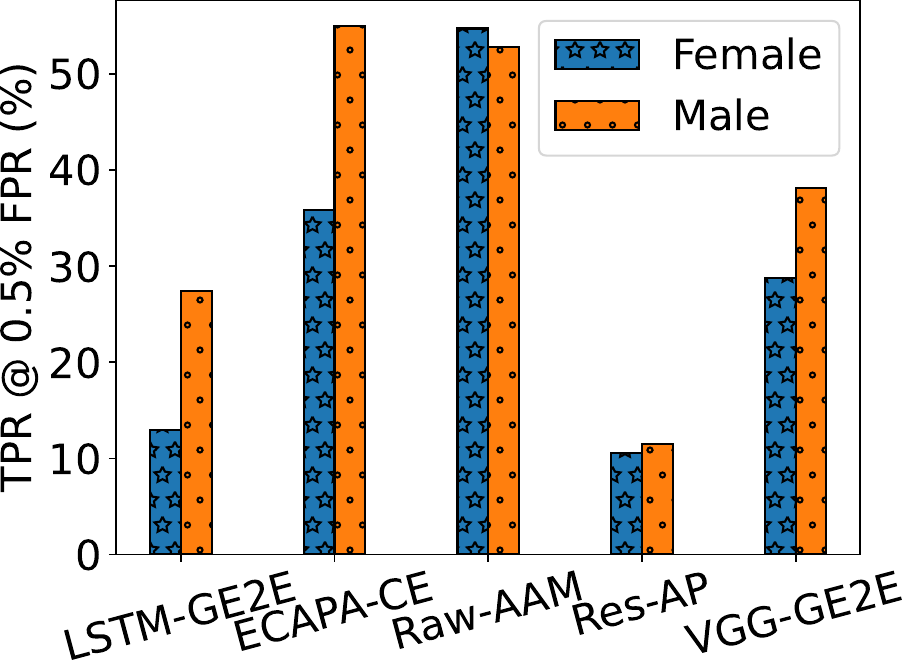}
    \end{minipage}
    \begin{minipage}{0.24\textwidth}
      \includegraphics[width=.95\textwidth]{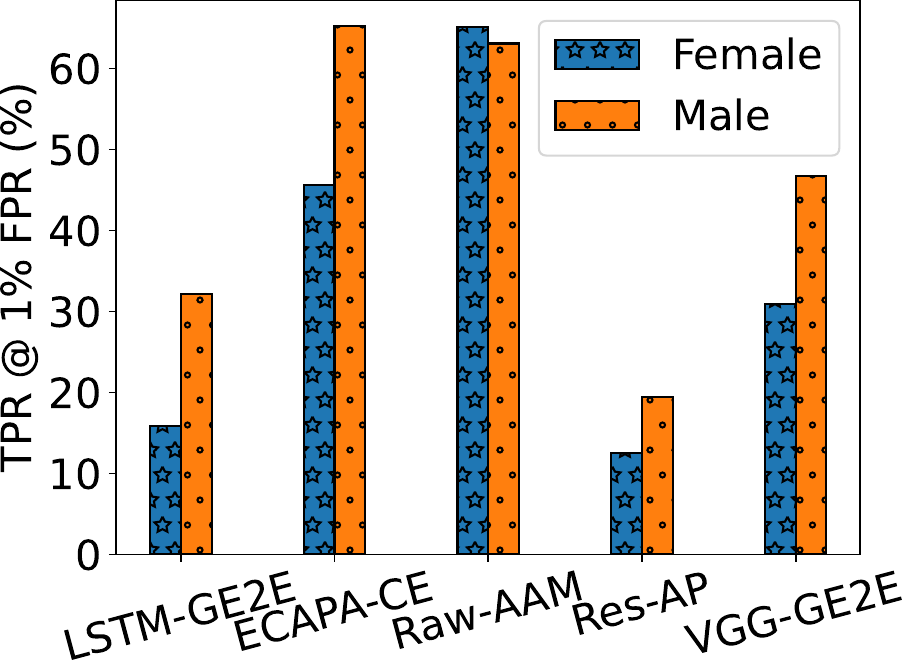}
    \end{minipage}
\vspace*{-2mm}
\caption{Comparison of the effectiveness of \attackname between female and male speakers.}
  \label{fig:gender_analysis}\vspace*{-1mm}
\end{figure*}

\smallskip \noindent {\bf \attackname for male and female speakers.} We compare the effectivess of \attackname between male and female speakers on the dataset VoxCeleb-2, and the results are shown in \figurename~\ref{fig:gender_analysis}. In most cases, \attackname achieves better inference performance for male speakers, indicating that it is more difficult to infer the speaker-level membership for female speakers. 

\begin{figure}[t]
    \centering
    \begin{subfigure}[b]{0.235\textwidth}
      \centering
      \includegraphics[width=1\textwidth]{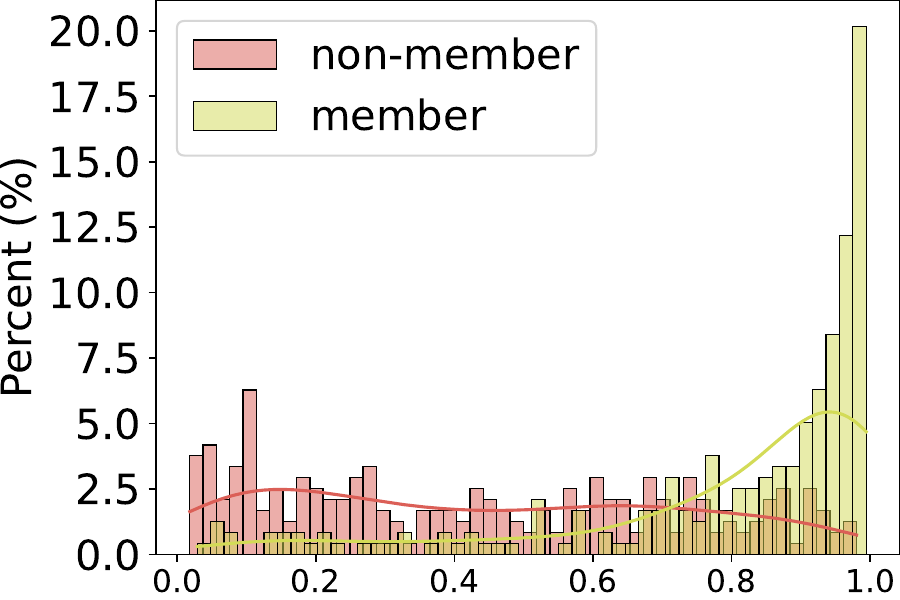}\vspace*{-2mm}
      \caption{Comparison of distribution of probability of being member between potential training and potential non-training speakers.}
      \label{fig:azure_analysis}
    \end{subfigure}
    \begin{subfigure}[b]{0.235\textwidth}
      \centering
      \includegraphics[width=1\textwidth]{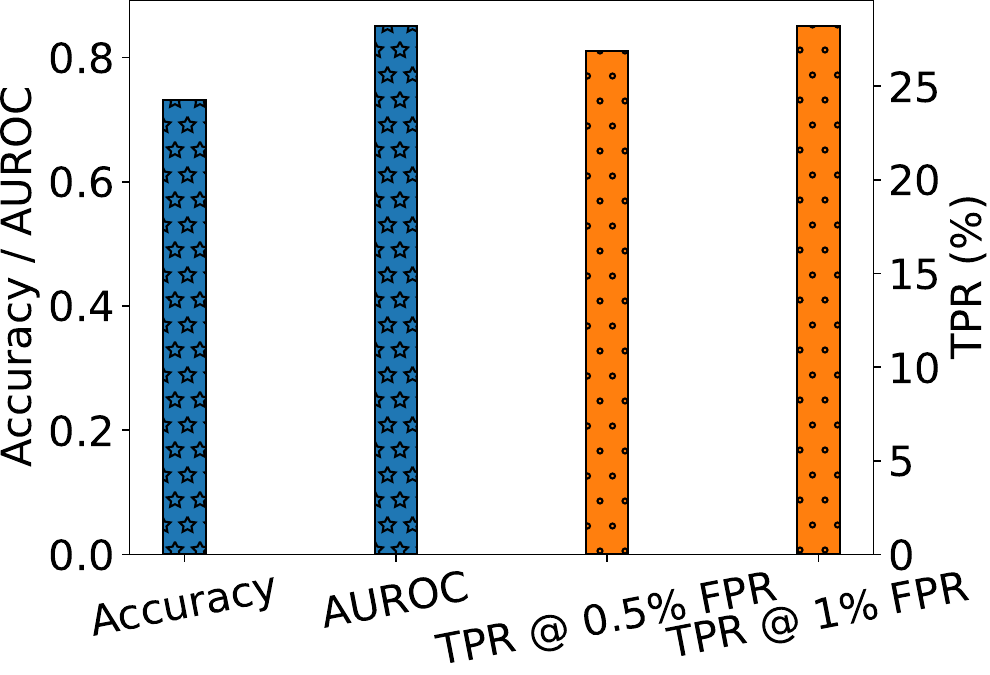}\vspace*{-2mm}
      \caption{Effectiveness of \attackname on the Microsoft Azure SRS.}
      \label{fig:azure_result}
    \end{subfigure}\vspace*{-3mm}
    \caption{The effectiveness of \attackname w.r.t. the overfitting level of the target SRS.}
    \label{fig:azure}\vspace*{-2mm}
  \end{figure}

\smallskip \noindent {\bf \attackname on commercial SRSs.} To demonstrate the practicability of \attackname, we evaluate its effectiveness on commercial SRSs. However, the training datasets of commercial SRSs are unknown, so we don't have the ground-truth required to calculate Accuracy, etc. We followed the practice of setting potential ground-truth~\cite{EncoderMI}. Since commercial SRS usually has its intended language, we used a dataset of that language as the members, and a dataset of other languages as the non-members. We consider the Microsoft Azure SRS~\cite{microsoft-azure-vpr} and regard the speakers of VoxCeleb-2 and KeSpeech as the potential members and non-members, respectively, since Azure SRS targets the English language. Since the query to Azure SRS is charged, we only consider the combination of intra-features for \attackname, set the number of inference voices $N=5$, do not apply voice chunk splitting, and utilize group enrollment and enrollment voice concatenation. We use the Res-AP model trained on the VoxCeleb-2 dataset as the shadow SRS. As shown in \figurename~\ref{fig:azure_analysis}, we observed that the potential members and non-members are statistically distinguishable w.r.t. the classifier’s probability, justifying the reasonableness of using potential ground-truth. As shown in \figurename~\ref{fig:azure_result}, SLMIA-SR performs effectively on Azure with 85\% AUROC and over 26\% TPR at 0.5\% FPR. We empahisze that \attackname will performs better on the Azure SRS if we use the inter-features, set a large $N$, or apply voice chunk splitting. 

\subsubsection{Effectiveness of Individual Components}\label{sec:effect-individual}
The success of \attackname may be attributed to the combination of several components,
including the feature extractor with numerous features, the mixing training strategy,
the voice-number dependent attack models, the voice chunk splitting strategy,
and the group enrollment and the voice concatenation techniques. 
Therefore,we perform additional experiments to understand the effectiveness and necessity of each component.

\noindent {\bf Settings.}
Since we have already considered all the combinations of the five target SRSs and the three datasets in the previous experiments,
here we target the VoxCeleb-2 dataset
since it contains much more speakers than LibriSpeech and the target SRSs trained on this dataset are generally less vulnerable to \attackname than those trained on LibriSpeech.
Also, we consider the target SRS LSTM-GE2E since it is lightweight and therefore computationally efficient,
but is not the most vulnerable to \attackname among the five target SRSs.

\begin{figure}[t]
    \centering
      \includegraphics[width=.46\textwidth]{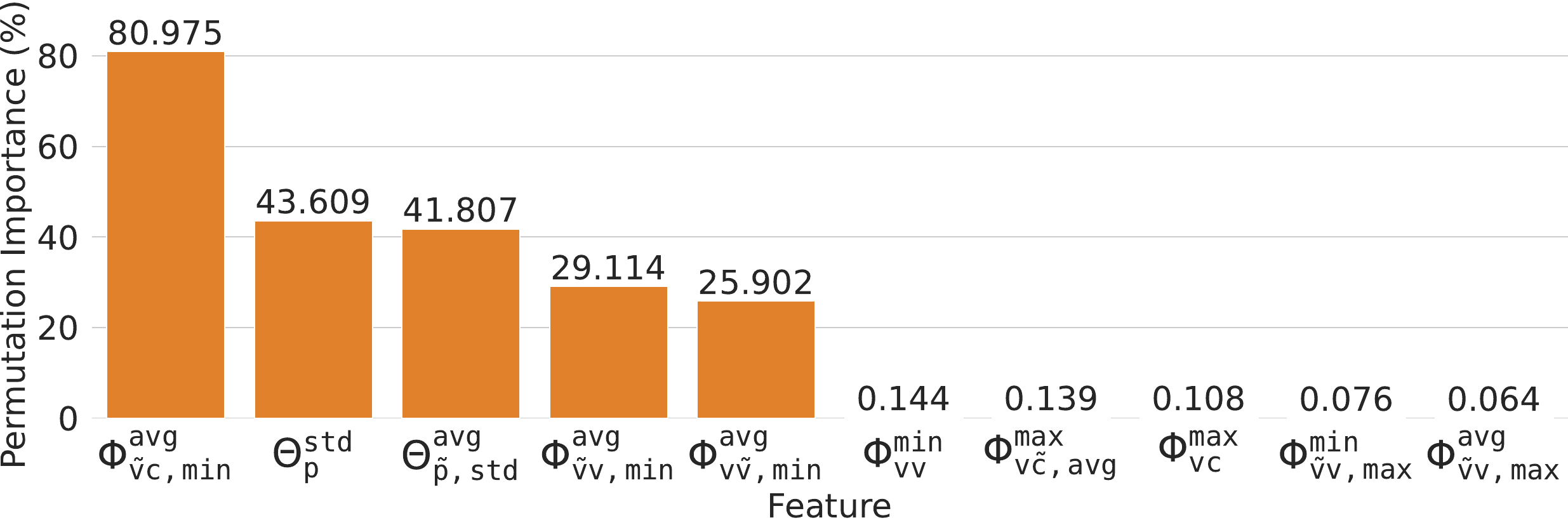}\vspace*{-2mm}
      \caption{Features with the top-5 highest and lowest PI}
      \label{fig:ALL-Ens-PI}\vspace*{-3mm}
\end{figure}

\begin{figure}[t]
  \begin{minipage}{0.23\textwidth}
    \centering
    \includegraphics[width=.95\textwidth]{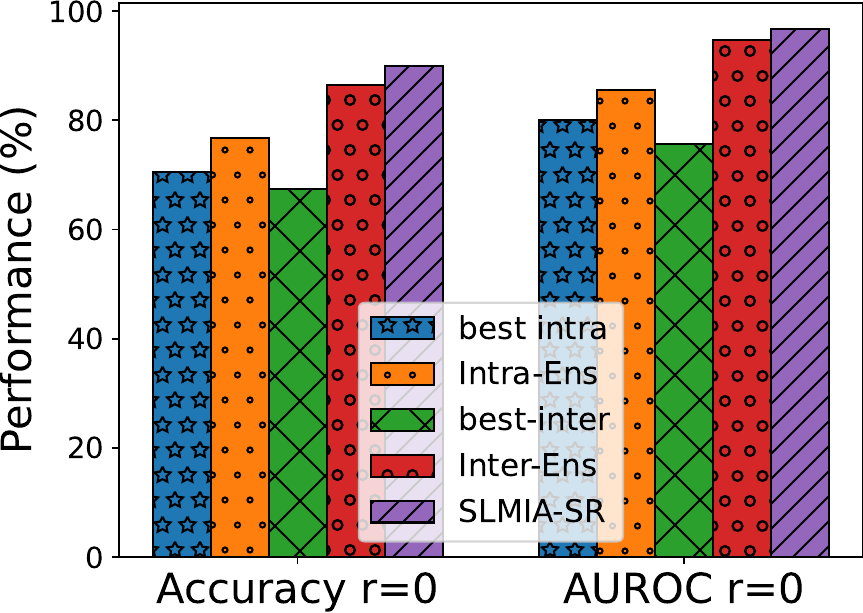}
  \end{minipage}
  \begin{minipage}{0.23\textwidth}
    \centering
    \includegraphics[width=.95\textwidth]{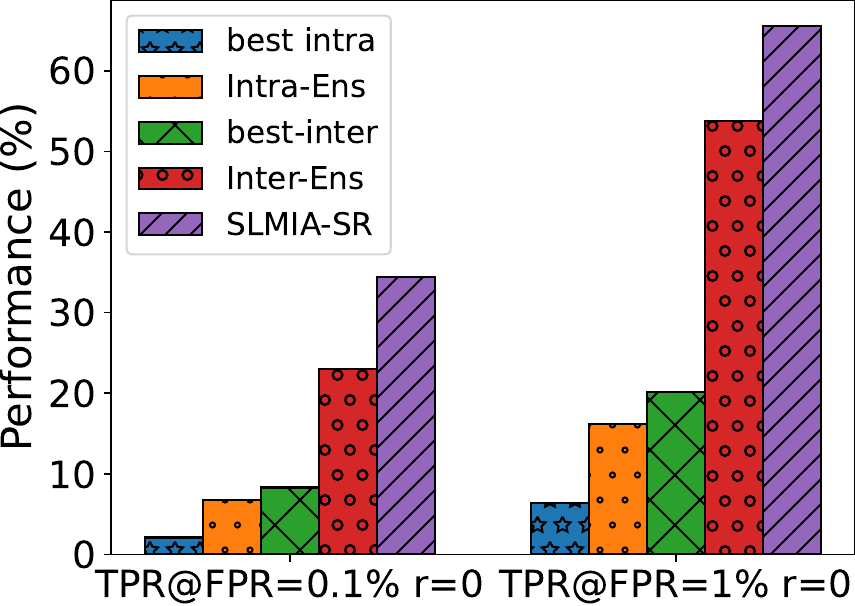}
  \end{minipage}\vspace*{-2mm}
  \caption{Comparison of \attackname with its variants.}
  \label{fig:comp-variant}\vspace*{-3mm}
\end{figure}

For comprehensive evaluation, we will also consider the following three variants of \attackname:
the attack using the threshold-based attack model with each single feature (denoted by the corresponding feature notation),
the attack using the classifier-based attack model with all the intra-features (denoted by Intra-Ens),
and the attack using the classifier-based attack model with all the inter-features (denoted by Inter-Ens).

\noindent {\bf Contribution of single feature.}
We conduct two experiments to understand the contribution of each feature to \attackname.

First, we utilize and report the permutation importance (PI)~\cite{PI} of each feature.
The PI of a feature to the classifier
is the change of a metric on the test data by permuting this feature across the test data.
We randomly permutate each feature 10 times and report 
the average change of all the metrics 
over 10 permutations.
The top-5 highest/lowest ones are depicted in \figurename~\ref{fig:ALL-Ens-PI}.
We find that all the features have positive PIs,
indicating that they all contribute to \attackname,
and the top-5 highest features 
have never been used in prior MIA,
which is one of the reasons why \attackname outperforms them.

Second, we compare \attackname with its variants,
and the results are shown in \figurename~\ref{fig:comp-variant}.
We find that Intra-Ens outperforms the best sole intra-feature,
Inter-Ens outperforms the best sole inter-feature,
and \attackname outperforms both Intra-Ens and Inter-Ens, in terms of all the metrics.
This further justifies the contribution of each feature
and demonstrates that these features can complement each other.

\begin{figure*}
  \centering
  \includegraphics[width=0.95\textwidth]{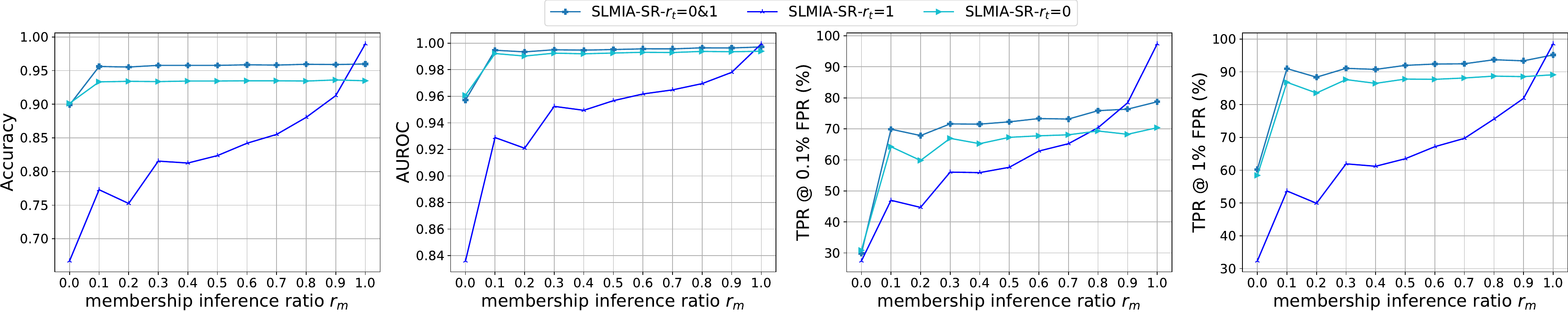}\vspace*{-2mm}
\caption{Effectiveness of \attackname w.r.t. the ratio $r$.}
\label{fig:alpha-exper}\vspace*{-2mm}
\end{figure*}

\begin{figure*}\centering
    \begin{minipage}{0.24\textwidth}
      \includegraphics[width=.95\textwidth]{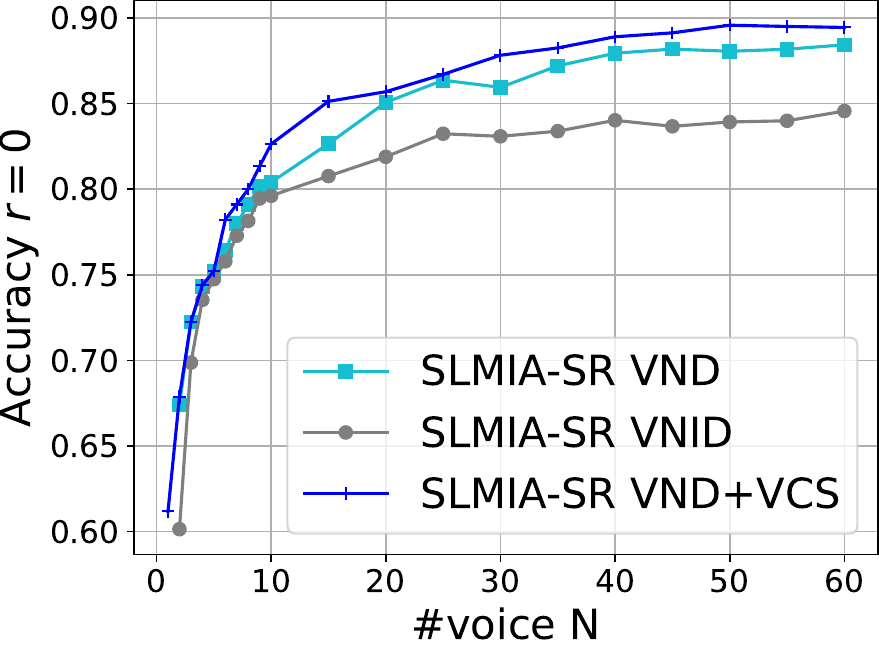}
    \end{minipage}
    \begin{minipage}{0.24\textwidth}
      \includegraphics[width=.95\textwidth]{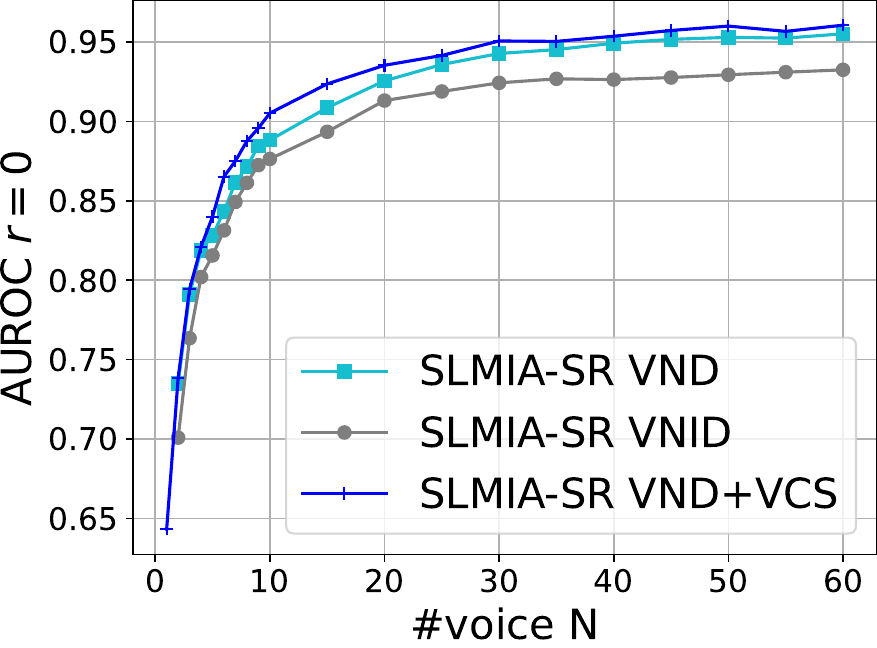}
    \end{minipage}
    \begin{minipage}{0.24\textwidth}
      \includegraphics[width=.95\textwidth]{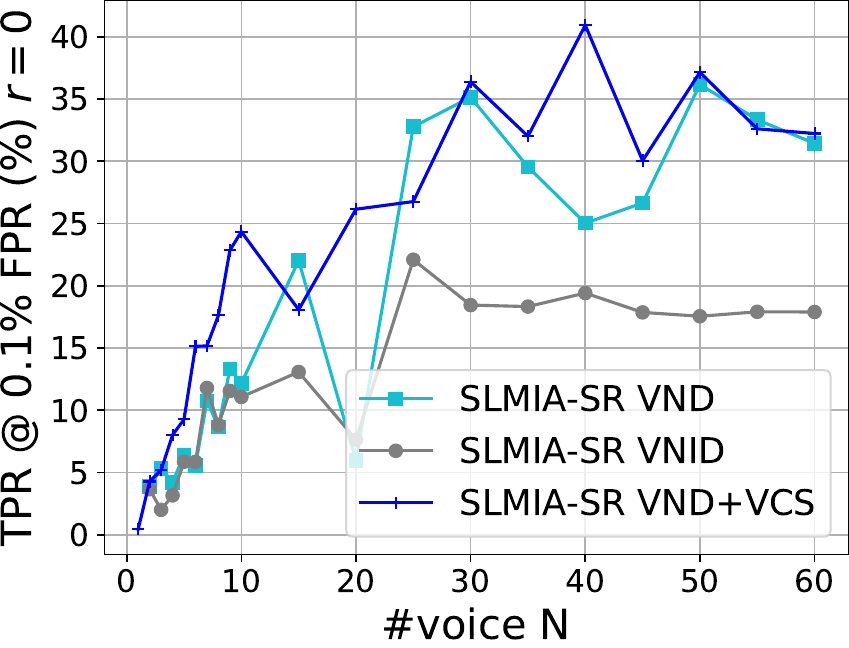}
    \end{minipage}
    \begin{minipage}{0.24\textwidth}
      \includegraphics[width=.95\textwidth]{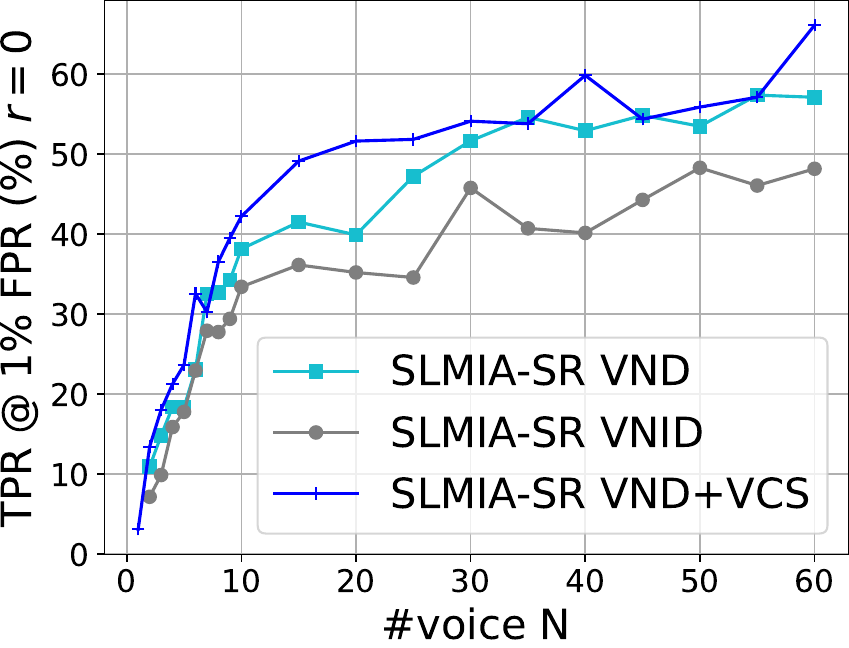}
    \end{minipage}
\vspace*{-2mm}\caption{Effectiveness of \attackname w.r.t. the number of voices $N$.}
  \label{fig:num-voice-exper}\vspace*{-2mm}
\end{figure*}

\begin{figure*}\centering
  \begin{minipage}{0.24\textwidth}
    \includegraphics[width=0.95\textwidth]{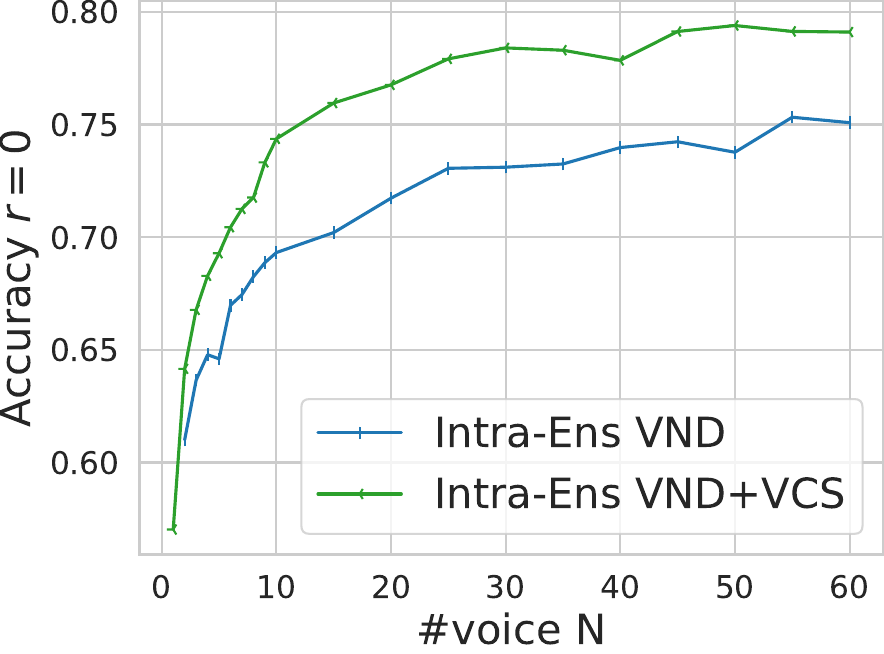}
  \end{minipage}
  \begin{minipage}{0.24\textwidth}
    \includegraphics[width=0.95\textwidth]{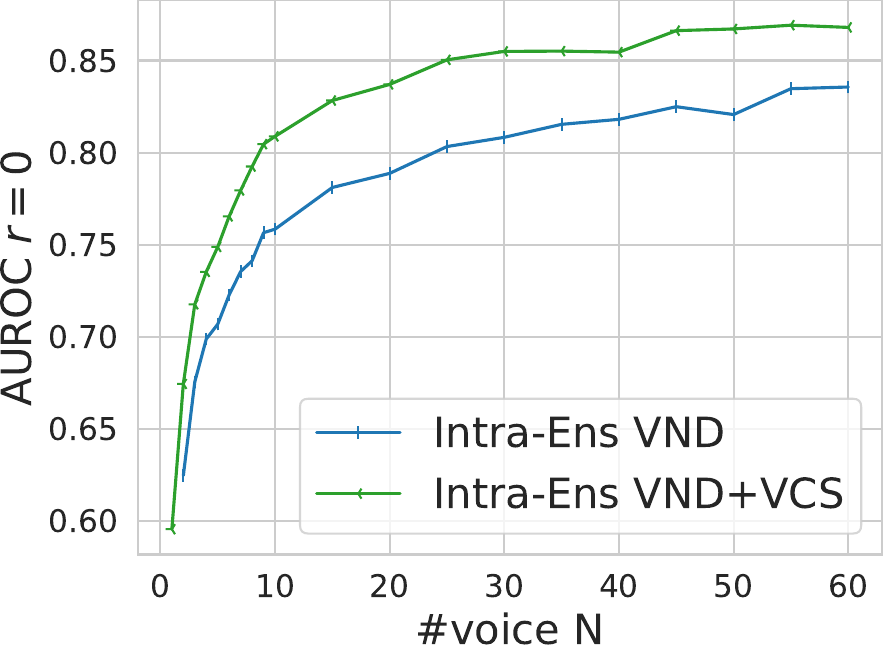}
  \end{minipage}
  \begin{minipage}{0.24\textwidth}
    \includegraphics[width=0.95\textwidth]{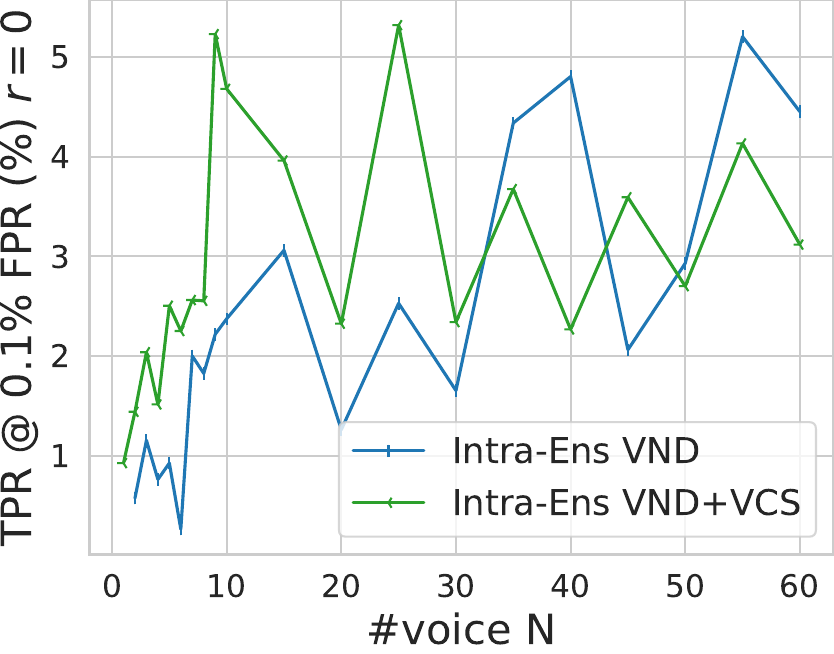}
  \end{minipage}
  \begin{minipage}{0.24\textwidth}
    \includegraphics[width=0.95\textwidth]{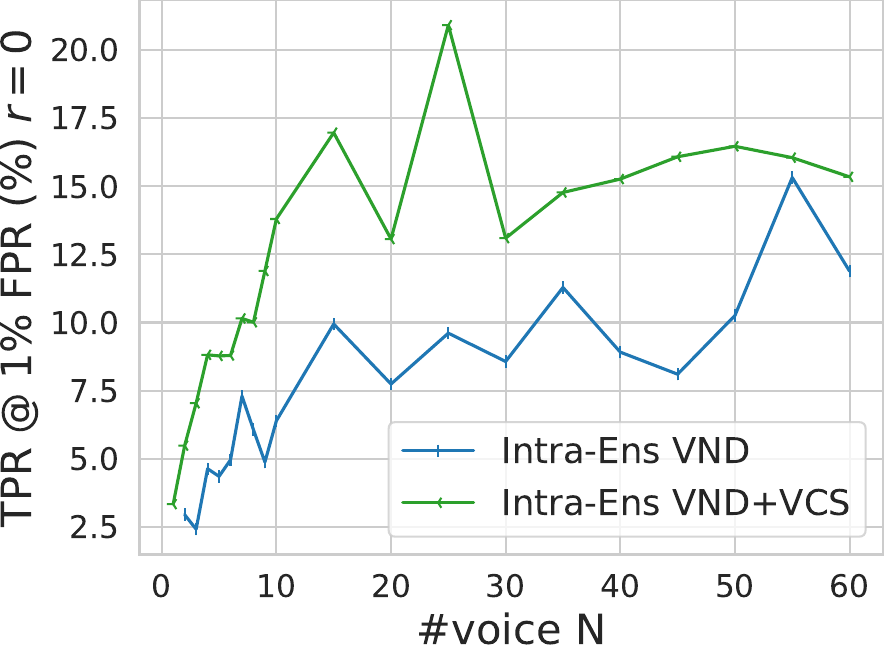}
  \end{minipage}\vspace*{-2mm}
\caption{Effectiveness of our voice chunk splitting.}
\label{fig:effectiveness-VCS-intra-ens}\vspace*{-2mm}
\end{figure*}

\noindent {\bf Effectiveness of mixing ratio training.}
Let $r_t$ (resp. $r_m$) denote the ratio $r$ used in attack model building
(resp. membership inference).
To understand the effect of the ratio $r_m$ and the effectiveness of our mixing training strategy,
we vary $r_m$ from 0 to 1 with step 0.1.
For each training speaker, $N\times r_m$ voices are randomly sampled from $\mathcal{V}^{t}_{tr}$
and $N\times (1-r_m)$ voices are randomly sampled from $\mathcal{V}^{t}_{ntr,tr}$,  forming the inference voices of the speaker.
The results are depicted in \figurename~\ref{fig:alpha-exper}, where
\attackname-$r_t$=$0\&1$, \attackname-$r_t$=$0$ and \attackname-$r_t$=$1$
denote \attackname with the attack models trained with both $r_t=0$ and $r_t=1$ (i.e., our mixing
ratio training strategy), with only $r_t=0$, and with only $r_t=1$, respectively.
We observe that the performance of \attackname increases with the ratio $r_m$.
Though \attackname-$r_t$=$0\&1$ performs worse than \attackname-$r_t$=$0$ (resp. \attackname-$r_t$=$1$) at $r_m=0$ (resp. $r_m=1$),
it performs much better at the other values of $r_m$.
This demonstrates the effectiveness of our mixing training strategy
in enhancing the generalizability of \attackname to different ratios.

\noindent {\bf Effectiveness of voice-number-dependent attack models.}
To understand the effect of the number $N$ of voices provided for membership inference
and the effectiveness of the voice-number-dependent attack models,
we vary $N$ from 1 to 10 with step 1 and from 15 to 60 with step 5.
We train both VND and VNID attack models (denoted by \attackname VND and \attackname VNID).
The results are depicted in \figurename~\ref{fig:num-voice-exper}.
We  observe that the performance of attack models increases with $N$,
probably because more voices help compute more precise features.
As expected, the VND attack models (i.e., \attackname VND) perform better than the single VNID attack model (i.e., \attackname VNID),
because features vary significantly with $N$ (cf. \figurename~\ref{fig:num-voice-feature-value}).
This indicates the effectiveness and necessity of building voice-number-dependent attack models.

\noindent {\bf Effectiveness of voice chunk splitting.}
In our voice chunk splitting approach, we set the window size $w$ to 3,200 milliseconds
and the window step $s$ to $\frac{w}{2}$, following the common practice in speech signal processing~\cite{rabiner1978digital}.
During splitting, when a chunk is shorted than the window size, it is padded with zero values to fulfill the window size
if its length is no shorter than 70\% of the window size, otherwise omitted.
The results are also depicted in \figurename~\ref{fig:num-voice-exper} (curve \attackname VND+VCS).
Compared with \attackname VND,
we observe that our voice chunk splitting improves the performance of \attackname VND
 with a little exception on TPR at 0.1\% FPR when $N>10$,
indicating that our voice chunk splitting can more reliably improve TPR at low FPR when $N$ is small.
For example, our voice chunk splitting improves the TPR at 0.1\% FPR from 12\% to 25\% when $N=10$.
To further understand the effectiveness of our voice chunk splitting,
we also apply it to Intra-Ens, which is less effective than \attackname according to \figurename~\ref{fig:comp-variant}.
As shown in \figurename~\ref{fig:effectiveness-VCS-intra-ens}, the improvement brought by our voice chunk splitting is more significant.

\noindent {\bf Effectiveness of enrollment voice concatenation and group enrollment.}
While both enrollment voice concatenation and group enrollment can effectively reduce the number of queries to the target SRS in the black-box scenario,
they may affect the performance of \attackname. To understand the effect, we set $N=10$, $M=20$, and $K=10$, the same as Setting-2.
The effect of the number $M$  of imposters and the number $K$ of voices per imposter refers to Appendix~\ref{sec:more-result-imposter-num-imposter-voice}.
We partition the features into fixed groups according to their similarity and distance types.
We train one VND attack model with our mixing training strategy
for each group of features.
Voice chunk splitting is not utilized since it will increase the number of queries in the black-box scenario.
The results are reported in \tablename~\ref{tab:reduce-query-exper}.
We can observe that our group enrollment technique reduces the number of queries by nearly half with no effect on the performance of \attackname,
which is not surprising, as group enrollment does not change the recognition scores.
Though our enrollment voice concatenation technique changes the recognition scores,
the performance of \attackname  decreases slightly and even increases in some cases,
indicating that the embedding of the concatenated voice can well approximate the centroid of voice embeddings.

\begin{table}[t]
    \centering\renewcommand\arraystretch{1.15}\setlength\tabcolsep{2pt}
    \caption{The effectiveness of enrollment voice concatenation and group enrollment.
    ``Recog'' denotes recognition.}\vspace*{-1mm}
    \scalebox{0.81}{\begin{threeparttable}
      \begin{tabular}{c|c|c|cc|cc|cccc|ccc}
      \hline
      \multicolumn{3}{c|}{\multirow{3}[6]{*}{}} & \multicolumn{2}{c|}{\multirow{2}{*}{\textbf{Accuracy}}} & \multicolumn{2}{c|}{\multirow{2}{*}{\textbf{AUROC}}} & \multicolumn{4}{c|}{\textbf{TPR @ x\% FPR}} & \multicolumn{3}{c}{\multirow{2}{*}{\textbf{\#Query (Target)}}} \\
      \multicolumn{3}{c|}{} & \multicolumn{2}{c|}{} & \multicolumn{2}{c|}{} & \multicolumn{2}{c}{\textbf{x=0.1}} & \multicolumn{2}{c|}{\textbf{x=1}} &  \multicolumn{3}{c}{} \\
      \multicolumn{3}{c|}{} & \boldmath{}\textbf{$r$=1}\unboldmath{} & \boldmath{}\textbf{$r$=0}\unboldmath{} & \boldmath{}\textbf{$r$=1}\unboldmath{} & \boldmath{}\textbf{$r$=0}\unboldmath{} & \boldmath{}\textbf{$r$=1}\unboldmath{} & \boldmath{}\textbf{$r$=0}\unboldmath{} & \boldmath{}\textbf{$r$=1}\unboldmath{} & \boldmath{}\textbf{$r$=0}\unboldmath{} &  {\bf Enroll} & {\bf Recog.} & {\bf Total} \\
      \hline
      \multirow{3}{*}{\textbf{Intra}} & \multirow{2}{*}{\makecell[c]{\textbf{Centroid-} \\ {\bf based}}} & \textbf{Baseline} & 0.797  & 0.684  & 0.959  & 0.747  & 9.9\% & 2.4\% & 38.9\% & 7.6\% & 10 & 10 & 20 \\
            &       & \textbf{Concat} & 0.786  & 0.681  & 0.950  & 0.742  & 11.6\% & 2.4\% & 31.1\% & 6.1\% & 1 & 10 & 11 \\
      \cline{2-14}     & \textbf{Pairwise} & \textbf{Baseline} & 0.798  & 0.695  & 0.961  & 0.759  & 8.5\% & 2.0\% & 37.2\% & 6.5\% & 9 & 45 & 54 \\
      \hline
      \multirow{9}{*}{\textbf{Inter}} & \multirow{2}{*}{\makecell[c]{\textbf{Centroid-} \\ {\bf centroid}}} & \textbf{Baseline} & 0.722  & 0.636  & 0.805  & 0.682  & 14.1\% & 2.5\% & 28.4\% & 10.4\% & 10 & 20 & 30 \\
            &       & \textbf{Concat} & 0.738  & 0.648  & 0.821  & 0.714  & 9.3\% & 2.3\% & 23.1\% & 9.3\% & 1 & 20 & 21 \\
  \cline{2-14}          & \multirow{2}{*}{\makecell[c]{\textbf{Centroid-} \\{\bf voice}}} & \textbf{Baseline} & 0.769  & 0.651  & 0.852  & 0.715  & 17.1\% & 4.0\% & 26.5\% & 8.5\% & 10 & 200 & 210 \\
            &       & \textbf{Concat} & 0.730  & 0.643  & 0.809  & 0.702  & 11.1\% & 2.8\% & 21.4\% & 8.2\% & 1 & 200 & 201 \\
  \cline{2-14}          & \multirow{4}{*}{\makecell[c]{\textbf{Voice-} \\{\bf centroid}}} & \textbf{Baseline} & {0.782} & {0.645} & {0.870} & {0.693} & {19.2\%} & {4.9\%} & {27.8\%} & {8.2\%} & 200 & 200 & 400 \\
            &       & \textbf{Group} & {0.782} & {0.645} & {0.870} & {0.693} & {19.2\%} & {4.9\%} & {27.8\%} & {8.2\%} & 200 & 10 & 210 \\
           &       & \textbf{Concat} & {0.794} & {0.651} & {0.878} & {0.696} & {17.4\%} & {4.0\%} & {29.6\%} & {8.4\%} & 20 & 200 & 220 \\
            &       & \makecell[c]{\textbf{Group+} \\ {\bf Concat}} & {0.794} & {0.651} & {0.878} & {0.696} & {17.4\%} & {4.0\%} & {29.6\%} & {8.4\%}   & 20 & 10   & 30 \\
\cline{2-14}          & \multirow{2}{*}{\textbf{Voice-voice}} & {\bf Baseline} & 0.830  & 0.675  & 0.918  & 0.731  & 12.3\% & 2.2\% & 36.0\% & 9.9\% & 10 & 2000 & 2010 \\
& & {\bf Group} & 0.830  & 0.675  & 0.918  & 0.731  & 12.3\% & 2.2\% & 36.0\% & 9.9\% & 10 & 200 & 210 \\
 \hline
      \end{tabular}%
      \begin{tablenotes}
        \item Note: (i) Refer to \tablename~\ref{tab:number-query} for the computation of the number of queries.
        (ii) As comparison, when the adversary has white-box access to the shadow SRS including the embeddings, 
        the total number of queries to the shadow SRS is $N+M*K=210$. (iii) The row w/o and w/ ``Group" show results for the speaker verification and identification tasks, respectively. They have the same effectiveness but different queries.
      \end{tablenotes}
    \end{threeparttable}
    }
    \label{tab:reduce-query-exper}%
  \end{table}%

\section{Ablation Study}\label{sec:ablation-study}
In this section, we perform ablation studies to understand the effects of the overfitting level of the target SRS
and the effects of the differences in dataset distribution and model architecture between the target and shadow SRSs.
Here we only incorporate all intra-features into \attackname (i.e., Intra-Ens)
since it requires fewer queries than incorporating all the inter-features (i.e., Inter-Ens),
according to the results in \tablename~\ref{tab:reduce-query-exper}.
We set the number $N$ of voices per target speaker to 40 since
according to \figurename~\ref{fig:num-voice-exper}, the accuracy and AUROC almost converge at $N\geq 40$.

\begin{figure}[t]
    \centering
    \begin{subfigure}[b]{0.235\textwidth}
      \centering
      \includegraphics[width=1\textwidth]{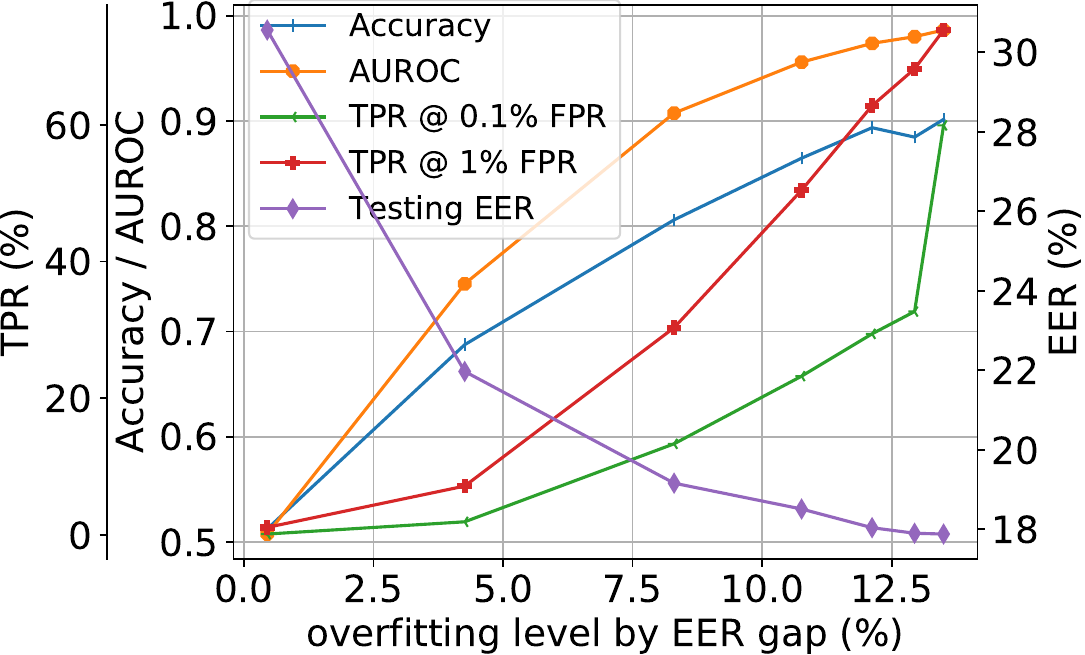}\vspace*{-2mm}
      \caption{r=1}
    \end{subfigure}
    \begin{subfigure}[b]{0.235\textwidth}
      \centering
      \includegraphics[width=1\textwidth]{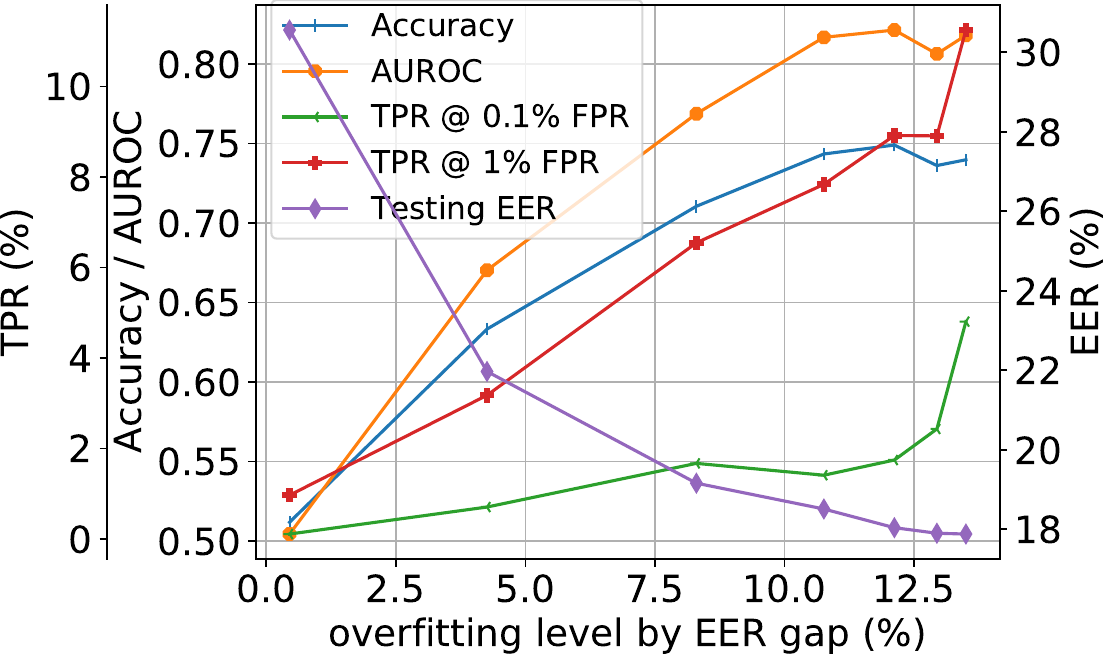}\vspace*{-2mm}
      \caption{r=0}
    \end{subfigure}\vspace*{-3mm}
    \caption{The effectiveness of \attackname w.r.t. the overfitting level of the target SRS.}
    \label{fig:overfitting-level}\vspace*{-2mm}
  \end{figure}

\noindent\textbf{Overfitting level of the target SRS.}
The same as in \cref{sec:effect-individual}, we use the LSTM-GE2E SRS and the VoxCeleb-2 dataset.
We measure the overfitting level of the target SRS by analyzing the gap between training EER and testing EER.
Specifically, we randomly sample 50,000 trials from the training voices $\mathcal{V}_{tr}^{t}$ of the training speakers $\mathcal{S}^{t}_{tr}$
and 50,000 trails from the non-training voices $\mathcal{V}^{t}_{ntr,ntr}$ of non-training speakers $\mathcal{S}^{t}_{ntr}$,
on which the training and testing EER are computed, respectively.
We control the EER gap by varying the number of training epochs
from 40 to 1800.
The results are reported in \figurename~\ref{fig:overfitting-level} under both $r=1$ and $r=0$.

We find that the effectiveness of \attackname generally increases with the overfitting level.
This is not surprising since the success of membership inference attacks lies in the overfitting of target models.
We also find that the testing EER decreases with the increase of the overfitting level,
indicating that the target SRS should have a large overfitting level to achieve low EER, which in turn benefits \attackname.
For example, to achieve about 18\% testing EER, which is still a moderate performance,
the target SRS has an overfitting level of about 12\% where \attackname achieves 0.85 AUROC and
2\% TPR at 0.1\% FPR even when $r=0$.
The results demonstrate that it is nontrivial to balance the recognition performance
and the resilience against \attackname.

\begin{figure}
  \begin{minipage}{0.235\textwidth}
    \centering
    \includegraphics[width=.95\textwidth]{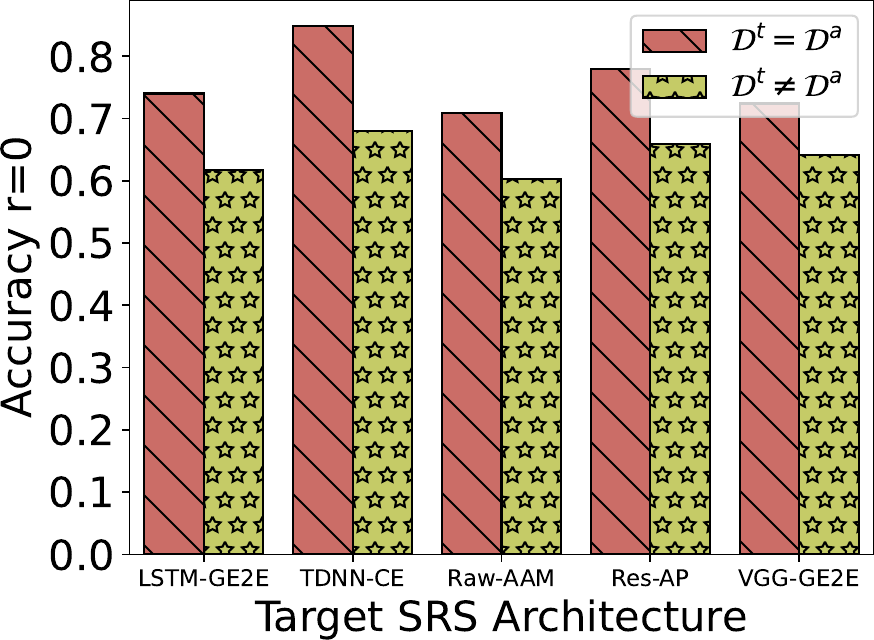}
  \end{minipage}
  \begin{minipage}{0.235\textwidth}
    \centering
    \includegraphics[width=.95\textwidth]{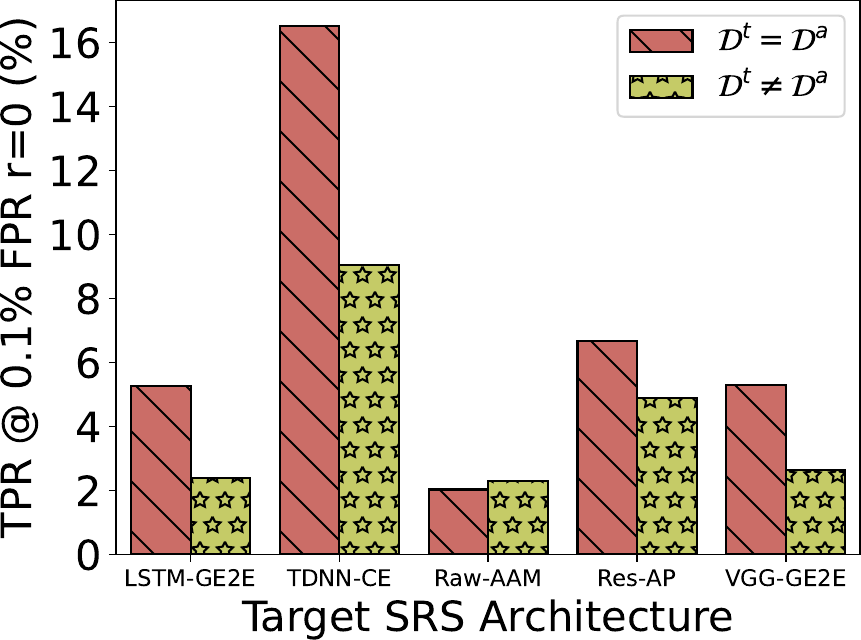}
  \end{minipage}\vspace*{-1mm}
  \caption{Effect of the dataset distribution.}
  \label{fig:dataset-distribution}\vspace*{-3mm}
\end{figure}

\noindent\textbf{Disjoint datasets.}
We relax the assumption in \cref{sec:threat-model} that the adversary has an auxiliary speaker dataset $\mathcal{S}^{a}$
which is sampled from the same distribution as the target SRS's training speaker dataset $\mathcal{S}^{t}_{tr}$.
We denote by $\mathcal{D}^a$ and $\mathcal{D}^{t}_{tr}$ the distributions of $\mathcal{S}^a$ and $\mathcal{S}^t_{tr}$, respectively.
We compare the effectiveness of \attackname between when $\mathcal{D}^a = \mathcal{D}^{t}_{tr}$ and when $\mathcal{D}^a \neq \mathcal{D}^{t}_{tr}$.
For $\mathcal{D}^a = \mathcal{D}^{t}_{tr}$, both $\mathcal{S}^{a}$ and $\mathcal{S}^{t}_{tr}$ are sampled from the VoxCeleb-2 dataset, but are disjoint from each other in the speakers and voices.
For $\mathcal{D}^a \neq \mathcal{D}^{t}_{tr}$, $\mathcal{S}^{a}$ and $\mathcal{S}^{t}_{tr}$ are respectively sampled from the LibriSpeech and VoxCeleb-2 datasets.
Recall that the training dataset of the shadow SRS is a subset of the auxiliary dataset $S^a$.
The accuracy and TPR at 0.1\% FPR when $r=0$ are shown in \figurename~\ref{fig:dataset-distribution},
while other results refer to Appendix~\ref{sec:more-results-ablation-dataset}.
We observe that \attackname still achieves good performance with at least 2\% TPR at 0.1\% FPR and 60\% accuracy when $r=0$,
a more challenging setting than $r=1$.
Aligning with prior works~\cite{loss-trajectory,membership-first},
the effectiveness of \attackname decreases with a dataset distribution shift in most cases,
probably because training datasets with different distributions make the target and shadow SRSs learn different speaker embedding mappings.

\begin{figure}
  \begin{minipage}{0.235\textwidth}
    \centering
    \includegraphics[width=.98\textwidth]{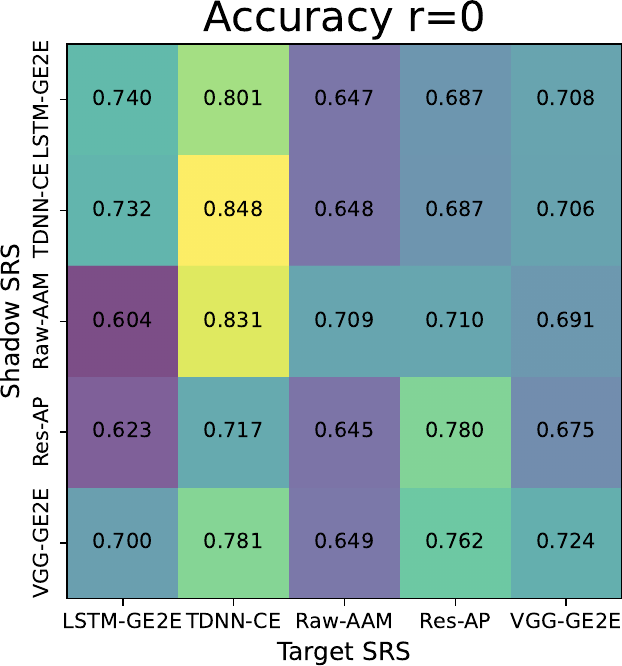}
  \end{minipage}
  \begin{minipage}{0.235\textwidth}
    \centering
    \includegraphics[width=.98\textwidth]{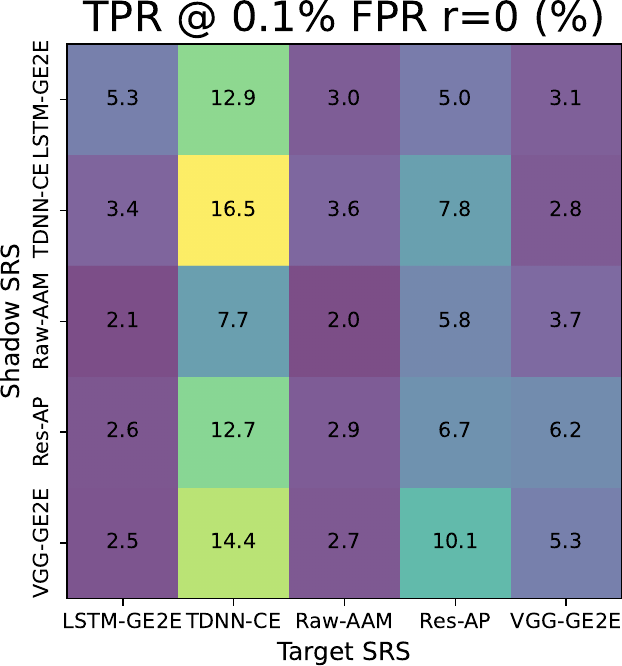}
  \end{minipage}\vspace*{-1mm}
  \caption{Effect of the architectures.}\vspace*{-3mm}
  \label{fig:archi}
\end{figure}

\noindent\textbf{Disjoint SRS architectures.}
We relax the assumption in \cref{sec:threat-model}
that the adversary knows the architecture of the target SRS and adopt the same architecture for the shadow SRS.
We consider all the $5\times 5$ pairs of the five architectures given in \tablename~\ref{tab:srs-info},
all of which are trained using the dataset VoxCeleb-2.
The results are shown in \figurename~\ref{fig:archi} and Appendix~\ref{sec:more-results-ablation-archi}.
Aligning with prior works~\cite{loss-trajectory,membership-first,FACE-AUDITOR},
the architectures of the target and shadow SRSs do affect the effectiveness of \attackname.
\attackname often achieves the best performance when the shadow SRS shares the same architecture
with the target SRS, especially in terms of accuracy.
However, interestingly, we find that the attack using a different shadow SRS architecture from the target SRS
may achieve higher TPR at 0.1\% FPR than the attack using the same shadow SRS architecture as the target SRS,
e.g., when the architecture of the target SRS is RawNet3, ResNetSE34V2, and VGGVox40.
Nevertheless, \attackname always achieves at least 60\% accuracy and 2\% TPR at 0.1\% FPR when $r=0$
which is more challenging than $r=1$.

\section{Discussion}\label{sec:discussion}
We discuss the attack practicability of \attackname, possible countermeasures against \attackname
as well as improvements and extensions of \attackname.

\noindent {\bf Attack practicability.} 
\attackname works for deployed SRS services. Firstly, the dataset distributions and architectures of deployed services are unknown. SLMIA-SR remains effective in this case (cf.~\cref{sec:ablation-study}).  
Secondly, SLMIA-SR only requires deployed services to return scores, which is reasonable~\cite{chen2019real, QFA2SR, Occam}, e.g., Microsoft Azure~\cite{microsoft-azure-vpr} and iFlytek SRSs~\cite{QFA2SR}. 
We have confirmed the effectiveness of \attackname against commercially deployed services in \cref{sec:add_eva}. 
Although few deployed services (e.g., Jingdong~\cite{QFA2SR}) hide scores, which can prevent our attack, this also reduces useful information available to benign developers (cf. discussion of countermeasures). 

\noindent {\bf Countermeasures.}
Both training and inference phase defenses could be utilized to prevent our attack \attackname.

{\it Training phase defenses.}
Since the effectiveness of \attackname lies in the overfitting of the target SRS,
techniques that alleviate overfitting could be used to defeat our attack,
e.g., early-stopping~\cite{EncoderMI}.
As shown in \figurename~\ref{fig:overfitting-level}, with the decrease in the number of training epochs of the target SRS,
the overfitting level and attack effectiveness decrease, consistent with the finding in \cite{EncoderMI}, while the testing EER increases.
This suggests that early-stopping, i.e., training the target SRS with less number of epochs,
can be used to prevent \attackname, at the cost of sacrificing the speaker recognition utility.

Another line of training phase defenses is differential privacy~\cite{first-principle, loss-trajectory, EncoderMI}.
In contrast to early-stopping which is an empirical defense,
differential privacy can provide provable membership privacy guarantees~\cite{EncoderMI}.
Differential privacy features a parameter called privacy budget $\epsilon$, the smaller, the more private.
To check feasibility, we adopt the differentially private stochastic gradient descent
(DP-SGD)~\cite{DP-SGD} to train a differential private SRS using the TDNN-CE and VoxCeleb-2 dataset, by first clipping the per-example gradient
to a pre-set maximal gradient norm $C$
and then adding noise to the aggregated gradient within a training batch
where the magnitude of the noise is positively correlated with a parameter $\sigma$.
We fix $C=5$ after tuning, and vary $\sigma$
from 0, 0.2, 0.5 to 0.8. The privacy budget $\epsilon$, performance of \attackname,
and  testing EER are reported in \tablename~\ref{tab:DP-SGD-result}.
We find that only clipping the per-example gradient with $C=5$ without adding noise ($\sigma=0$)
cannot prevent our attack.
While \attackname becomes almost a random guesser using DP-SGD with $\sigma=0.2$,
the testing EER increases by more than 10\%.
These results demonstrate that DP-SGD can prevent \attackname
at the cost of sacrificing the speaker recognition utility.

\begin{table}[t]
    \centering\setlength\tabcolsep{5pt}
    \caption{Results of DP-SGD.  ``Standard'' denotes training using SGD without differential privacy.}\vspace*{-2mm}
 \scalebox{1}{
      \begin{tabular}{ccccc}
     \hline
      \boldmath{}$\sigma$\unboldmath{} & \boldmath{}$\epsilon$\unboldmath{} & {\bf testing EER} & {\bf Accuracy} & {\bf TPR @ 0.1\% FPR} \\ \hline
    {\bf Standard} & $\infty$ & 6.56\% & 0.906 & 26.3\% \\ \hline
      {\bf 0} & $\infty$ & 7.88\% & 0.842 & 17.6\%\\
      {\bf 0.2} & 398 & 17.31\% & 0.521 & 0.4\% \\
      {\bf 0.5} & 3.87 & 28.03\% & 0.506 & 0.1\% \\
      {\bf 0.8} & 0.614 & 32.61\% & 0.502 & 0.1\% \\
     \hline
      \end{tabular}
    }
    \label{tab:DP-SGD-result}
  \end{table}

{\it Inference phase defenses.}
Since \attackname in the black-box scenario requires recognition scores from a target SRS to compute features,
a straightforward inference phase defense
is to hide scores and only output the recognition results, i.e., ``accept'' or ``reject''
for speaker verification, and the identified speaker or ``reject'' for speaker identification.
However, such defense reduces the useful information for third-party developers,
e.g., without scores, they cannot fine-tune the threshold of SRSs
to achieve better recognition performance in their specific applications.
It may also be circumvented
by decision-only membership inference attacks~\cite{label-only-MIA-adver-attack}.

\noindent {\bf Removing shadow SRS training.}
Due to the lack of knowledge about the training and non-training speakers of the target SRS,
\attackname trains a shadow SRS based on which an attack model is built in a supervised manner.
When some  non-training speakers of the target SRS are available to the adversary,
the adversary can either tune the threshold to satisfy a specific false positive rate
or training an out-of-distribution classifier, thus removing the requirement of shadow SRS training,
such as the attack in \cite{self-supervised-speech-membership}.
A possible way to obtain such non-training speakers is artificial speaker generation
with text-to-speech~\cite{SV2TTS-github}, similar to the random image generation~\cite{data-model-independent-MIA}.
The adversary first randomly samples an embedding in the voice embedding space,
and then synthesizes a voice from that embedding.
Due to the remarkably large voice embedding space,
the synthesized speaker is highly likely to be a non-training speaker of the target SRS.

\noindent {\bf Decision-only MIA.}
\attackname in the black-box scenario requires recognition scores from a target SRS to compute features.
However, in practice, some SRSs may only output recognition results instead of recognition scores,
e.g., ``accept'' or ``reject'' for speaker verification.
The adversary can utilize adversarial attacks to achieve decision-only membership inference~\cite{label-only-MIA-adver-attack}.
The assumption is that the voices of training speakers are more robust against adversarial perturbation
than that of non-training speakers. Hence, the adversary can exploit the average minimal perturbation
to change the recognition result of each voice of the target speaker as the feature
to perform speaker-level membership inference.
We leave this interesting problem for future work.

\noindent {\bf Extending to other biometric recognition.}
Speaker recognition is one kind of biometric recognition that recognizes a person's identity from biometric data.
Other biometric recognition, e.g., face and fingerprint recognition,
share similar pipelines (training, enrollment, and recognition phases),
architecture (the final fully connected layer is removed after training),
training paradigms (classification-based or verification-based losses),
and training objectives (pulling the data of the same subject together and pushing different subjects away)
with speaker recognition~\cite{MASBZ23}. Therefore, \attackname may be extended to other biometric recognition,
which is worthy to explore in the future.

\section{Conclusion}\label{sec:conclusion}

We proposed \attackname, the \emph{first} membership inference attack against speaker recognition.
Instead of considering voice-level membership inference that determines whether \emph{some} given voices were contained in the training of a target SRS,
\attackname features speaker-level membership inference
to determine whether \emph{any} voices of a target speaker were contained in the training of a target SRS,
where the voices of training speakers used for membership inference are not required to get involved in the training of the target SRS.
We showed that prior MIA designed for embedding models are unsatisfactory to SRSs.
Thus, we designed and studied a large number of features to characterize the differences between training and non-training speakers,
introduced a mixing ratio training strategy to improve the generalizability of attack models,
voice-number-dependent attack models and voice chunk splitting to enhance
attack effectiveness,  group enrollment and enrollment voice concatenation techniques
to reduce queries probed to the target SRS in the black-box scenario.
Extensive experiments demonstrate the effectiveness of \attackname, the proposed techniques and features.
Our work sheds light on future research in the area of private speaker recognition.

\section*{Acknowledgments}
We thank the anonymous reviewers for their constructive feedbacks. 
This work is supported by the National Natural Science Foundation of China (NSFC) under Grant No.\ 62072309,  CAS Project for Young Scientists in Basic Research (YSBR-040), and ISCAS New Cultivation Project (ISCAS-PYFX-202201).

\bibliographystyle{IEEEtran}
\bibliography{ref}


\appendix

\subsection{\attackname vs. Prior MIA on SR embedding models}\label{sec:oursvspriorMIA}
In \tablename~\ref{tab:comp}, we compare our attack \attackname with
four recent promising MIA designed for embedding models: Li et al.~\cite{person-re-identification-MIA},
Tseng et al.~\cite{self-supervised-speech-membership},
EncoderMI~\cite{EncoderMI},
and FaceAuditor~\cite{FACE-AUDITOR}, regarding the focused tasks, MIA level, feature, adversarial capacity, attack model, etc.
We also apply and generalize those MIA to speaker-level membership inference against speaker recognition and compare their performance with \attackname.

\begin{itemize}
  \item {\bf Li et al.~\cite{person-re-identification-MIA}:} it utilized the intra-features $\Theta_{\tt c}^{\tt avg}$ and $\Theta_{\tt p}^{\tt avg}$
  as inputs to a binary classifier-based attack model.
  \item {\bf Tseng et al.~\cite{self-supervised-speech-membership}:} it utilized the intra-feature $\Theta_{\tt p}^{\tt avg}$ for a threshold-based attack model
  and improved the attack by replacing the pre-defined cosine similarity in $\Theta_{\tt p}^{\tt avg}$ with the similarity produced by a neural network. We report the better result among the basic and the improved attacks.
  \item {\bf EncoderMI~\cite{EncoderMI}:} it proposed three different attack models, EncoderMI-T, EncoderMI-V, and EncoderMI-S.
  We consider the two most effective variants: EncoderMI-T and EncoderMI-V.
  EncoderMI-T is a threshold-based attack model that utilizes the intra-feature $\Theta_{\tt p}^{\tt avg}$,
  and EncoderMI-V is a classifier-based attack model that utilizes the sorted set of similarities contributing to $\Theta_{\tt p}^{\tt avg}$.
  \item {\bf FaceAuditor~\cite{FACE-AUDITOR}:} it proposed classifier-based attack models that utilize either the similarities contributing to feature $\Theta_{\tt p}^{\tt avg}$ (denoted by FaceAuditor-S)
  or the similarities contributing to $\Theta_{\tt c}^{\tt avg}$ and $\Phi_{\tt vc}^{\tt avg}$ (denoted by FaceAuditor-P/R).
\end{itemize}

\begin{table*}[htbp]
  \centering
  \caption{Comparison of \attackname with the membership inference attacks on embedding models.
  }
  \resizebox{1.0\textwidth}{!}{\begin{threeparttable}
    \begin{tabular}{c|c|c|c|c|c|c|c|c|c}
    \hline
    \multirow{2}[4]{*}{} & \multirow{2}{*}{\textbf{Task}} & \multirow{2}{*}{\textbf{Level}} & \multicolumn{2}{c|}{\textbf{Feature}} & {\bf Adversarial} & {\textbf{Attack}} & {\bf Shadow} & \textbf{Non-Member} & \textbf{TPR@}  \\
\cline{4-5}          &       &       & \multicolumn{1}{c|}{\textbf{Aspect}} & \multicolumn{1}{l|}{\textbf{Num}} &  {\bf Capacity} & {\bf Model} & {\bf Model}    &  {\bf Set of Target}     &   {\bf 0.1\% FPR} \\
    \hline
    \textbf{LRL-MIA~\cite{person-re-identification-MIA}} & \makecell[c]{person \\ re-identification} & user  & intra-similarity & 2   & embedding  & classifier  & \cmark & \xmark & 1.5\% \\  
    \hline
    \multicolumn{1}{c|}{\textbf{EncoderMI-T~\cite{EncoderMI}}$^\dag$} & {contrastive} & \multirow{2}{*}{example} & \multirow{2}{*}{intra-similarity} & \multirow{2}{*}{1$^\sharp$} & \multirow{2}{*}{embedding} & threshold & \multirow{2}{*}{\cmark} & \multirow{2}{*}{\xmark} & 1.6\% \\  
\multicolumn{1}{c|}{\textbf{EncoderMI-V~\cite{EncoderMI}}$^\dag$} &    learning   &       &       &       &    & classifier  &      & & 2.0\% \\ 
    \hline
    \textbf{TLK-MIA~\cite{self-supervised-speech-membership}}$^\S$ & \makecell[c]{self-supervised \\ speech} & \makecell[c]{utterance \\ speaker$^{\natural}$} & intra-similarity & 1   &  embedding & threshold  & \xmark & \cmark & 1.6\% \\  
    \hline
    \multicolumn{1}{c|}{\textbf{FaceAuditor-S~\cite{FACE-AUDITOR}}$^\ddag$} & \multirow{2}[3]{*}{\makecell[c]{face \\ recognition}} & \multirow{2}[4]{*}{user} & intra-similarity & 1$^\sharp$   &  \multirow{2}[4]{*}{score} & \multirow{2}[3]{*}{classifier}  & \multirow{2}[4]{*}{\cmark} & \multirow{2}[4]{*}{\xmark} & 1.2\% \\  
\multicolumn{1}{c|}{\textbf{FaceAuditor-P/R~\cite{FACE-AUDITOR}}$^\ddag$} &       &       & \makecell[c]{intra-similarity \\ \& inter-dissimilarity} & 2$^\sharp$   &  &      &  & &  1.6\% \\ 
    \hline
    \makecell[c]{{\bf \attackname} \\ (\textbf{Ours})} & \makecell[c]{speaker \\ recognition} & speaker & \makecell[c]{intra-similarity \\ \& inter-dissimilarity} & \makecell[c]{103}   & \makecell[c]{embedding \\ \& score} & classifier & \cmark & \xmark & 33.5\% \\ 
    \hline
  \end{tabular}%
  \begin{tablenotes}
  \scriptsize
    \item Note:
    (1) $\dag$: We compared with the most two effective versions of EncoderMI~\cite{EncoderMI}, excluding EncoderMI-S.
    (2) $\S$: We compared with the better one between the basic attack and the improved attack of \cite{self-supervised-speech-membership}.
    (3) $\ddag$: FaceAuditor~\cite{FACE-AUDITOR} was available online when we were preparing this manuscript.
    (4) ``Level'': Example- and utterance-level predicts whether a given example is used for training,
    while user- and speaker-level determines whether \emph{any} example of a given user/speaker is involved in training.
    $\natural$: Although \cite{self-supervised-speech-membership} designed speaker-level MIA, in their evaluation, the voices provided for MI of training speakers come from training voices,
    thus deviating from our definition of speaker-level MIA which should reliably reach the ``member'' decision for a training speaker even if all of her/his voices provided for MI are different from training voices.
    (5) ``Aspect'': the properties that features quantify.
    Prior works covered three unique features ($\Theta_{\tt c}^{\tt avg}$, $\Theta_{\tt p}^{\tt avg}$, $\Phi_{\tt vc}^{\tt avg}$),
    which are the strict subset of our 103 features. $\sharp$: EncoderMI-V and FaceAuditor used the raw set of similarities contributing to the features as input to attack models.
    (6) ``embedding'': the adversary can obtain the embedding of any input. ``score'': the adversary can only obtain the output recognition score.
    (7) ``classifier'': binary classifier-based attack models. ``threshold'': attack models predicting by thresholding a pre-set threshold.
    (8) ``Shadow Model'': whether the adversary has to train a shadow model to build the attack model.
    (9) ``Non-Member Set of Target'': whether the adversary is aware of a dataset in which each data is the non-member of the target model.
    With this knowledge, \cite{self-supervised-speech-membership} did not need to train a shadow model.
    (10) ``TPR@0.1\% FPR'' denotes the true positive rate when the false positive rate is 0.1\%. The reported TPR are obtained on target SRS GE2E-LSTM and dataset VoxCeleb-2.
    More results refer to \tablename~\ref{tab:overall-comp-s1} and \tablename~\ref{tab:overall-comp-s2}.
  \end{tablenotes}
\end{threeparttable}
  }
  \label{tab:comp}%
\end{table*}%

\subsection{Comparison between the white-box and black-box scenarios regarding the centroid-centroid inter-features}\label{sec:c-c-w-b-comp}
\begin{table}
  \centering
  \caption{Performance comparison between white-box and black-box scenarios regarding the centroid-centroid inter-features.}
  \begin{tabular}{c|c|c|c|cc}
    \hline
    \multicolumn{2}{c|}{\multirow{2}{*}{}} & \multirow{2}{*}{\bf Accuracy} & \multirow{2}{*}{\bf AUROC} & \multicolumn{2}{c}{\bf TPR @ x\% FPR} \\
    \multicolumn{1}{c}{} & & & & {\bf x=0.1} & {\bf x=1} \\
    \hline
    \multirow{1}{*}{\bf All} & {\bf White-box} &  0.753 & 0.834 & 4.0\% & 14.7\% \\
    \cline{2-6}
    \multirow{1}{*}{\bf features} & {\bf Black-box} &  0.749 & 0.819 & 5.1\%& 14.8\% \\
    \hline
    \multirow{1}{*}{\bf Centroid-centroid} & {\bf White-box} &  0.636 & 0.682 & 2.5\% & 10.4\% \\
    \cline{2-6}
    \multirow{1}{*}{\bf inter-features} & {\bf Black-box} &  0.648 & 0.714 & 2.3\% & 9.3\% \\
    \hline
  \end{tabular}
  \label{tab:c-c-w-b-comp}
\end{table}

As mentioned in \cref{sec:feature-extractor}, our feature extractor demonstrates unity in white-box and black-box scenarios.
The minor difference lies in the computation of the centroid-centroid inter-features.
In the black-box scenario, to compute such group of features, we apply enrollment voice concatenation
to obtain one longer and concatenated voice for each imposter with multiple voices.
Here we check whether this difference leads to a significant performance gap between the two scenarios.
We adopt the LSTM-GE2E target SRS and the dataset VoxCeleb-2, set the number of voices of target speakers $N=10$,
the number of imposters $M=20$ and the number of voices per imposter $K=10$, the same as Setting-1 in \cref{sec:evaluation}.
We consider two cases, namely, all features are used and only the centroid-centroid inter-features are used.
The results are shown in \tablename~\ref{tab:c-c-w-b-comp}.
We find that there is no obvious performance gap between white-box and black-box scenarios,
and sometimes the attack even performs better in the black-box scenario than in the white-box scenario.
It is because for each imposter, the embedding of the concatenated voice (used in the black-box scenario)
can well approximate the centroid embedding of the voice embeddings (used in the white-box scenario).

\begin{algorithm}[h]
  \footnotesize
    \caption{Determining an upper bound $N'$}
    \label{al:decide-attack-model-num}
    \KwIn{shadow SRS $SR^{s}$;
           datasets $\boldsymbol{\mathcal{V}} = \{\mathcal{V}^{s}_{tr},
          \mathcal{V}^{s}_{ntr,tr}, \mathcal{V}^{s}_{ntr,ntr}\}$;
          features $\Psi=\{\cdots,\psi_i,\cdots\}$;  \#voice step $s$; T-test significance value $\alpha$}
    \KwOut{an upper bound $N'$}
    $\mathcal{C} \gets \emptyset$ \\
    \For{$\psi \in \Psi$}{
      \For{$\mathcal{V}\in \boldsymbol{V}$}{
        $n_1\gets 1$;       $n_2\gets 1+s$ \\
          \While{True}{
          $\mathcal{D}_1 \gets$ $\psi$ of $\mathcal{V}$ on $SR^s$ with $n_1$ voices per speaker \\
            $\mathcal{D}_2 \gets$ $\psi$ of $\mathcal{V}$ on $SR^s$ with $n_2$ voices per speaker \\
            $\eta \gets $ T-test($\mathcal{D}_1$, $\mathcal{D}_2$) \textcolor{blue}{\Comment{compute $p$-value}} \\
            \If(\textcolor{blue}{\Comment{$n_1$ attack models suffice for $\psi$ on $\mathcal{V}$}}){$\eta \geq \alpha$}{
              $\mathcal{C} \gets \mathcal{C} \cup \{n_1\}$ \\
              {\bf break}
            }
            $n_1\gets n_1+1$;   $n_2\gets n_2+1$ \\
        }
      }
    }
    \Return{maximum of $\mathcal{C}$}
  \end{algorithm}
  
\subsection{Algorithm to bound the number of attack models}\label{sec:bound-num-attack-models}
Alg.~\ref{al:decide-attack-model-num} iterates the given set $\Psi$ of features
(single feature for threshold-based attack model and multiple features for classifier-based attack model)
and three datasets ($\mathcal{V}^{s}_{tr}$, $\mathcal{V}^{s}_{ntr,tr}$, and $\mathcal{V}^{s}_{ntr,ntr}$).
For each feature $\psi\in\Phi$ and dataset $\mathcal{V}$,
it first computes two sets of features $D_1$ and $D_2$ from $n_1$ and $n_2$ randomly sampled voices per speaker
from the dataset $\mathcal{V}$, where $n_1$ and $n_2$ are initialized by $1$ and $1+s$ with the given step size $s$.
Then, it computes the $p$-value $\eta$ by performing a two-samples T-test~\cite{two-sample-T-test} on $D_1$ and $D_2$
with the null hypothesis $H_0$: \textit{$D_1$ and $D_2$ have the same mean}.
If the $p$-value is no smaller than the pre-set significance value $\alpha$,
we accept $H_0$, indicating that there is no obvious statistic difference
of the feature $\psi$ between $N=n_1$ and $N=n_2$, hence it suffices to build $n_1$ attack models for the feature $\psi$ on the dataset $\mathcal{V}$.
In this case, $n_1$ is recorded in the set $\mathcal{C}$.
Otherwise, we increase $n_1$ and $n_2$ by $1$ and repeat the above T-test until
$H_0$ is accepted. Finally, Alg.~\ref{al:decide-attack-model-num} returns the maximal number in the set
$\mathcal{C}$  as the upper bound $N'$.

\subsection{Results of the Effect of the Number of Imposters and Imposter Voices}\label{sec:more-result-imposter-num-imposter-voice}
\begin{figure}
    \begin{minipage}{0.23\textwidth}
      \centering
      \includegraphics[width=1\textwidth]{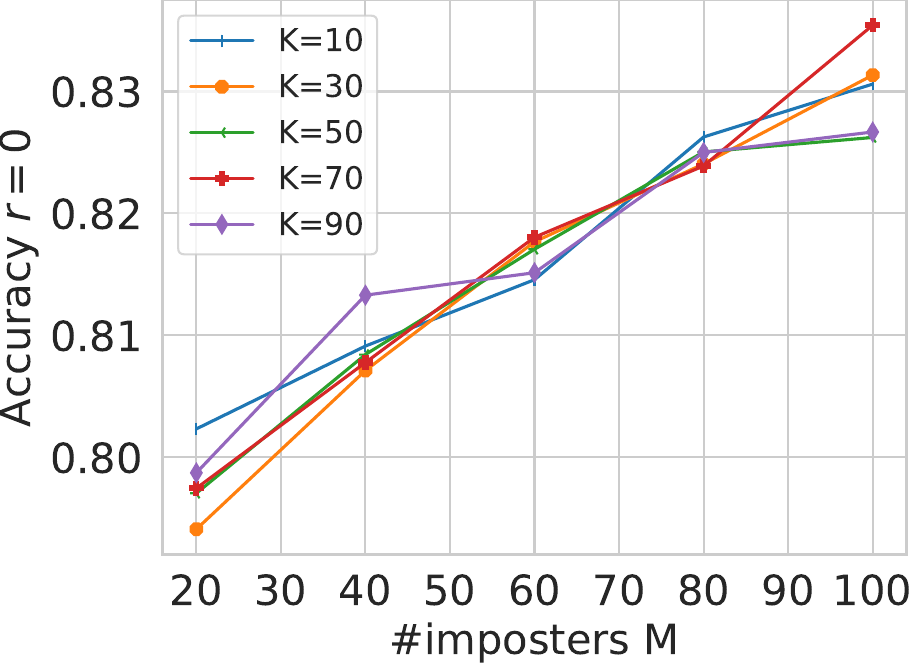}
    \end{minipage}
    \hfill
    \begin{minipage}{0.23\textwidth}
        \centering
        \includegraphics[width=1\textwidth]{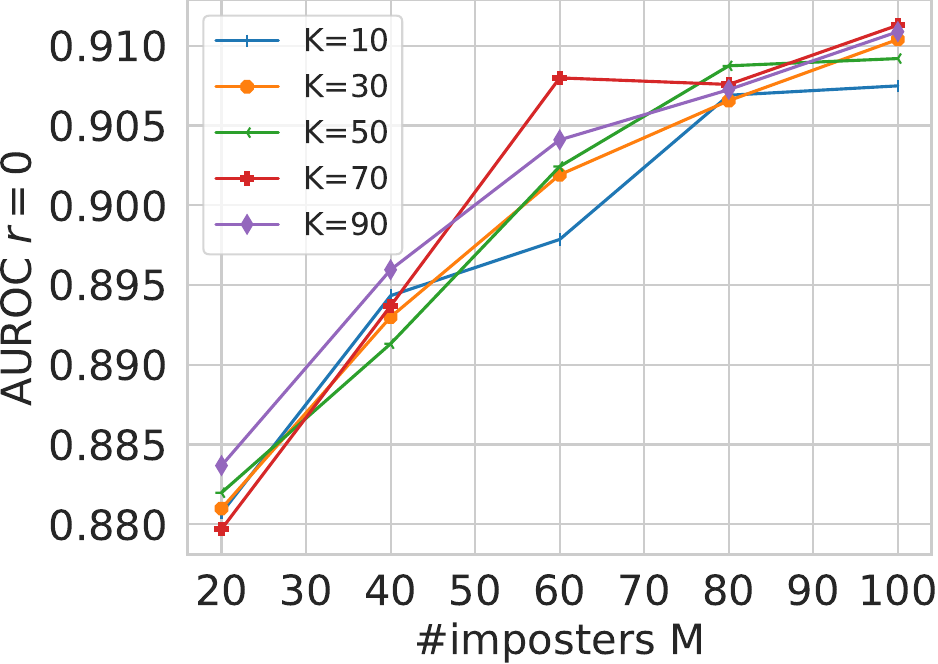}
      \end{minipage}
      \hfill
    \begin{minipage}{0.23\textwidth}
      \centering
      \includegraphics[width=1\textwidth]{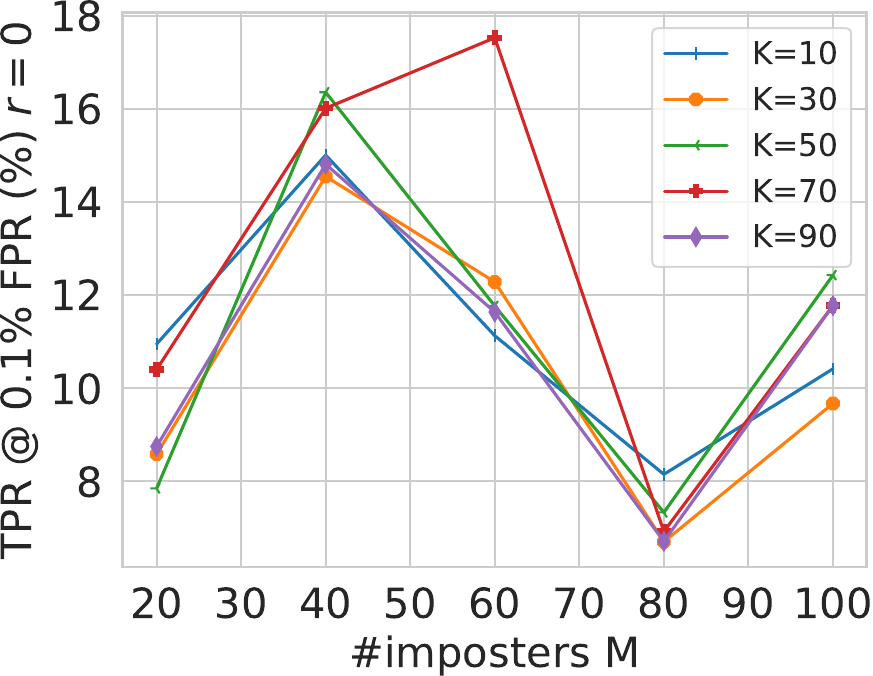}
    \end{minipage}
    \hfill
    \begin{minipage}{0.23\textwidth}
        \centering
        \includegraphics[width=1\textwidth]{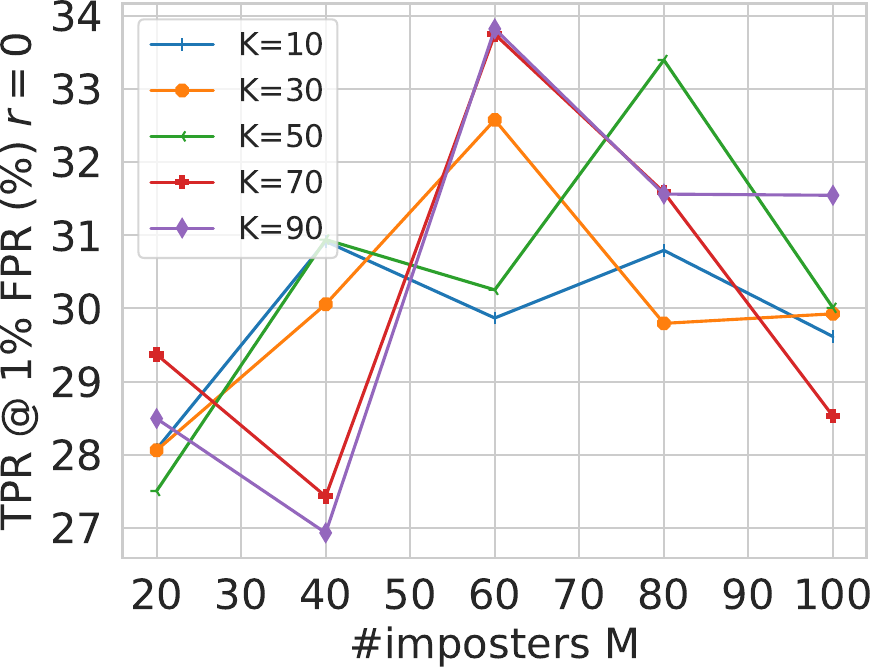}
      \end{minipage}
    \caption{Effects of the number $M$ of imposters and the number $K$ of voices per imposter on the performance of Inter-Ens.}
    \label{fig:impact-imposter-num-voice-num-more-result}
  \end{figure}

\figurename~\ref{fig:impact-imposter-num-voice-num-more-result} demonstrates the effects
of the number $M$ of imposters and the number $K$ of voices per imposter on the effectiveness of \attackname with the attack models trained using all inter-features (i.e., Inter-Ens).
Intra-features are omitted as they do not require imposters.

\noindent {\bf Effect of the number of imposters.}
We evaluate the effect of the number $M$ of imposters by varying $M$ from 20 to 100 with step 20.
The number $N$ of voices per target speaker is set to 40 in attack model training and membership inference, as
\attackname almost converges at $N=40$ in terms of accuracy and AUROC (cf. \figurename~\ref{fig:num-voice-exper}).
We only use VND attack models trained with our mixing training strategy.

We find that accuracy (as well as AUROC) generally increases with the number $M$ of imposters, because more imposters
make the statistics of the sets of distances more precise,
leading to more discriminative inter-features.
However, TPR at 0.1\% FPR does not monotonically increase with the number $M$ of imposters,
indicating that using more imposters improves accuracy and AUROC at the cost of more queries,
but is not necessarily helpful for TPR.

\noindent {\bf Effect of the number of imposters' voices.}
To understand the effect of the number of imposters' voices,
we vary the number $K$ of voices per imposter from 10 to 90 with step 20,
while the other settings are the same as above.
We surprisingly find that both accuracy and TPR at 0.1\% FPR do not monotonically increase
with the number $K$.
For example, when $M=20$, $K=10$ yields higher accuracy and TPR at 0.1\% FPR than $K>10$.

  \subsection{More Results of Ablation Study on the Disjoint Datasets}\label{sec:more-results-ablation-dataset}
\begin{figure}
    \begin{minipage}{0.23\textwidth}
        \centering
        \includegraphics[width=1\textwidth]{figure/dataset-distribution-Accuracy-r=0-3.pdf}
      \end{minipage}
      \hfill
      \begin{minipage}{0.23\textwidth}
        \centering
        \includegraphics[width=1\textwidth]{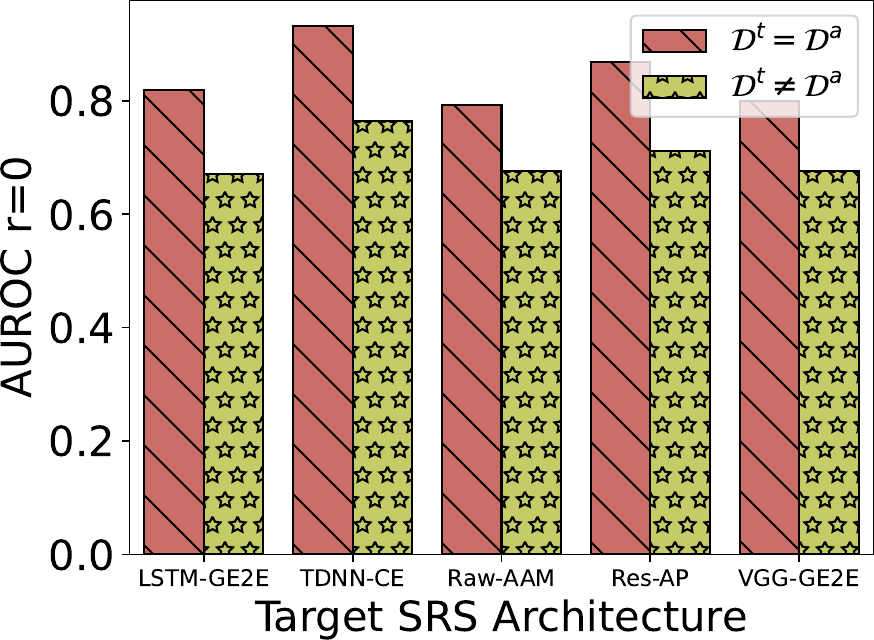}
      \end{minipage}
      \hfill
      \begin{minipage}{0.23\textwidth}
        \centering
        \includegraphics[width=1\textwidth]{figure/dataset-distribution-TPR-0-1-FPR-r=0-3.pdf}
      \end{minipage}
      \hfill
      \begin{minipage}{0.23\textwidth}
        \centering
        \includegraphics[width=1\textwidth]{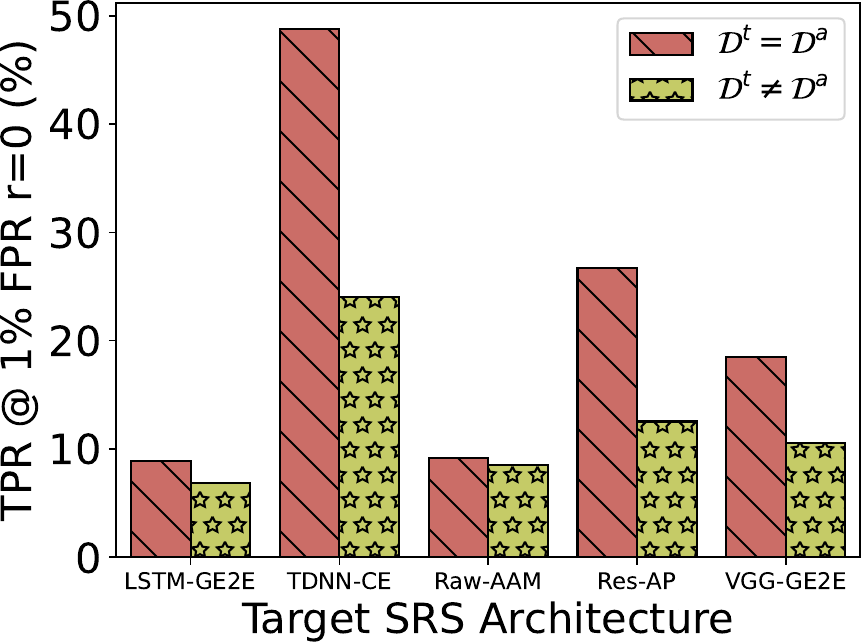}
      \end{minipage}
    \caption{Effect of the dataset distribution of the target and shadow SRSs
    on the effectiveness of \attackname.}
    \label{fig:more-results-ablation-dataset}
  \end{figure}

  The results are shown in \figurename~\ref{fig:more-results-ablation-dataset}.
  We observe that \attackname still achieves good performance with at least 2\% TPR at 0.1\% FPR and 60\% accuracy when $r=0$,
a more challenging setting than $r=1$.
Aligning with previous works~\cite{loss-trajectory,membership-first},
the effectiveness of \attackname decreases with a dataset distribution shift in most cases,
probably because training datasets with different distributions make the target and shadow SRSs learn different speaker embedding mappings.

  \subsection{More Results of Ablation Study on Disjoint Architectures}\label{sec:more-results-ablation-archi}
\begin{figure}
    \centering
    \begin{minipage}{0.23\textwidth}
      \centering
      \includegraphics[width=1\textwidth]{figure/archi-Accuracy-r=0-3.pdf}
    \end{minipage}
    \hfill
    \begin{minipage}{0.23\textwidth}
        \centering
        \includegraphics[width=1\textwidth]{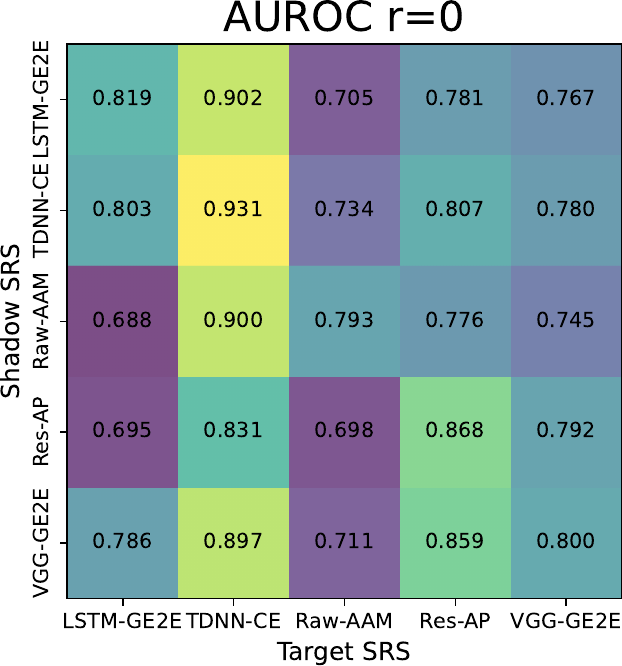}
      \end{minipage}
      \hfill
      \begin{minipage}{0.23\textwidth}
        \centering
        \includegraphics[width=1\textwidth]{figure/archi-TPR-0-1-FPR-r=0-3.pdf}
      \end{minipage}
      \hfill
      \begin{minipage}{0.23\textwidth}
        \centering
        \includegraphics[width=1\textwidth]{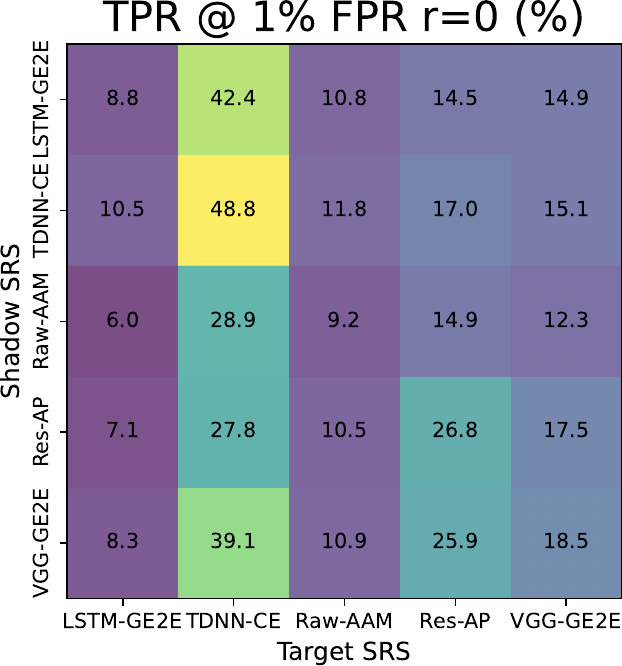}
      \end{minipage}
    \caption{Effect of the architectures of the target and shadow SRSs on the effectiveness of \attackname.}
    \label{fig:more-results-ablation-archi}
  \end{figure}

  The results are reported in \figurename~\ref{fig:more-results-ablation-archi}.
Aligning with previous works~\cite{loss-trajectory,membership-first,FACE-AUDITOR},
the architectures of the target and shadow SRSs do affect the effectiveness of \attackname.
 In general, \attackname achieves the best attack performance when the shadow SRS shares the same architecture
with the target SRS, especially in terms of the average accuracy and AUROC.
However, interestingly, we find that the attack using a different shadow SRS architecture from the target SRS
may achieve higher TPR at 0.1\% FPR and 1\% FPR than the attack using the same shadow SRS architecture as the target SRS,
e.g., when the architecture of the target SRS is Raw-AAM, Res-AP, and VGG-GE2E.
Nevertheless, \attackname always achieves at least 60\% accuracy and 2\% TPR at 0.1\% FPR when $r=0$
which is more challenging than $r=1$.

\end{document}